\documentclass[12pt]{article}
\pdfoutput=1
\usepackage{amssymb}
\usepackage{amsmath}
\usepackage{latexsym}
\usepackage[usenames]{color}
\usepackage{fancybox}
\usepackage{simplewick}
\usepackage{comment}
\usepackage{cite}
\usepackage{framed}
\definecolor{shadecolor}{rgb}{0.9,0.9,0.95}
\usepackage{setspace}
\usepackage[normalem]{ulem}
\definecolor{darkgreen}{rgb}{0,0.5,0}
\definecolor{darkblue}{cmyk}{0.9,0.9,0,0}
\definecolor{darkred}{rgb}{0.6,0,0.3}

\usepackage{graphicx}
\usepackage{tikz}
\usepackage[setpagesize=false,pagebackref=false, linktocpage, bookmarksopen=true, colorlinks=true, linkcolor=darkblue,citecolor=darkblue,urlcolor=black]{hyperref}
\usepackage{hyperref}

\newcommand{\id}{{\bf 1}}
\newcommand{\tr}{{\rm tr}}
\renewcommand{\Im}{{\rm Im}}
\renewcommand{\Re}{{\rm Re}}
\renewcommand{\thefootnote}{\arabic{footnote}}

\def\eqref#1{(\ref{#1})}

\def\beq{\begin{equation}}
\def\eeq{\end{equation}}

\newcommand{\rD}{\mathrm{D}}

\newcommand{\rF}{\mathrm{F}}

\newcommand{\rN}{\mathrm{N}}

\newcommand{\rT}{\mathrm{t}}

\numberwithin{equation}{section}
\begin{document}
\thispagestyle{empty}

\renewcommand{\thefootnote}{\fnsymbol{footnote}}
\setcounter{page}{1}
\setcounter{footnote}{0}
\setcounter{figure}{0}
%%%%%%%%%%%%%%%%%%%%%%%%%%%%%%%%%%%%%%%%%%%%%%%%%%%%%%%%%%%%%%%%%%%%%%%%%%%%%%%%%%%%%%%%%%%%%%%%%%%
\begin{flushright}
{\tt CERN-TH-2020-019, UMTG-303, USTC-ICTS/PCFT-20-04
}
\end{flushright}
\vspace{0.7cm}
\begin{center}
\Large{\textbf{Cylinder partition function of
the 6-vertex model from algebraic geometry}}

\vspace{1.3cm}

\normalsize{\textrm{Zoltan Bajnok$^1$, Jesper Lykke
Jacobsen$^{2,3,4}$, Yunfeng Jiang$^{5}$,\\ Rafael I. Nepomechie$^{6}$, Yang Zhang$^{7,8}$}}
\\ \vspace{1cm}
\footnotesize{\textit{
$^{1}$Wigner Research Centre for Physics, Konkoly-Thege Mikl\'os u. 29-33, 1121 Budapest, Hungary
\\
$^{2}$Institut de Physique Th\'eorique, Paris Saclay, CEA, CNRS, 91191 Gif-sur-Yvette, France
\\
$^{3}$Laboratoire de Physique de l'\'Ecole Normale Sup\'erieure, ENS, Universit\'e PSL, \\
CNRS, Sorbonne Universit\'e, Universit\'e de Paris, F-75005 Paris, France \\
$^{4}$Sorbonne Universit\'e, \'Ecole Normale Sup\'erieure, CNRS, \\
Laboratoire de Physique (LPENS), F-75005 Paris, France\\
$^{5}$CERN Theory Department, Geneva, Switzerland\\
$^{6}$Physics Department, P.O. Box 248046, University of Miami, Coral
Gables, FL 33124 USA\\
$^{7}$Peng Huanwu Center for Fundamental Theory, Hefei, Anhui 230026, China\\
$^{8}$Interdisciplinary Center for Theoretical Study, University of Science and Technology of China,
Hefei, Anhui 230026, China
}
\vspace{1cm}
}

\par\vspace{1.0cm}

\textbf{Abstract}\vspace{2mm}
\end{center}
\noindent
We compute the exact partition function of the isotropic 6-vertex
model on a cylinder geometry with free boundary conditions,
for lattices of intermediate size, using
Bethe ansatz and algebraic geometry.  We perform the computations in
both the open and closed channels.  We also consider the partial
thermodynamic limits, whereby in the open (closed) channel,
the open (closed) direction is kept small while the other direction
becomes large. We compute the zeros of the partition function in the two
partial thermodynamic limits, and compare with the condensation
curves.
\setcounter{page}{1}
\renewcommand{\thefootnote}{\arabic{footnote}}
\setcounter{footnote}{0}
\setcounter{tocdepth}{2}
\newpage
\tableofcontents

\section{Introduction}
Computing partition functions of integrable vertex models at
\emph{intermediate} lattice size is a hard problem.  For small lattice
size, the partition function can be computed simply by brute force.
For large lattice size, where the thermodynamic limit is a good
approximation, various methods are available, including the Wiener-Hopf
method \cite{VegaWoynarovich85, Woynarovich87, Hamer:1987ei}, non-linear integral equations
\cite{PearceKlumper91,DestriDeVega94} and a distribution approach \cite{GranetJacobsenSaleur18}.
At intermediate lattice size, brute force is no longer an option, and the
thermodynamic approximation is inaccurate.  In a previous work
\cite{Jacobsen:2018pjt}, three of the authors developed an efficient
method to compute the \emph{exact} partition function of the 6-vertex
model \emph{analytically} for intermediate lattice size.  They
considered the 6-vertex model at the isotropic point on the torus,
\emph{i.e.} with \emph{periodic} boundary conditions in both
directions.  The method is based on the rational $Q$-system
\cite{Marboe:2016yyn} and computational algebraic geometry (AG).  The
algebro-geometric approach to Bethe ansatz was initiated in
\cite{Jiang:2017phk}, with the general goal of exploring the structure
of the solution space of Bethe ansatz equations (BAE) and developing
new methods to obtain analytic results in integrable models.  The
simplest example for such a purpose is the BAE of the
$SU(2)$-invariant Heisenberg XXX spin chain with periodic boundary
conditions.  It is an interesting question to generalize these methods
to more sophisticated cases such as higher-rank spin chains, quantum
deformations and non-trivial boundary conditions.

In the current work, we take one step forward in this direction and
consider the partition function of the 6-vertex model on the cylinder.
Namely, we take one direction of the lattice to be periodic and impose
free \emph{open} boundary conditions in the other direction.  This
set-up has several new features compared to the torus geometry
already considered in \cite{Jacobsen:2018pjt}.

First of all, to consider open boundary conditions for the vertex
model, we put the model on a diagonal square lattice where each
square is rotated by $45^{\circ}$, as is shown in
figure~\ref{fig:annulus}.
The partition function on such a lattice can be formulated in terms of
a \emph{diagonal-to-diagonal} transfer matrix \cite{Owczarek:1989},
which does \emph{not} commute for different values of the spectral
parameter. Nevertheless, the $R$-matrix approach (the so-called Quantum Inverse Scattering
Method) can be applied by reformulating this transfer matrix
in terms of an inhomogeneous \emph{double-row} transfer matrix \cite{Sklyanin:1988yz}
with suitable alternating inhomogeneities \cite{Destri:1991zm, Yung:1994td}.
It turns out that these inhomogeneities \emph{depend on the
spectral parameter}.  As a result, the BAE depend on a \emph{free
parameter}; hence, the Bethe roots are functions of this parameter, instead
of pure numbers.  In general,
this new feature makes it significantly more difficult to solve
the BAE. However, in the algebro-geometric approach, there is no
extra difficulty, because the computations are purely
algebraic and analytic --  there is not much qualitative difference
between manipulating numbers and algebraic expressions.  Therefore, the
AG computations can be adapted to cases with free parameters
straightforwardly, which further demonstrates the power of our method.

Secondly, in the torus case, the computation of the partition function
can be done in two directions which are equivalent.  For the cylinder
case, however, the computations of the partition function in the two
directions or channels are quite different.  In the open channel, we
need to diagonalize the transfer matrix corresponding to \emph{open}
spin chains.  The partition function is given by the sum over traces
of powers of the open-channel transfer matrix, similarly to the torus case.  In the
closed channel, we diagonalize transfer matrices corresponding to
\emph{closed} spin chains, and the open boundaries become non-trivial
boundary states.  The partition function is thus given by a matrix
element, between boundary states, of powers of the closed-channel
transfer matrix.

For a given lattice size, the final
results should be the same in both channels.  Nevertheless, we may
consider different limits.  In the open (closed) channel, we can take the
lattice size in the open (closed) direction to be finite and let the other
direction tend to infinity.  This \emph{partial thermodynamic} limit has
been studied in the torus case \cite{Jacobsen:2018pjt}.  In the
cylinder case, there are two
different partial thermodynamic limits (``long narrow straw'' and
``short wide pancake''), which we study in detail in this
paper.
In these limits, it follows from the Beraha-Kahane-Weiss theorem
\cite{BKW} that the zeros of the partition function condense on
certain curves in the complex plane of the spectral parameter.

The rest of the paper is structured as follows.  In
section~\ref{sec:setup}, we give the set-up of the vertex model and
its reformulation in terms of a diagonal-to-diagonal transfer matrix.
In sections~\ref{sec:open} and \ref{sec:close}, we discuss the
computation of the partition function in the two different channels,
using Bethe ansatz and algebraic geometry.  Section~\ref{sec:AG} is
devoted to some general discussions on the algebro-geometric
computations for the BAE/QQ-relation with a free parameter.  In
section~\ref{sec:analyticalResult} we present the partition functions
which can be written in closed forms for any $M$ or $N$ in the two
channels.  These include $M=1$ in the open channel and $N=1,2,3$ in
the closed channel.  In section~\ref{sec:zero}, we compute the zeros
of the partition function close to the two partial thermodynamic
limits (i.e., for very large aspect ratios) and compare them with the
condensation curves, which we compute from the transfer matrix
spectra. In appendix~\ref{app:AGbasic}, we give a brief introduction to the basic notions of computational algebraic geometry which are used in this paper. In appendix~\ref{app:AGdetail}, we give more details on the
algebro-geometric computations.  Appendix~\ref{app:overlap} to
\ref{app:parity} contain the proofs of some statements in the main
text.  We collect explicit exact results in
appendix~\ref{eq:exactsmallMN} for small $M$ and $N$, where we can
perform the computation in both channels and make consistency
checks.\par

Some of the results we obtained are too large to be presented in the
paper. They can also be downloaded from the webpage ($576$ MB in
compressed form):
\begin{quotation}
\url{http://staff.ustc.edu.cn/~yzhphy/integrability.html}
%\url{http://staff.ustc.edu.cn/~yzhphy/partition_function.zip}
\end{quotation}

\section{Set-up}
\label{sec:setup}
\begin{figure}[h!]
\begin{center}
\includegraphics[scale=0.4]{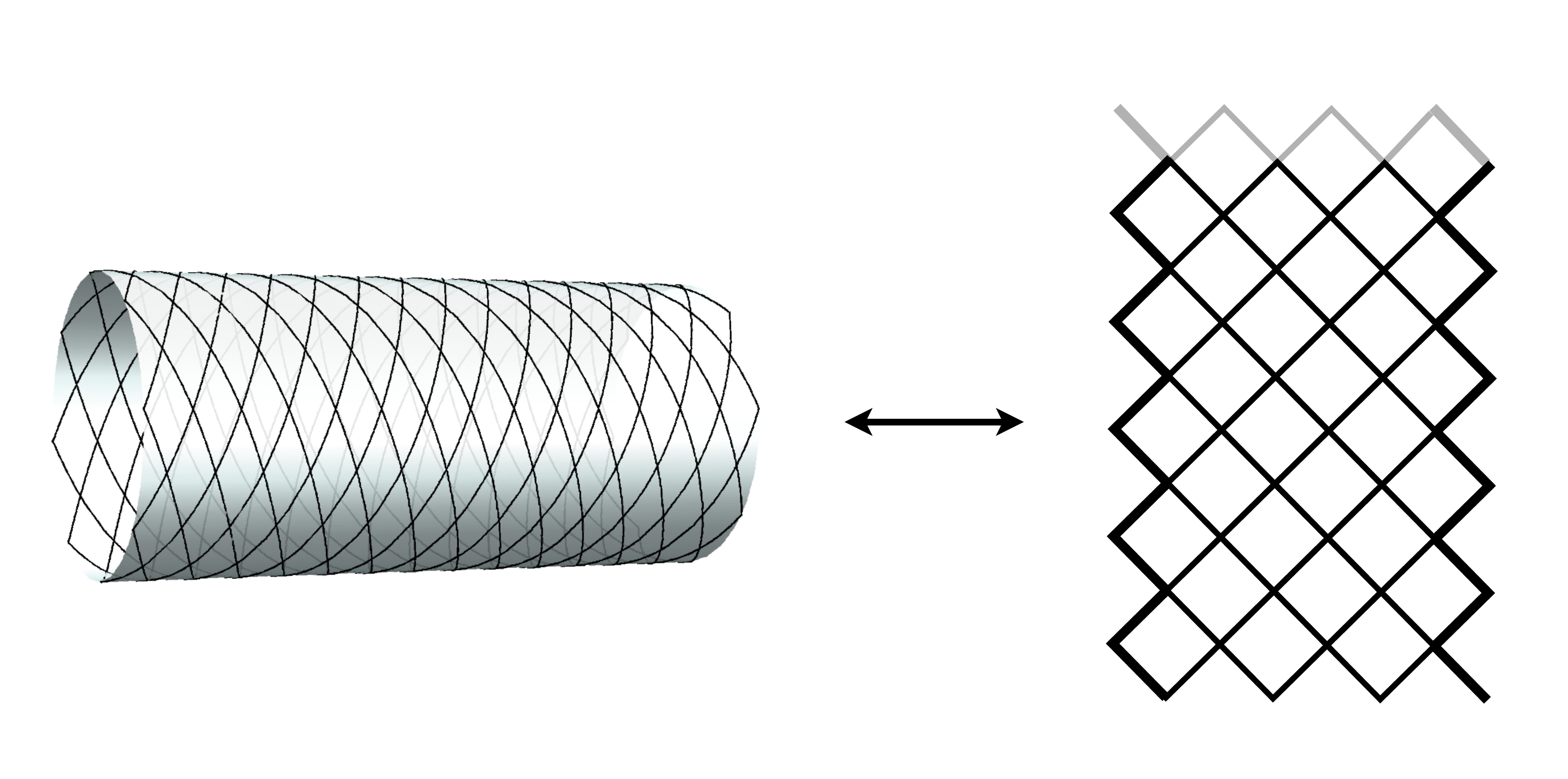}
\caption{The 6-vertex model on a cylinder.
In the open channel, there are $2M+1$ sites in the horizontal
direction with free boundary conditions, and $2N$ sites in
the vertical direction with periodic boundary conditions.}
\label{fig:annulus}
\end{center}
\end{figure}

We consider the 6-vertex model at the isotropic point on a
$(2M+1)\times 2N$ medial lattice, for positive integers $M$ and $N$.
We impose periodic boundary conditions in the vertical direction,
and free boundary conditions in the horizontal direction; the
geometry under consideration is a cylinder,
as is shown in figure~\ref{fig:annulus}.
The partition function on the lattice can be computed in two different channels.

\begin{figure}[htb]
	\centering
	\includegraphics[width=0.3\textwidth]{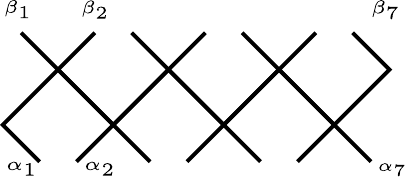}
	\caption{Diagonal-to-diagonal transfer matrix}
	\label{fig:DiagToDiag}
\end{figure}

\paragraph{Open channel.} In the open channel, we define the diagonal-to-diagonal transfer matrix
\begin{align}
\label{eq:openT}
\mathrm{t}_{D}(u)=\check{R}_{23}(u)\,
\check{R}_{45}(u)\cdots\check{R}_{2M,2M+1}(u)\, \check{R}_{12}(u)\,
\check{R}_{34}(u)\cdots\check{R}_{2M-1,2M}(u) \,,
\end{align}
shown in figure~\ref{fig:DiagToDiag}; by convention the direction of propagation (the ``imaginary time'' direction)
is upwards in our figures.
The subscripts label the spaces being acted upon, and $\check{R}$ is related to the standard $R$-matrix of the
isotropic 6-vertex model by
\begin{align}
\label{defcheckR}
\check{R}_{jk}(u)=P_{jk}\, R_{jk}(u) \,,
\end{align}
where $P$ is the permutation operator. Written explicitly, the $R$-matrix is given by
\begin{align}
R(u)=u+i P=\left(
            \begin{array}{cccc}
              a(u) & 0 & 0 & 0 \\
              0 & b(u) & c(u) & 0 \\
              0 & c(u) & b(u) & 0 \\
              0 & 0 & 0 & a(u) \\
            \end{array}
          \right) \,,
\label{Rmat}		
\end{align}
with the Boltzmann weights
\begin{align}
a(u)=u+i\,,\qquad b(u)=u\,,\qquad c(u)=i \,.
\end{align}
The partition function is given by
\begin{align}
\label{eq:openZ}
Z(u,M,N)=\mathrm{tr}\,\big[\mathrm{t}_D(u)^N\big] \,.
\end{align}
For small values of $M$ and $N$, the results can be directly computed by brute force
from the definition, for example
\begin{align}
Z(u,1,1) & =  2 (u+2 i)^2 \,, \nonumber \\
Z(u,1,2) & =  2 \left(u^4+8 i u^3-12 u^2-8 i u+4\right) \,, \label{Zexamples} \\ \nonumber
Z(u,2,2) & =  2 \left(u^8+16 i u^7-76 u^6-184 i u^5+268 u^4+256 i
u^3-160 u^2-64 i u+16\right) \,.
\end{align}

\begin{figure}[htb]
	\centering
	\includegraphics[height=0.4\textwidth]{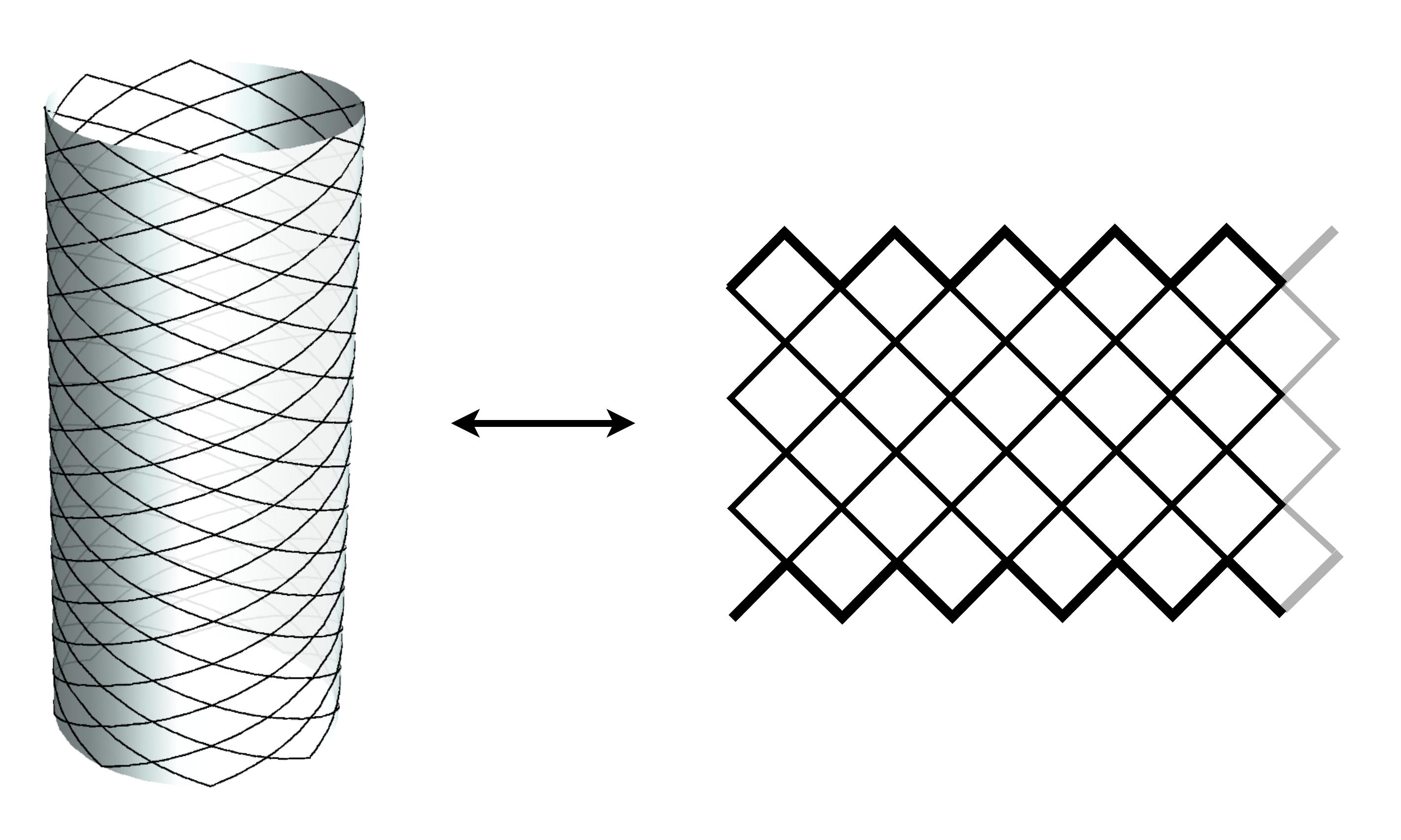}
	\caption{In the closed channel, there are $2N$ sites in
the horizontal direction with periodic boundary conditions, and
$2M+1$ sites in the vertical direction with free boundary
conditions.}
	\label{fig:crossed}
\end{figure}

\paragraph{Closed channel.}
In the closed channel, the graph is rotated by $90^{\circ}$ degrees,
as shown in figure \ref{fig:crossed}.
The rotated $R$-matrix which will be denoted by
$R^{\text{c}}$ takes the following form
\begin{align}
\label{Rtilde}
R^{\text{c}}(u)=\left(
                       \begin{array}{cccc}
                         b(u) & 0 & 0 & 0 \\
                         0 & a(u) & c(u) & 0 \\
                         0 & c(u) & a(u) & 0 \\
                         0 & 0 & 0 & b(u) \\
                       \end{array}
                     \right) \,.
\end{align}
Notice that this $R$-matrix does not satisfy the Yang-Baxter equation.
In the closed channel, the partition function is no longer given by a
trace, since periodic boundary conditions are not imposed in the vertical
direction.  Instead, the open boundary conditions give rise to non-trivial
boundary states in the closed channel.  The partition function in the
closed channel is given by
\begin{align}
\label{eq:closedZ}
Z^{\text{c}}(u,M,N)=\langle\Psi_0|U^{\dagger}\,\tilde{\mathrm{t}}_D(u)^M|\Psi_0\rangle \,,
\end{align}
where $U$ is the one-site shift operator
\begin{align}
U=P_{12}P_{23}\cdots P_{2N-1,2N} \,,
\end{align}
and $\tilde{\rT}_D(u)$ is defined as
\begin{align}
\label{eq:closedT}
\tilde{\rT}_D(u)=\check{R}^{\text{c}}_{12}(u)\,
\check{R}^{\text{c}}_{34}(u)\cdots
\check{R}^{\text{c}}_{2N-1,2N}(u)\, \check{R}^{\text{c}}_{23}(u)\,
\check{R}^{\text{c}}_{45}(u)\cdots\check{R}^{\text{c}}_{2N-2,2N-1}(u)\, \check{R}^{\text{c}}_{2N,1}(u)\,,
\end{align}
where $\check{R}^{\text{c}}_{ij}(u)=P_{ij}R^{\text{c}}_{ij}(u)$. The boundary state $|\Psi_0\rangle$ is given by
\begin{align}
\label{bdrystate}
|\Psi_0\rangle=|\psi_0\rangle^{\otimes N},\qquad |\psi_0\rangle=|\uparrow\,\rangle\otimes|\downarrow\,\rangle+|\downarrow\,\rangle\otimes|\uparrow\,\rangle \,,
\end{align}
where we have used the notation
\begin{align}
|\uparrow\,\rangle\equiv{1 \choose 0}\,, \qquad
|\downarrow\,\rangle\equiv{0 \choose 1} \,.
\end{align}
The result for the partition function of course does not depend on how we
perform the computation, so we have
\begin{align}
Z(u,M,N)=Z^{\text{c}}(u,M,N) \,.
\end{align}
To verify the correctness of our various computations (see below),
we have explicitly checked this identity for small value of $M$ and $N$.

Our goal is to compute analytic expressions of $Z(u,M,N)$ explicitly for different
intermediate values of $M$ and $N$.  When both $M$ and $N$ are large, the
system can be well approximated by the computation in the
thermodynamic limit.  Here we instead focus on the interesting intermediate
case where we keep one of $M$, $N$ to be finite (namely, the one that determines
the dimension of the transfer matrix) and the other to be
large.  For finite $M$ ($\le 10$) and large $N$ (around a few hundred
to thousands), we perform the computation in the open channel using
(\ref{eq:openZ}); whereas for finite $N$ and large $M$, we work in the closed
channel using (\ref{eq:closedZ}).  We discuss the computation of the
partition function in both channels from the perspective of Bethe ansatz and algebraic
geometry.

\section{Partition function in the open channel}
\label{sec:open}
In this section, we discuss the computation of the partition function
in the open channel using Bethe ansatz and algebraic geometry.  Using
this method, we are able to compute the partition function for finite
$M\le 10$ and large $N$ (ranging from a few hundred to thousands).

\subsection{Reformulation and Bethe ansatz}
In order to apply the $R$-matrix machinery, the first step is to
re-express the diagonal-to-diagonal transfer matrix $\rT_{D}(u)$
(\ref{eq:openT}) in terms of an integrable open-chain transfer matrix
with $2M+1$ sites and with inhomogeneities $\{ \theta_{j} \}$
\cite{Sklyanin:1988yz}.  Let us define
\begin{align}
\rT(u;\{\theta_j\})=\tr_{a} K^{+}(u)\, T_{a}^{(2M+1)}(u; \{ \theta_{j} \})\,
K^{-}(u)\, \widehat{T}_{a}^{(2M+1)}(u; \{ \theta_{j} \}) \,,
\label{transfer}
\end{align}
where the monodromy matrices are given by
\begin{align}
T_{a}^{(l)}(u; \{ \theta_{j}\}) &= R_{a1}(u-\theta_{1}) \ldots R_{a\,
l}(u-\theta_{l})\,, \nonumber \\
\widehat{T}_{a}^{(l)}(u; \{ \theta_{j}\}) &= R_{a\, l}(u+\theta_{l}) \ldots
R_{a 1}(u+\theta_{1})\,.
\label{monodromy}
\end{align}
For our isotropic problem, the $K$-matrices are simply $K^{+}(u) = K^{-}(u) =
\mathbb{I}$.

The eigenvalues $\Lambda(u; \{ \theta_{j}\})$ of the transfer matrix
$\rT(u; \{ \theta_{j}\})$ (\ref{transfer}), which can be obtained using
algebraic Bethe ansatz \cite{Sklyanin:1988yz}, are given by
\begin{align}
\Lambda(u; \{ \theta_{j}\}) &= \frac{2(u+i)}{(2u+i)}
\left[\prod_{j=1}^{2M+1}(u-\theta_{j}+i)(u+\theta_{j}+i)\right]
\prod_{k=1}^{K}\frac{(u-u_{k} - \frac{i}{2})(u+u_{k} - \frac{i}{2})}
{(u-u_{k} + \frac{i}{2})(u+u_{k} + \frac{i}{2})}
\nonumber \\
&+ \frac{2u}{(2u+i)}
\left[\prod_{j=1}^{2M+1}(u-\theta_{j})(u+\theta_{j})\right]
\prod_{k=1}^{K}\frac{(u-u_{k} + \frac{3i}{2})(u+u_{k} + \frac{3i}{2})}
{(u-u_{k} + \frac{i}{2})(u+u_{k} + \frac{i}{2})} \,,
\label{Lambdagen}
\end{align}
where the $\{ u_{k} \}$ are solutions of the BAE
\begin{align}
\prod_{j=1}^{2M+1}\frac{(u_{k} - \theta_{j} + \frac{i}{2})(u_{k} + \theta_{j} + \frac{i}{2})}
{(u_{k} - \theta_{j} - \frac{i}{2})(u_{k} + \theta_{j} -
\frac{i}{2})} = \prod_{j=1; j\ne k}^{K}\frac{(u_{k}-u_{j}+i)(u_{k}+u_{j}+i)}
{(u_{k}-u_{j}-i)(u_{k}+u_{j}-i)} \,.
\label{BEgen}
\end{align}

\begin{figure}[h!]
\begin{center}
\includegraphics[scale=0.9]{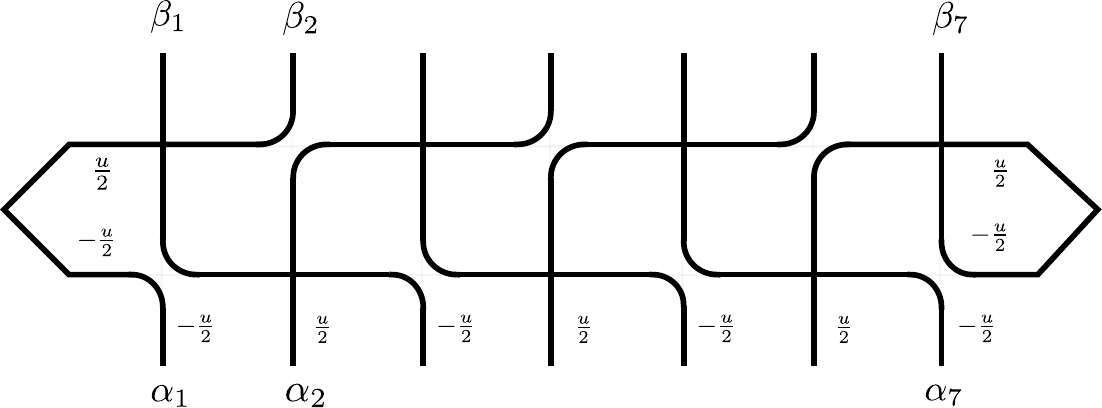}
\caption{The $R$-matrices with zero argument act as permutation operators depicted with avoiding lines. This transforms the double-row transfer matrix of the spin-chain geometry into the diagonal-to-diagonal transfer matrix of the boundary vertex model.
}
\label{fig:bdry}
\end{center}
\end{figure}

The key point (due to Destri and de Vega \cite{Destri:1991zm}) is to choose {\em alternating
spectral-parameter-dependent} inhomogeneities as follows
\begin{align}
\theta_{j} = \theta_{j}(u) = (-1)^{j} u \,, \qquad j = 1, \ldots, 2M+1 \,.
\label{DDV}
\end{align}
One can then show \cite{Yung:1994td} that the diagonal-to-diagonal transfer matrix $\rT_{D}(u)$ is given by \footnote{We note that $\rT_{D}(u)$ does
{\em not} commute with $\rT_{D}(v)$.}
\begin{align}
\rT_{D}(u)  = \frac{1}{i^{2M+1} (u+2i)} \rT(\tfrac{u}{2};
\{ \theta_{j}(\tfrac{u}{2})\} ) \,,
\label{digDR}
\end{align}
noting that half of the $R$-matrices become proportional to permutation operators, and the spin chain geometry is transformed into the vertex one, see Figure~\ref{fig:bdry}.
Specifying in (\ref{Lambdagen})-(\ref{BEgen}) the inhomogeneities as in (\ref{DDV}),
it follows that the eigenvalues $\Lambda_{D}(u)$ of $\rT_{D}(u)$ are
given by
\begin{align}
\Lambda_{D}(u) &= \Lambda_{D, K}(u) = \frac{1}{i^{2M+1} (u+2i)} \Lambda(\tfrac{u}{2};
\{ \theta_{j}(\tfrac{u}{2}) \} ) \\
&= (u+i)^{2M}
\prod_{k=1}^{K}\frac{(\frac{u}{2}-u_{k} - \frac{i}{2})(\frac{u}{2}+u_{k} - \frac{i}{2})}
{(\frac{u}{2}-u_{k} + \frac{i}{2})(\frac{u}{2}+u_{k} + \frac{i}{2})}
\,,
\label{LambdaD}
\end{align}
where the $\{ u_{k} \}$ are solutions of the Bethe equations
\begin{align}
\left[\frac{(u_{k} -\frac{u}{2} + \frac{i}{2})(u_{k} + \frac{u}{2} + \frac{i}{2})}
{(u_{k} - \frac{u}{2} - \frac{i}{2})(u_{k} + \frac{u}{2} -
\frac{i}{2})}\right]^{2M+1} &= \prod_{j=1; j\ne k}^{K}\frac{(u_{k}-u_{j}+i)(u_{k}+u_{j}+i)}
{(u_{k}-u_{j}-i)(u_{k}+u_{j}-i)}.
\label{BE}
\end{align}
Here $k = 1, \ldots, K$ and $K = 0, 1, \ldots, M \,$. Note that the BAE (\ref{BE}) depend on the spectral parameter $u$,
which is an unusual feature.

We observe that $\rT_{D}(u)$ has $su(2)$ symmetry
\begin{align}
\left[ \rT_{D}(u) \,, \vec S \right] = 0 \,, \qquad \vec S =
\sum_{j=1}^{2M+1}\tfrac{1}{2} \vec \sigma_{j} \,.
\end{align}
The Bethe states are $su(2)$ highest-weight states, with spin
\begin{align}
\label{eq:szminusK}
s = s^{z} = \frac{1}{2}(2M+1) - K \,.
\end{align}
For a given value of $K$, the corresponding eigenvalue therefore has degeneracy
\begin{align}
\label{su2degeneracy}
2s+1 = 2M + 2 - 2K \,.
\end{align}
We conclude that the partition function (\ref{eq:openZ}) is given by
\begin{align}
\label{eq:explicitZ1}
Z(u,M, N) = \sum_{K=0}^{M} \sum_{\text{sol}(M, K)} (2M + 2 - 2K)\, \Lambda_{D, K}(u)^{N} \,,
\end{align}
where $\Lambda_{D, K}(u)$ is given by (\ref{LambdaD}). Here
$\text{sol}(M, K)$ stands for physical solutions $\{ u_{1}, \ldots,
u_{K} \}$ of the BAE (\ref{BE})
with $2M+1$ sites and $K$ Bethe roots. The number $\mathcal{N}(M,K)$ of such solutions has been conjectured to
be given by \cite{Gainutdinov:2015vba}
\begin{align}
\label{dimsol-open}
\mathcal{N}(M,K) = {2M+1 \choose K} - {2M+1 \choose K-1} \,.
\end{align}

In order to find the explicit expressions for the partition function
(\ref{eq:explicitZ1}), we need to find the eigenvalues
$\Lambda_{D,K}(u)$.  They depend on the values of rapidities which are
solutions of the BAE (\ref{BE}).  We encounter two difficulties.
Firstly, the solution set of the BAE (\ref{BE}) contains some redundancy, since not all solutions
are physical; therefore one needs to impose extra selection rules \cite{Gainutdinov:2015vba}.
Secondly, generally Bethe equations are a complicated system of algebraic
equations, which cannot be solved analytically.  What is worse,
our BAE (\ref{BE}) depend on a free parameter $u$, which means that the
Bethe roots are functions of $u$, thereby making the BAE
even harder than usual to solve.\par

In order to overcome these two difficulties, we need new tools, namely
the rational $Q$-system and computational algebraic geometry.  These
methods have been applied successfully in computing the torus
partition function of the 6-vertex model \cite{Jacobsen:2018pjt}.  The
BAE can be reformulated as a set of $QQ$-relations, with appropriate
boundary conditions \cite{Marboe:2016yyn}.  The benefit of working
with the $Q$-system is twofold.  Firstly, it is much more efficient
to solve the rational $Q$-system than to directly solve the BAE.
Secondly, all the solutions of the $Q$-system are physical, so there
is no need to impose further selection rules
\cite{Nepomechie:2013mua,Granet:2019knz}.
The rational $Q$-system, which was first
developed for isotropic (XXX) spin chains with periodic boundary
conditions \cite{Marboe:2016yyn}, was recently generalized to
anisotropic (XXZ) spin chains and to spin chains with certain open
boundary conditions \cite{Granet:2019knz, Bajnok:2019zub}.  We briefly
review the $Q$-system for open boundary conditions in
section~\ref{sec:openQsystem}.

Turning to the second difficulty, finding all solutions of
the BAE (or of the corresponding $Q$-system) is in general only possible numerically.  However, it
was realized in \cite{Jiang:2017phk} that if the goal is to \emph{sum over}
all the solutions of the BAE/$Q$-system for some \emph{rational function
$f(\{u_j\})$} of the Bethe roots, then it can be done without
knowing all the solutions explicitly.  The idea is based on computational
algebraic geometry.  The solutions of the BAE/$Q$-system form a
finite-dimensional linear space called the \emph{quotient ring}.  The
dimension of the quotient ring is the number of physical solutions of
the BAE/$Q$-system.  A basis of the quotient ring can be constructed
by standard methods using a Gr\"obner basis.  Once a basis for the
quotient ring is known, one can construct the \emph{companion matrix}
for the function $f(\{u_j\})$, which is a finite-dimensional
representation of this function in the quotient ring.  Taking the
trace of the companion matrix gives the sought-after sum.  For
a more detailed introduction to these notions and explicit examples in the
context of toroidal boundary conditions, we
refer to the original papers \cite{Jiang:2017phk,Jacobsen:2018pjt}
and the textbooks \cite{CLO1, CLO2}.

The same strategy can be applied to the open boundary conditions.  The new
feature that appears in this case is the dependence on a free
parameter $u$.  While this creates extra difficulty for numerical
computations, it does not cost more effort in the algebro-geometric approach.
The reason is that the constructions of the Gr\"obner basis, the basis for the
quotient ring and companion matrices are purely algebraic; and it does not
make much qualitative difference whether we have to manipulate numbers or algebraic
expressions.%
\footnote{In practice, due to implementations of the
algorithm in packages, the efficiencies for manipulating numbers and
algebraic expressions can be different.}

\subsection{BAE and $Q$-system}
\label{sec:openQsystem}
In this section, we review the rational $Q$-system for the
$SU(2)$-invariant XXX spin chain with open boundary conditions \cite{Bajnok:2019zub}.  Let
us first consider the BAE with generic
inhomogeneities $\{\theta_{j}\}$ (\ref{BEgen})
\begin{align}
\label{eq:inhomu}
\prod_{l=1}^L\frac{(u_j-\theta_l+\tfrac{i}{2})(u_j+\theta_l+\tfrac{i}{2})}
{(u_j-\theta_l-\tfrac{i}{2})(u_j+\theta_l-\tfrac{i}{2})}=\prod_{k\ne
j}^K\frac{(u_j-u_k+i)(u_j+u_k+i)}{(u_j-u_k-i)(u_j+u_k-i)} \,,
\end{align}
where $L=2M+1$.
For given value of $L$ and $K$, we consider a two-row Young tableau with
number of boxes $(L-K,K)$.  At each vertex of the Young tableau, we
associate a $Q$-function denoted by $Q_{a,s}$. The BAE (\ref{eq:inhomu}) can be
obtained from the following $QQ$-relations
\begin{align}
v\, Q_{a+1,s}(v)\, Q_{a,s+1}(v) \propto Q^{+}_{a+1,s+1}(v)\,
Q^{-}_{a,s}(v) - Q^{-}_{a+1,s+1}(v)\, Q^{+}_{a,s}(v) \,,
\label{QQ}
\end{align}
where $f^{\pm}(v) := f(v\pm \frac{i}{2})$, and
the $Q$-functions $Q_{a,s}(v)$ are {\em even} polynomials of $v$
\begin{align}
Q_{a,s}(v) = v^{2 M_{a,s}} + \sum_{k=0}^{M_{a,s}-1} c_{a,s}^{(k)}\,
v^{2k} \,,
\end{align}
where $M_{a,s}$ is the number of boxes in the Young tableau to the right
and top of the vertex $(a,s)$.
The boundary conditions are chosen such that $Q_{2,s}=1$, $Q_{1,s>K}=1$
and
\begin{align}
Q_{0,0}(v) =&\, \prod_{j=1}^{L}(v-\theta_{j})(v+\theta_{j}) \,,
\nonumber \\
Q_{1,0}(v) =&\, Q(v)=\prod_{k=1}^{K}(v-u_k)(v+u_k) \,.
\label{Q00}
\end{align}
Here $Q_{1,0}(v)$ is the usual Baxter $Q$-function, whose zeros are the Bethe
roots.  Comparing to the periodic $QQ$-relations \cite{Marboe:2016yyn}, the main differences
are an extra factor $v$ that appears on the left-hand side of
(\ref{QQ}), and the degree of the polynomial of $Q_{a,s}$ which is twice
the one for the periodic case. More details can be found in section
4.2 of \cite{Bajnok:2019zub}.\par

For the Bethe equations (\ref{BE}), corresponding to the alternating
inhomogeneities (\ref{DDV}), we simply have\footnote{Recall
that the argument of the double-row transfer matrix $\rT$ in (\ref{digDR})
is $\frac{u}{2}$, rather than $u$. \label{foot:half}}
\begin{align}
Q_{0,0}(v) = \left[(v-\tfrac{u}{2})(v+\tfrac{u}{2})\right]^{2M+1} \,.
\end{align}
To solve the $Q$-system, we impose the condition that all the
$Q_{a,s}$ functions are polynomials.  This requirement generates a set
of algebraic equations called \emph{zero remainder conditions} (ZRC)
for the coefficients $c_{a,s}^{(k)}$.  In principle, one can then
solve the ZRC's and find $Q_{a,s}$, in particular the main
$Q$-function $Q_{1,0}$.  The zeros of $Q_{1,0}$ are the Bethe roots
$\{ u_{k} \}$, which are functions of the parameter $u$.\par

After finding the $Q$-functions, the next step is to find the
eigenvalues $\Lambda_{D}$ (\ref{LambdaD}), which in terms of
$Q$-functions are given simply by
\begin{align}
\label{eq:lambda}
\Lambda_{D}(u)=
(u+i)^{2M}\frac{Q(\frac{u}{2}-\frac{i}{2})}{Q(\frac{u}{2}+\frac{i}{2})} \,.
\end{align}
Plugging these into (\ref{eq:explicitZ1}), we finally obtain the partition function.

\subsection{Algebraic geometry}
In this subsection, we give the main steps for the algebro-geometric computation of the partition function:
\begin{enumerate}
\item Generate the set of zero remainder conditions (ZRC) from the rational $Q$-system;
\item Compute the Gr\"obner basis of the ZRC;
\item Construct the quotient ring of the ZRC;
\item Compute the companion matrix for the eigenvalues $\Lambda_{D,K}(u)$ (\ref{LambdaD}) which will be denoted by $\mathbb{T}_{M,K}(u)$;
\item Compute the matrix power of $\mathbb{T}_{M,K}(u)$ and take the trace
\begin{align}
Z(u,M,N)=\sum_{K=0}^M(2M+2-2K)\,\tr\left[\mathbb{T}_{M,K}(u)\right]^N \,.
\end{align}
\end{enumerate}
Most steps listed above can be done straightforwardly, adapting the corresponding working of \cite{Jacobsen:2018pjt}. The only step
that requires some additional work is step 4.  The variables of ZRC are
$c_{a,s}^{(k)}$ which are coefficients of the $Q$-functions.  From
these variables, it is easy to construct the companion matrix of the
$Q$-function.  For fixed $M$ and $K$, we denote the companion matrix
by $\mathbb{Q}_{M,K}$.  To find the companion matrix of $\Lambda_D$,
which is essentially the companion matrix of $\Lambda$
(\ref{eq:lambda}) up to some multiplicative factors, the most direct
way is to use homomorphism property of the companion matrix and
write
\begin{align}
\mathbb{T}_{M,K}(u)=&\,(u+i)^{2M}\frac{\mathbb{Q}_{M,K}(\tfrac{u}{2}-\tfrac{i}{2})}{\mathbb{Q}_{M,K}(\tfrac{u}{2}+\tfrac{i}{2})} \,,
\end{align}
where $\mathbb{T}_{M,K}(u)$ is the companion matrix for $\Lambda_D(u)$ with fixed $M$ and $K$.
Unfortunately, this method involves taking the inverse of the matrix $\mathbb{Q}_{M,K}(u+\tfrac{i}{2})$ analytically,
which can be slow when the dimension of the matrix is large.

We find that a much more efficient way is to use the following $TQ$-relation
\begin{align}
\label{eq:openTQ}
u\,T(u-\tfrac{i}{2})Q(u)=&\,(u+\tfrac{i}{2})\left[(u+\tfrac{i}{2})^2-(\tfrac{z}{2})^2\right]^{L}Q(u-i)\\\nonumber
&+(u-\tfrac{i}{2})\left[(u-\tfrac{i}{2})^2-(\tfrac{z}{2})^2\right]^{L}Q(u+i).
\end{align}
In our case, we need to take $L=2M+1$ and $z=u$. To solve the $TQ$ relation (\ref{eq:openTQ}), we make the following ansatz for the two polynomials
\begin{align}
\label{eq:ansatzTQ}
T(u)=&\,t_{2L} u^{2L}+t_{2L-1} u^{2L-1}+\cdots+t_0,\\\nonumber
Q(u)=&\,u^{2K}+s_{K-1}u^{2(K-1)}+\cdots+s_0.
\end{align}
Notice that $Q(u)$ is an even polynomial and only even powers of $u$ appear, which is not the case for $T(u)$. Plugging the ansatz (\ref{eq:ansatzTQ}) into (\ref{eq:openTQ}), we obtain a system of algebraic equations for the coefficients $\{t_0,t_1,\cdots,t_{2L},s_0,\cdots,s_{K-1}\}$. In fact, solving these set of algebraic equations is yet another way to find the Bethe roots. For our purpose, we only solve the equation \emph{partially}, namely we view $\{s_0,\cdots,s_{K-1}\}$ as parameters and solve $\{t_k\}$ in terms of $\{s_j\}$. This turns out to be much simpler since the equations are linear. We find that $t_k(\{s_j\})$ are polynomials in the variables $\{s_j\}$. From ZRC and algebro-geometric computations, we can find the companion matrix of $s_j$ which we denote by $\mathbf{s}_j$. Replacing $s_j$ by $\mathbf{s}_j$ and the products by matrix multiplication in $t_k(\{s_j\})$, we find the companion matrix $\mathbf{t}_k=t_k(\{\mathbf{s}_j\})$. Then the companion matrix of the eigenvalues of the transfer matrix is given by
\begin{align}
\mathbb{T}_{M,K}(u)=\mathbf{t}_{2L}\,u^{2L}+\mathbf{t}_{2L-1}\,u^{2L-1}+\cdots+\mathbf{t}_0.
\end{align}
More details on the implementation of the algebro-geometric computations are given in appendix~\ref{app:AGdetail}.

Using the AG approach, we have computed the partition
functions for $M$ up to $6$, with $N$ up to $2048$. We also calculated some partition functions
with higher $M$ and lower $N$. The results for $2\le M,N\le 6$ are
given in Appendix \ref{eq:exactsmallMN}.

\section{Partition function in the closed channel}
\label{sec:close}
In this section, we compute the partition function in the closed
channel.  There are both simplifications and complications due to the
presence of non-trivial boundary states.  Indeed, the presence of
boundary states imposes selection rules for the allowed solutions of
the BAE. Firstly, it restricts to the states with zero total spin.
This implies that the length of the spin chain must be even, which we
denote by $2N$; and the only allowed number of Bethe roots is $K=N$.
In contrast, for the periodic (torus) case \cite{Jacobsen:2018pjt},
one must consider all the sectors $K=0, 1,\ldots, N$.  Moreover,
the Bethe roots must form Cooper-type pairs (\ref{paired}), which leads to
significant simplification in the computation of the Gr\"obner
basis and quotient ring.

This simplification comes with a price.  Recall that the
partition function in the closed channel
takes the form of a matrix element given by
(\ref{eq:closedZ}).  To evaluate this matrix element, we need the
overlaps between the boundary states and the Bethe states.  These
overlaps are a new feature, which is not present in the open channel.
They are complicated functions of the rapidities, which makes the computation
of the companion matrix more difficult.

\subsection{Reformulation and Bethe ansatz}
To compute the expression (\ref{eq:closedZ}) for the partition
function in the closed channel, the first step is to rewrite
$\tilde{\rT}_D$ (\ref{eq:closedT}) in terms of integrable
\emph{closed-chain} transfer matrices.  To this end, we observe that
${R}^{\text{c}}(u)$ (\ref{Rtilde}) is related to $R(u)$ by
\begin{equation}
{R}^{\text{c}}_{12}(u) = -\sigma_{1}^{z}\, R_{12}(\tilde{u})\, \sigma_{1}^{z} =
-\sigma_{2}^{z}\, R_{12}(\tilde{u})\, \sigma_{2}^{z} \,,
\end{equation}
where $\tilde{u}$ is the `crossing transformed' spectral parameter defined by
\begin{align}
\tilde{u} = -u-i \,.
\label{utilde}
\end{align}
The corresponding ``checked'' $R$-matrices are therefore related by
\begin{equation}
\label{eq:crossingR}
\check{{R}}^{\text{c}}_{12}(u) = -\sigma_{2}^{z}\, \check{R}_{12}(\tilde{u})\,
\sigma_{2}^{z} \,.
\end{equation}
For later convenience, we define
\begin{align}
\tilde{V}^{(1)}=&\,\check{R}^{\text{c}}_{23}(u)\,
\check{R}^{\text{c}}_{45}(u)\cdots
\check{R}^{\text{c}}_{2N-2,2N-1}(u)\,
\check{R}^{\text{c}}_{2N,1}(u)\,, \nonumber \\
\tilde{V}^{(2)}=&\,\check{R}^{\text{c}}_{12}(u)\,
\check{R}^{\text{c}}_{34}(u)\cdots \check{R}^{\text{c}}_{2N-1,2N}(u)
\,,
\end{align}
in terms of which $\tilde{\rT}_D$ (\ref{eq:closedT}) is given by
\begin{align}
\tilde{\rT}_D(u)=\tilde{V}^{(2)}(u)\, \tilde{V}^{(1)}(u)
\label{tcDnew} \,.
\end{align}
It follows from (\ref{eq:crossingR}) that
\begin{align}
\tilde{V}^{(1)}(u) &= (-1)^{N}\, \Omega^{(2)}\, V^{(1)}(\tilde{u})\,
\Omega^{(1)} \,, \nonumber \\
\tilde{V}^{(2)}(u) &= (-1)^{N}\, \Omega^{(1)} \, V^{(2)}(\tilde{u})\,  \Omega^{(2)}\,,
\label{VtildeVrltn}
\end{align}
where
\begin{align}
\Omega^{(1)} = \sigma^{z}_{1}\sigma^{z}_{3} \cdots \sigma^{z}_{2N-1}
\,, \qquad
\Omega^{(2)} = \sigma^{z}_{2}\sigma^{z}_{4} \cdots \sigma^{z}_{2N}
\,,
\end{align}
and the $V^{(i)}$ are the same as the corresponding
$\tilde{V}^{(i)}$, but with $R$'s instead of ${R}^{\text{c}}$'s:
\begin{align}
V^{(1)}(u) &= \check{R}_{23}(u)\, \check{R}_{45}(u)\, \ldots
\check{R}_{2N-2,2N-1}(u)\, \check{R}_{2N,1}(u) \,, \nonumber \\
V^{(2)}(u) &=\check{R}_{12}(u)\, \check{R}_{34}(u)\, \ldots
\check{R}_{2N-1,2N}(u)\,.
\label{V1V2}
\end{align}
Let us now introduce integrable inhomogeneous closed-chain transfer matrices of length $2N$
\begin{align}
\tau(u; \{ \theta_{j}\} ) &= \tr_{a} T_{a}^{(2N)}(u; \{ \theta_{j}\})  \,,
\nonumber \\
\widehat{\tau}(u; \{ \theta_{j}\} ) &= \tr_{a} \widehat{T}_{a}^{(2N)}(u; \{ \theta_{j}\})  \,,
\label{transferclosed}
\end{align}
where the monodromy matrices are defined in (\ref{monodromy}). Using crossing symmetry
\begin{align}
R_{12}(\tilde{u}) = -\sigma_{1}^{y}\, R_{12}^{t_{1}}(u)\, \sigma_{1}^{y}
\,,
\end{align}
where `$t_1$' stands for transposition in the first quantum space, one can show that
\begin{align}
\tau(\tilde{u} ; \{ \theta_{j}\} ) = \widehat{\tau}(u; \{ \theta_{j}\} ) \,.
\label{tauhatid}
\end{align}
The transfer matrices (\ref{transferclosed})
can be diagonalized by algebraic Bethe ansatz. We define
the operators $A$, $B$, $C$, $D$ as matrix elements of the monodromy matrix
(\ref{monodromy}) as usual
\begin{align}
T_a^{(2N)}(u-\tfrac{i}{2};\{\theta_j\})=
\left(\begin{array}{cc}
A(u) & B(u) \\
C(u) & D(u) \\
\end{array}\right) \,,
\label{ABCD}
\end{align}
but note the shift in the spectral parameter.
We consider the reference state or pseudovacuum
\begin{align}
|0\rangle=|\uparrow\rangle^{\otimes 2N} \,.
\label{refstate}
\end{align}
The Bethe states and their duals are constructed by acting with $B$-
and $C$-operators on the reference state\footnote{It
should be kept in mind that the Bethe states depend on the
inhomogeneities $\{ \theta_{j}\}$; in order to lighten the notation,
this dependence is not made explicit.}
\begin{align}
|\mathbf{u}\rangle=B(u_1)\cdots B(u_K)|0\rangle\,, \qquad
\langle\mathbf{u}|=\langle0|C(u_1)\cdots C(u_K) \,.
\label{Bethestates}
\end{align}
These states are eigenstates of the transfer matrix
$\tau(u;\{\theta_j\})$ (\ref{transferclosed})
\begin{align}
\tau(u;\{\theta_j\})|\mathbf{u}\rangle=\Lambda_c(u;\{\theta_j\})|\mathbf{u}\rangle \,,
\end{align}
with eigenvalues $\Lambda_c(u;\{\theta_j\})$ given by
\begin{align}
\Lambda_c(u;\{\theta_j\})=\prod_{j=1}^{2N}(u-\theta_j+i)\prod_{k=1}^K\frac{u-u_k-\tfrac{i}{2}}{u-u_k+\tfrac{i}{2}}
+\prod_{j=1}^{2N}(u-\theta_j)\prod_{k=1}^K\frac{u-u_k+\tfrac{3i}{2}}{u-u_k+\tfrac{i}{2}} \,,
\label{Lambdagenclosed}
\end{align}
provided that the rapidities $\mathbf{u}=\{u_1,\cdots,u_K\}$ satisfy the BAE
\begin{align}
\label{BEgenclosed}
\prod_{j=1}^{2N}\frac{u_k-\theta_j+\tfrac{i}{2}}{u_k-\theta_j-\tfrac{i}{2}}=\prod_{j=1; j\ne k}^K\frac{u_k-u_j+i}{u_k-u_j-i} \,.
\end{align}
In contradistinction with \eqref{BE} these BAE do not depend on the spectral parameter $u$, as is usually the case.
The eigenvalues $\widehat{\Lambda}_c(u; \{ \theta_{j}\})$ of $\widehat{\tau}(u; \{ \theta_{j}\})$ are given by
\begin{align}
\widehat{\Lambda}_c(u; \{ \theta_{j}\}) = \Lambda_c(\tilde{u}; \{\theta_{j}\}) \,,
\label{hatLambdagenclosed}
\end{align}
as follows from (\ref{tauhatid}).\par

\begin{figure}[h!]
\begin{center}
\includegraphics[scale=0.9]{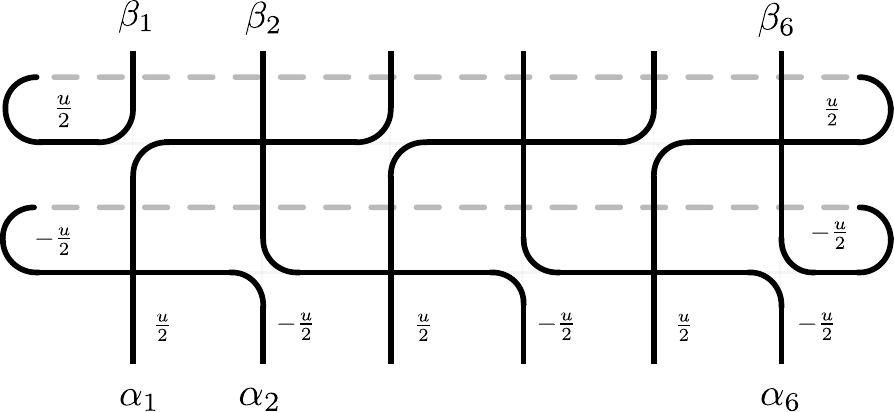}
\caption{The $R$-matrices with zero argument act as permutation operators depicted with avoiding lines. This transforms the product of periodic transfer matrices of the spin-chain geometry into the periodic diagonal-to-diagonal transfer matrix of the vertex model.}
\label{fig:periodic}
\end{center}
\end{figure}

To make contact with $\tilde{\rT}_D$, we again choose alternating spectral-parameter-dependent inhomogeneities
\begin{align}
\theta_j=\theta_j(u)=(-1)^{j+1}u,\qquad j=1,\ldots,2N \,.
\label{alt2}
\end{align}
The $V$'s (\ref{V1V2}) can then be related to the closed-chain
transfer matrices (\ref{transferclosed}) by
\begin{align}
V^{(2)}(u)\,
V^{(1)}(u)=(-1)^N\,\tau(\tfrac{u}{2};\{\theta_j(\tfrac{u}{2})\})\,
\widehat{\tau}(\tfrac{u}{2};\{\theta_j(\tfrac{u}{2})\}) \,,
\end{align}
which is similar to (3.6), see Figure \ref{fig:periodic}.
In view of the relations (\ref{VtildeVrltn}) between $\tilde V$'s and
$V$'s, we conclude that $\tilde{\rT}_D$ (\ref{tcDnew}) is given by
\begin{align}
\tilde{\rT}_D(u) & = \tilde{V}^{(2)}(u)\, \tilde{V}^{(1)}(u) \nonumber \\
& = \Omega^{(1)} \, V^{(2)}(\tilde{u})\, V^{(1)}(\tilde{u})\, \Omega^{(1)} \nonumber \\
& =  (-1)^{N} \Omega^{(1)} \, \tau(\tfrac{\tilde{u}}{2}; \{
\theta_{j}(\tfrac{\tilde{u}}{2})\})\, \widehat{\tau}(\tfrac{\tilde{u}}{2}; \{
\theta_{j}(\tfrac{\tilde{u}}{2})\})\, \Omega^{(1)} \,.
\end{align}
The expression (\ref{eq:closedZ}) for the partition function in the closed channel can therefore be recast as
\begin{align}
Z^{\text{c}}(u,M,N) &= (-1)^{M N} \langle \Psi_{0} | U^{\dagger}
\Omega^{(1)} \left[ \tau(\tfrac{\tilde{u}}{2}; \{
\theta_{j}(\tfrac{\tilde{u}}{2})\})\, \widehat{\tau}(\tfrac{\tilde{u}}{2}; \{
\theta_{j}(\tfrac{\tilde{u}}{2})\})\right]^{M}
\Omega^{(1)} |\Psi_{0} \rangle \nonumber \\
&= (-1)^{(M+1) N} \langle \Phi_{0} | U^{\dagger}
\left[ \tau(\tfrac{\tilde{u}}{2}; \{
\theta_{j}(\tfrac{\tilde{u}}{2})\})\, \widehat{\tau}(\tfrac{\tilde{u}}{2}; \{
\theta_{j}(\tfrac{\tilde{u}}{2})\})\right]^{M}
|\Phi_{0} \rangle \,,
\label{Zcrossed2}
\end{align}
where $|\Phi_{0} \rangle$ is the so-called dimer state
\begin{align}
|\Phi_0\rangle=\Omega^{(1)}|\Psi_0\rangle=(-1)^N\,\Omega^{(2)}|\Psi_0\rangle=|\phi_0\rangle^{\otimes N},\qquad
|\phi_0\rangle=|\uparrow\,\rangle\otimes|\downarrow\,\rangle-|\downarrow\,\rangle\otimes|\uparrow\,\rangle,
\label{dimer}
\end{align}
and we have also used the fact that $U^{\dagger}\Omega^{(1)}U=\Omega^{(2)}$. We now insert in (\ref{Zcrossed2}) the completeness relation in
terms of Bethe states (which are $SU(2)$ highest-weight states) and
their lower-weight descendants
\begin{align}
\sum_{K} \sum_{\text{sol}_{\rm c}(N, K)}
\frac{1}{{\cal N}(\mathbf{u})}| \mathbf{u}\rangle \langle \mathbf{u}| + \ldots = \id \,,
\end{align}
where
$\text{sol}_{\rm c}(N, K)$ stands for physical solutions $\mathbf{u}$ of the
closed-chain BAE (\ref{BEgenclosed}) with $2N$ sites and $K$ Bethe roots. Moreover,
the normalization factor is given by
\begin{align}
{\cal N}(\mathbf{u}) =  \langle \mathbf{u}|\mathbf{u}\rangle  \,,
\label{normalization}
\end{align}
and the ellipsis denotes the descendant terms.  However, these
descendant terms do not contribute to the matrix element
(\ref{Zcrossed2}), since the dimer state is annihilated by the spin
raising and lowering operators
\begin{align}
S^{\pm}\,  |\Phi_{0} \rangle = 0 \,, \qquad \text{where} \qquad
S^{\pm} = S^{x} \pm i S^{y}\,, \qquad \vec S = \sum_{j=1}^{2N}
\tfrac{1}{2}\vec \sigma_{j} \,.
\end{align}
Moreover, in view of the fact
\begin{align}
0 = \langle \mathbf{u} | S^{z} |\Phi_{0} \rangle =
(N-K) \langle \mathbf{u} | \Phi_{0} \rangle \,,
\end{align}
the overlap $\langle \mathbf{u} | \Phi_{0} \rangle$
vanishes unless $K=N$. The matrix element (\ref{Zcrossed2})
therefore reduces to
\begin{align}
Z^{\text{c}}(u,M,N) &=
(-1)^{(M+1) N} \sum_{\text{sol}_{c}(N, N)}
\frac{\langle \Phi_{0} | U^{\dagger} | \mathbf{u} \rangle
 \langle \mathbf{u} | \Phi_{0} \rangle}{{\cal N}(\mathbf{u})}
\left[ \Lambda_{\rm c}(\tfrac{\tilde{u}}{2}; \{
\theta_{j}(\tfrac{\tilde{u}}{2})\})\, \widehat{\Lambda}_{c}(\tfrac{\tilde{u}}{2}; \{
\theta_{j}(\tfrac{\tilde{u}}{2})\})\right]^{M}\,,
\label{Zcrossed3}
\end{align}
where the sum runs over all physical solutions of the closed-chain BAE
(\ref{BEgenclosed}) in the $K=N$ sector. We note that the expressions
involving the eigenvalues are given by
\begin{align}
\Lambda_{\rm c}(\tfrac{\tilde{u}}{2}; \{\theta_{j}(\tfrac{\tilde{u}}{2})\})
&= (-i u)^{N}
\prod_{k=1}^{N} \frac{-\frac{u}{2}-u_{k}-i}{-\frac{u}{2}-u_{k}} \,, \nonumber\\
\widehat{\Lambda}_{c}(\tfrac{\tilde{u}}{2}; \{\theta_{j}(\tfrac{\tilde{u}}{2})\})	
&= (-i u)^{N}
\prod_{k=1}^{N} \frac{\frac{u}{2}-u_{k}+i}{\frac{u}{2}-u_{k}} \,,	
\label{Lambdasimpler}
\end{align}
as follows from (\ref{Lambdagenclosed}) and
(\ref{hatLambdagenclosed}).
The normalization factor
(\ref{normalization}) is given by the Gaudin formula
\cite{Gaudin:1981cyg, Korepin:1982gg}
\begin{align}
{\cal N}(\mathbf{u}) &= (-1)^{N}
\prod_{j=1}^{N}
\left[\left(u_{j} + \tfrac{\tilde{u}}{2}\right)^{2} + \tfrac{1}{4}
\right]^{N}  \left[\left(u_{j} - \tfrac{\tilde{u}}{2}\right)^{2} + \tfrac{1}{4}
\right]^{N}
\left[ \prod_{j, k=1\,; j \ne
k}^{N}\frac{u_{j}-u_{k}-i}{u_{j}-u_{k}}\right]\,
{\rm det}{}_{N}\left(G_{jk}\right) \,,
\label{Gaudin}
\end{align}
where
\begin{align}
G_{jk} = \delta_{jk}\left\{
N \left[ K_{\frac{1}{2}}(u_{j} -\tfrac{\tilde{u}}{2}) +
K_{\frac{1}{2}}(u_{j} +\tfrac{\tilde{u}}{2}) \right]
- \sum_{l=1}^{N} K_{1}(u_{j} - u_{l}) \right\} +  K_{1}(u_{j} - u_{k}) \,,
\end{align}
and
\begin{align}
K_{a}(u) = \frac{2a}{u^{2} + a^{2}}\,.
\label{Ka}
\end{align}
Overlaps similar to $\langle \mathbf{u}| \Phi_{0} \rangle$ have been studied extensively, see \emph{e.g.} \cite{Pozsgay:2013,
Brockmann:2014a, Brockmann:2014b, Piroli:2017sei, Pozsgay:2018ybn,
deLeeuw:2015hxa, Buhl-Mortensen:2015gfd, deLeeuw:2016umh}, see also \cite{Tsuchiya:1998, Kozlowski:2012fv}. The cases of even and odd $N$ must be analyzed separately.

\subsection{Even $N$}
Let us first consider even values of $N$.
Interestingly, only Bethe states with
``paired''  Bethe roots of the form
\begin{align}
\{ u_{1} \,, -u_{1} \,, \ldots \,, u_{\frac{N}{2}} \,,
-u_{\frac{N}{2}} \}
\label{paired}
\end{align}
have non-zero overlaps \cite{Brockmann:2014b,  Piroli:2017sei}. Such Bethe states
have even parity, see (\ref{paritypairedevenN}) below.
For such Bethe states, the overlaps are given by (see Appendix \ref{app:overlap})
\begin{align}
\langle \mathbf{u} | \Phi_{0} \rangle
=\left( \tfrac{\tilde{u}}{2} + \tfrac{i}{2} \right)^{N}\,
\left[\prod_{j=1}^{\frac{N}{2}}\frac{1}{u_{j}
\sqrt{u_{j}^{2}+\frac{1}{4}}} \right]
\sqrt{ \frac{ {\rm det}{}_{\frac{N}{2}}\left(G^{+}_{jk}\right)}
 {{\rm det}{}_{\frac{N}{2}}\left(G^{-}_{jk}\right)}} \sqrt{{\cal N}(\mathbf{u})}\,,
\label{overlap}
\end{align}
where
\begin{align}
G^{\pm}_{jk} = \delta_{jk}\left\{
N \left[K_{\frac{1}{2}}(u_{j} -\tfrac{\tilde{u}}{2}) +
K_{\frac{1}{2}}(u_{j} +\tfrac{\tilde{u}}{2}) \right]
- \sum_{l=1}^{\frac{N}{2}} K^{(+)}_{1}(u_{j}\,, u_{l}) \right\} +
K^{(\pm)}_{1}(u_{j}\,, u_{k}) \,,
\end{align}
with
\begin{align}
K^{(\pm)}_{a}(u,v) = K_{a}(u-v) \pm K_{a}(u+v) \,,
\label{Kpma}
\end{align}
and $K_{a}(u)$ is defined in (\ref{Ka}). We remark that, for these
states,
\begin{align}
{\rm det}{}_{N}\left(G_{jk}\right)  =
{\rm det}{}_{\frac{N}{2}}\left(G^{+}_{jk}\right) \,
{\rm det}{}_{\frac{N}{2}}\left(G^{-}_{jk}\right) \,.
\end{align}
Moreover, we show in Appendix \ref{app:proof} the relation
\begin{align}
\langle \Phi_{0} | U^{\dagger} | \mathbf{u} \rangle  =
i^{N} (\tilde{u} + i)^{-N}  \Lambda_{\rm c}(\tfrac{\tilde{u}}{2}; \{
\theta_{j}(\tfrac{\tilde{u}}{2})\}) \langle \mathbf{u} | \Phi_{0}
\rangle \,.
\label{claim}
\end{align}
The matrix element (\ref{Zcrossed3}) therefore reduces to
\begin{align}
\label{eq:evenNZ}
Z^{\text{c}}(u,M,N) &=
\frac{i^{N} (-1)^{(M+1) N}}{2^{2N}}  (\tilde{u}+i)^{N}
\sum_{\text{sol}(u_{1}, \ldots, u_{\frac{N}{2}})}
\left[\prod_{j=1}^{\frac{N}{2}}\frac{1}{u_{j}^{2}
\left(u_{j}^{2}+\frac{1}{4}\right)} \right]
\frac{ {\rm det}{}_{\frac{N}{2}}\left(G^{+}_{jk}\right)}
 {{\rm det}{}_{\frac{N}{2}}\left(G^{-}_{jk}\right)} \nonumber \\
& \qquad\qquad\times \Lambda_{\rm c}(\tfrac{\tilde{u}}{2}; \{
\theta_{j}(\tfrac{\tilde{u}}{2})\})
\left[ \Lambda_{\rm c}(\tfrac{\tilde{u}}{2}; \{
\theta_{j}(\tfrac{\tilde{u}}{2})\})\, \widehat{\Lambda}_{c}(\tfrac{\tilde{u}}{2}; \{
\theta_{j}(\tfrac{\tilde{u}}{2})\})\right]^{M}\,, \nonumber \\
&= 2^{-2N} u^{2N(M+1)}
\sum_{\text{sol}(u_{1}, \ldots, u_{\frac{N}{2}})}
\left[\prod_{j=1}^{\frac{N}{2}}\frac{1}{u_{j}^{2}
\left(u_{j}^{2}+\frac{1}{4}\right)} \right]
\frac{ {\rm det}{}_{\frac{N}{2}}\left(G^{+}_{jk}\right)}
 {{\rm det}{}_{\frac{N}{2}}\left(G^{-}_{jk}\right)} \nonumber \\
& \qquad\qquad\times \left[ \prod_{k=1}^{\frac{N}{2}}
\left(\frac{\frac{u}{2}-u_{k}+i}{\frac{u}{2}-u_{k}}\right)
\left(\frac{\frac{u}{2}+u_{k}+i}{\frac{u}{2}+u_{k}}\right)\right]^{2M+1} \,,
\end{align}
where we have used (\ref{Lambdasimpler}) to pass to the second
equality, and
the sum is over all physical solutions of the BAE (\ref{BEgenclosed})
with paired Bethe roots (\ref{paired}), see (\ref{closedBAEeven}) below.

\subsection{Odd $N$}
For odd values of $N$, the only Bethe states with non-zero overlaps have
one 0 Bethe root, and all the other Bethe roots form pairs; \emph{i.e.}, the
Bethe roots are of the form
\begin{align}
\{ u_{1} \,, -u_{1} \,, \ldots \,, u_{\frac{N-1}{2}} \,,
-u_{\frac{N-1}{2}}\,,  0 \} \,.
\label{pairedodd}
\end{align}
Such Bethe states have odd parity, see (\ref{paritypairedoddN}) below.
The overlaps are now given by
\begin{align}
\langle \mathbf{u}| \Phi_{0} \rangle
= -\left( \tfrac{\tilde{u}}{2} + \tfrac{i}{2} \right)^{N}\,
\left[\prod_{j=1}^{\frac{N-1}{2}}\frac{1}{u_{j}
\sqrt{u_{j}^{2}+\frac{1}{4}}} \right]
\sqrt{ \frac{ {\rm det}{}_{\frac{N+1}{2}}\left(H_{jk}\right)}
 {{\rm det}{}_{\frac{N-1}{2}}\left(G^{-}_{jk}\right)}} \sqrt{{\cal N}(\mathbf{u})}\,,
\end{align}
where $H$ is the block matrix
\begin{align}
H = \left( \begin{array}{cc}
           G^{+} & 2 C \\
	   C^{t}     & D
	   \end{array} \right)_{\frac{N+1}{2} \times \frac{N+1}{2}} \,,
\label{Hmat}	
\end{align}
and $G^{\pm}$ are now the $\frac{N-1}{2} \times \frac{N-1}{2}$ matrices  given by
\begin{align}
G^{\pm}_{jk} &= \delta_{jk}\left\{
N \left[K_{\frac{1}{2}}(u_{j} -\tfrac{\tilde{u}}{2}) +
K_{\frac{1}{2}}(u_{j} +\tfrac{\tilde{u}}{2}) \right]
- K_{1}(u_{j})
- \sum_{l=1}^{\frac{N-1}{2}} K^{(+)}_{1}(u_{j}\,, u_{l}) \right\}
\nonumber\\
&\qquad\qquad\qquad\qquad +
K^{(\pm)}_{1}(u_{j}\,, u_{k}) \,, \qquad\qquad j\,, k = 1, \ldots \,,
\frac{N-1}{2} \,,
\end{align}
with $K^{(\pm)}_{a}(u,v)$ and $K_{a}(u)$ defined as before, see (\ref{Kpma}), (\ref{Ka}).
Moreover,
in (\ref{Hmat}),
$C$ is an $\frac{N-1}{2}$-component column vector,  $C^{t}$ is the
corresponding row vector, and $D$ is a scalar, which are given by
\begin{align}
C_{j} &= K_{1} (u_{j}) \,, \qquad j = 1, \ldots \,, \frac{N-1}{2} \,,
\nonumber \\
D &= 2 N K_{\frac{1}{2}}(\tfrac{\tilde{u}}{2}) -
2\sum_{l=1}^{\frac{N-1}{2}} K_{1}(u_{l}) \,.
\end{align}
We remark that, for these states,
\begin{align}
{\rm det}{}_{N}\left(G_{jk}\right)  =
{\rm det}{}_{\frac{N+1}{2}}\left(H_{jk}\right) \,
{\rm det}{}_{\frac{N-1}{2}}\left(G^{-}_{jk}\right) \,.
\end{align}
Moreover,
\begin{align}
\langle \Phi_{0} | U^{\dagger} | \mathbf{u} \rangle  =
i^{N}(\tilde{u} + i)^{-N} \Lambda_{\rm c}(\tfrac{\tilde{u}}{2}; \{
\theta_{j}(\tfrac{\tilde{u}}{2})\}) \langle \mathbf{u}| \Phi_{0} \rangle \,.
\end{align}
The matrix element (\ref{Zcrossed3}) now reduces to
\begin{align}
\label{eq:oddNZ}
Z^{\text{c}}(u,M,N) &=
\frac{i^{N} (-1)^{(M+1) N}}{2^{2N}}  (\tilde{u}+i)^{N}
\sum_{\text{sol}(u_{1}, \ldots, u_{\frac{N-1}{2}})}
\left[\prod_{j=1}^{\frac{N-1}{2}}\frac{1}{u_{j}^{2}
\left(u_{j}^{2}+\frac{1}{4}\right)} \right]
\frac{ {\rm det}{}_{\frac{N+1}{2}}\left(H_{jk}\right)}
 {{\rm det}{}_{\frac{N-1}{2}}\left(G^{-}_{jk}\right)} \nonumber \\
& \qquad\qquad\times \Lambda_{\rm c}(\tfrac{\tilde{u}}{2}; \{
\theta_{j}(\tfrac{\tilde{u}}{2})\})
\left[ \Lambda_{\rm c}(\tfrac{\tilde{u}}{2}; \{
\theta_{j}(\tfrac{\tilde{u}}{2})\})\, \widehat{\Lambda}_{c}(\tfrac{\tilde{u}}{2}; \{
\theta_{j}(\tfrac{\tilde{u}}{2})\})\right]^{M}\,, \nonumber \\
&=
2^{-2N} u^{2N(M+1)}
\sum_{\text{sol}(u_{1}, \ldots, u_{\frac{N-1}{2}})}
\left[\prod_{j=1}^{\frac{N-1}{2}}\frac{1}{u_{j}^{2}
\left(u_{j}^{2}+\frac{1}{4}\right)} \right]
\frac{ {\rm det}{}_{\frac{N+1}{2}}\left(H_{jk}\right)}
 {{\rm det}{}_{\frac{N-1}{2}}\left(G^{-}_{jk}\right)} \nonumber \\
& \qquad\qquad\times \left[ \left(\frac{u+2i}{u}\right) \prod_{k=1}^{\frac{N-1}{2}}
\left(\frac{\frac{u}{2}-u_{k}+i}{\frac{u}{2}-u_{k}}\right)
\left(\frac{\frac{u}{2}+u_{k}+i}{\frac{u}{2}+u_{k}}\right)\right]^{2M+1} \,,
\end{align}
where we have used (\ref{Lambdasimpler}) to pass to the second
equality, and
the sum is over all physical solutions of the BAE (\ref{BEgenclosed})
with paired Bethe roots (\ref{pairedodd}), see (\ref{closedBAEodd}) below.

\subsection{BAE and $Q$-system}
We now summarize the BAE and $Q$-systems in the closed channel.  They
are special cases of those for the spin chain with periodic boundary
condition with length $2N$ and magnon number $N$.
\paragraph{Even $N$.} For the paired Bethe roots (\ref{paired}), the closed-chain Bethe
equations (\ref{BEgenclosed}) reduce to open-chain-like Bethe equations
\begin{align}
\left(\frac{u_{k} - \tfrac{\tilde{u}}{2} + \frac{i}{2}}
{u_{k} - \tfrac{\tilde{u}}{2} - \frac{i}{2}}\right)^{N}
\left(\frac{u_{k} + \tfrac{\tilde{u}}{2} + \frac{i}{2}}
{u_{k} + \tfrac{\tilde{u}}{2} - \frac{i}{2}}\right)^{N}
&= \left(\frac{u_{k}+\frac{i}{2}}{u_{k}-\frac{i}{2}}\right)
\prod_{j=1; j\ne
k}^{\frac{N}{2}}\left(\frac{u_{k}-u_{j}+i}{u_{k}-u_{j}-i}\right)
\left(\frac{u_{k}+u_{j}+i}{u_{k}+u_{j}-i}\right) \,,
\label{closedBAEeven}
\end{align}
where $k=1,\ldots,\frac{N}{2}$. The corresponding $QQ$-relations are
\begin{align}
Q_{a+1,s}(v)\, Q_{a,s+1}(v) \propto Q^{+}_{a+1,s+1}(v)\,
Q^{-}_{a,s}(v) - Q^{-}_{a+1,s+1}(v)\, Q^{+}_{a,s}(v) \,,
\label{QQclosed}
\end{align}
where $Q_{a,s}(v)$ are {\em even} polynomial functions of $v$. In particular, the main $Q$-function is given by
\begin{align}
Q_{1,0}(v)
=\prod_{j=1}^{\frac{N}{2}}(v-u_j)(v+u_j)= \sum_{k=0}^{\frac{N}{2}}
c_{1,0}^{(2k)} v^{2k} =
v^N+c_{1,0}^{(N-2)}v^{N-2}+\cdots+c_{1,0}^{(0)} \,.
\label{Q10even}
\end{align}
Moreover,
\begin{align}
Q_{0,0}(v) =
\left[(v-\tfrac{\tilde{u}}{2})(v+\tfrac{\tilde{u}}{2})\right]^{N} \,.
\label{Q00sclosed}
\end{align}
Therefore, to obtain the ZRC for this case, we can simply take the ZRC for the generic periodic case and add the following constraints
\begin{align}
c_{1,0}^{(2k+1)}=0,\qquad k=0,1,\ldots,\tfrac{N}{2}-1.
\end{align}

\paragraph{Odd $N$.} For odd $N$, the nonzero paired Bethe roots (\ref{pairedodd}) satisfy the
open-chain-like Bethe equations
\begin{align}
\left(\frac{u_{k} - \tfrac{\tilde{u}}{2} + \frac{i}{2}}
{u_{k} - \tfrac{\tilde{u}}{2} - \frac{i}{2}}\right)^{N}
\left(\frac{u_{k} + \tfrac{\tilde{u}}{2} + \frac{i}{2}}
{u_{k} + \tfrac{\tilde{u}}{2} - \frac{i}{2}}\right)^{N}
&= \left(\frac{u_{k}+\frac{i}{2}}{u_{k}-\frac{i}{2}}\right)
\left(\frac{u_{k}+i}{u_{k}-i}\right)
\prod_{j=1; j\ne
k}^{\frac{N-1}{2}}\left(\frac{u_{k}-u_{j}+i}{u_{k}-u_{j}-i}\right)
\left(\frac{u_{k}+u_{j}+i}{u_{k}+u_{j}-i}\right) \,,
\label{closedBAEodd}
\end{align}
where $k = 1, \ldots, (N-1)/2$.  The corresponding $QQ$-relations are
again given by (\ref{QQclosed}), with $Q_{0,0}(v)$ given by
(\ref{Q00sclosed}).  The $Q$-functions are odd polynomials in this
case.  In particular, the main $Q$-function takes the form
\begin{align}
Q_{1,0}(v)=v\prod_{j=1}^{\frac{N-1}{2}}(v-u_j)(v+u_j)=\sum_{k=0}^{\frac{N-1}{2}}
c_{1,0}^{(2k+1)} v^{2k+1}
=v^N+c_{1,0}^{(N-2)}v^{N-2}+\cdots c_{1,0}^{(1)}v \,.
\label{Q10odd}
\end{align}
Therefore, to obtain the ZRC in this case, we take the general ZRC for
the generic periodic case and impose the conditions
\begin{align}
c_{1,0}^{(2k)}=0,\qquad k=0,1,\ldots,(N-1)/2.
\end{align}

For both even and odd values of $N$, we conjecture that the number
$\mathcal{N}(N)$ of such physical solutions of the BAE
(\ref{closedBAEeven}), (\ref{closedBAEodd}) is given simply by
\begin{align}
\mathcal{N}(N) ={N\choose \lfloor N/2 \rfloor} \,,
\label{closedconj}
\end{align}
where $\lfloor x \rfloor$ denotes the integer part of $x$.
The first 10 values are given by
\begin{align}
\{1,2,3,6,10,20,35,70,126,252\}\,,
\end{align}
which we checked by explicit computations.

\subsection{Algebraic geometry}
The procedure for algebro-geometric computations follows the same steps
as in the open channel.  As we mentioned before, the computation of
the Gr\"obner basis and quotient ring is simpler.  The complication comes
from computing the companion matrices.  The companion matrix of the
transfer matrices $\Lambda_{\rm c}(v;\{\theta_j(u)\})$ can be constructed
similarly from the $TQ$ relation
\begin{align}
Q(v)T(v-\tfrac{i}{2})=\left[(v+\tfrac{i}{2})^2-u^2\right]^N Q(v-i)+\left[(v-\tfrac{i}{2})^2-u^2\right]^NQ(v+i).
\end{align}
The most complicated part is the ratio of determinants in
(\ref{eq:evenNZ}) and (\ref{eq:oddNZ}).  These are complicated
functions in terms of rapidities $\mathbf{u}$.  As in the open
channel, the natural variables that enter the AG computation are
$c_{a,s}^{(k)}$.  Therefore, in order to construct the companion
matrices of the ratio of determinants, we need to first convert it to
be functions $c_{a,s}^{(k)}$.  This can be done because the ratio of
determinants are \emph{symmetric} rational functions.\par
\paragraph{Even $N$.} For even $N$, after expanding the determinant the result can be written in the form
\begin{align}
\frac{\rN(u_1,\ldots,u_{N/2})}{\rD(u_1,\ldots,u_{N/2})} \,,
\end{align}
where $\rN(u_1,\ldots,u_{N/2})$ and $\rD(u_1,\ldots,u_{N/2})$ are
\emph{symmetric polynomials} in
$\{u_1^{2},\ldots,u_{N/2}^{2}\}$.  By the
fundamental theorem of symmetric polynomials, they can be written in
terms of elementary symmetric polynomials of
$\{u_1^{2},\ldots,u_{N/2}^{2}\}$,
which we denote by $\{s_0,s_1,\ldots,s_{N/2-1}\}$:
\begin{align}
s_0=&\,u_1^{2} u_2^{2} \cdots u_{\frac{N}{2}}^{2} \,,\\\nonumber
\cdots\\\nonumber
s_{\frac{N}{2}-2}=&\,u_1^{2} u_2^{2} +u_1^{2} u_3^{2}
+\ldots+u_{\frac{N}{2}-1}^{2} u_{\frac{N}{2}}^{2} \,, \\\nonumber
s_{\frac{N}{2}-1}=&\,u_1^{2}+u_2^{2}+\ldots+u_{\frac{N}{2}-1}^{2} \,.
\end{align}
They are related to the coefficients $c_{1,0}^{(2k)}$ in
(\ref{Q10even}) as
\begin{align}
c_{1,0}^{(2k)}=(-1)^{\frac{N}{2}+k}s_k \,, \qquad k=0,1,\ldots,\tfrac{N}{2}-1 \,.
\end{align}

\paragraph{Odd $N$.} For odd $N$, the result can be written as
\begin{align}
\frac{\rN(u_1,\ldots,u_{\frac{N-1}{2}})}{\rD(u_1,\ldots,u_{\frac{N-1}{2}})}
\end{align}
Similarly, we can do the symmetry reduction and write the result in terms of the elementary symmetric polynomials
\begin{align}
s_0=&\,u_1^{2} u_2^{2}\cdots u_{\frac{N-1}{2}}^{2} \,,\\\nonumber
\cdots\\\nonumber
s_{\frac{N-1}{2}-2}=&\,u_1^{2} u_2^{2} +u_1^{2} u_3^{2}
+\ldots+u_{\frac{N-1}{2}-1}^{2} u_{\frac{N-1}{2}}^{2} \,,\\\nonumber
s_{\frac{N-1}{2}-1}=&\,u_1^{2}+u_2^{2}+\ldots +u_{\frac{N-1}{2}-1}^{2} \,.
\end{align}
They are related to the coefficients $c_{1,0}^{(2k+1)}$ in
(\ref{Q10odd}) as
\begin{align}
c_{1,0}^{(2k+1)}=(-1)^{\frac{N-1}{2}+k}s_k,\qquad
k=0,1,\ldots,\tfrac{N-1}{2}-1 \,.
\end{align}

There are two sources of complication worth mentioning. Firstly, computing the determinant explicitly and performing the symmetric reduction is straightforward in principle, but becomes cumbersome very quickly. It would be desirable to have a simpler form for these quantities. Secondly, the companion matrix of the quantity $1/\rD$ is the inverse of the companion matrix of $\rD$. Computing the inverse of a matrix analytically is also straightforward, but it has a negative impact on the efficiency of the computations when the dimension of the matrix becomes large. For the eigenvalues of the transfer matrix, we saw in \eqref{eq:openTQ} that the problem of computing inverses can be circumvented by using the $TQ$-relations. For the expression of the overlaps, it is not clear whether we can find better means to compute the companion matrix of the ratio $\rN/\rD$ so as to avoid taking matrix inverses.

Using the algebro-geometric approach in the closed channel, we
computed partition functions for $N$ up to 7 and $M$ up to 2048.
The results for $2\le M,N\le 6$ are
listed in Appendix \ref{eq:exactsmallMN}.

%%%%%%%%%%%%%%%%%%%%%%%%%%%%%%%%%%%%%%%%%%%%%%%%%%%%%%%
\section{Algebraic equation with free parameters}
\label{sec:AG}
%%%%%%%%%%%%%%%%%%%%%%%%%%%%%%%%%%%%%%%%%%%%%%%%%%%%%%%
In this section, we discuss the Gr\"obner basis of the ZRC in the
closed channel in more detail.  This will demonstrate further the
power of the algebro-geometric approach for algebraic equations,
especially for cases with free parameters.\par

The system of algebraic equations we consider depends on a parameter
$u$.  This means that the coefficients of the equations are no longer
pure numbers, but functions of $u$.  As a result, the solutions also
depend on the parameter $u$.  As we vary the parameter $u$, the
solutions also change.  One important question is if there are any
special values $u$ where the solution space changes drastically.  To
understand this point, let us consider the following simple equation
for $x$ whose coefficients depend on the free parameter $u$
\begin{align}
(u^2-1)x^2+u x-1=0.
\end{align}
At generic values of $u$, this is a quadratic equation with two
solutions.  However, when $u=\pm 1$, the leading term vanishes and the
equation become linear.  The number of solutions becomes one.
Therefore at these `singular' points, the structure of the solution
space changes drastically.\par

A similar phenomena occurs in the BAE of the Heisenberg spin chain.
Consider for a moment the more general XXZ spin chain and take the
anisotropy parameter ({\em alias} quantum group deformation parameter)
$q$ as the free parameter of the BAE. It is well-known that the
solution space is very different between generic $q$ and $q$ being a
root of unity.  The traditional way to see this is by studying
representation theory of the $U_q(\mathfrak{sl}(2))$ symmetry of the
spin chain \cite{Pasquier:1989kd}.  A more straightforward way to see
this fact is by the algebro-geometric approach.  We can compute the
Gr\"obner basis of the corresponding BAE/$Q$-system and analyze
the coefficients as functions of $q$.  We shall discuss this
problem in more detail in a future publication.

Related to the current work, we consider the ZRC in the closed channel
for the XXX spin chain. Here the free parameter is the inhomogeneity
$u$. We want to know whether there are special singular points of $u$
where the structure of solution space changes drastically. Recall that
from elementary algebraic geometry, the number of solutions equals the linear dimension of the quotient
ring. Furthermore, the quotient ring dimension is completely
determined by the leading terms of the Gr\"obner basis. Therefore, to this end, we compute the Gr\"obner basis explicitly. For $N=3$, the ideal can be written as $\langle g_1,g_2,g_3\rangle$ where the elements of the Gr\"obner basis $g_i$ are given by
\begin{align}
\label{eq:Grn3}
g_1=&\,192 s_1^3+(192u^2-208)s_1^2-(192u^4+288u^2-4)s_1-192u^6+48u^4+60u^2+9 \,, \\\nonumber
g_2=&\,s_2 \,, \\\nonumber
g_3=&\,s_0 \,.
\end{align}
Here we have chosen the ordering
\begin{align}
s_0\prec s_1\prec s_2.
\end{align}
We see from (\ref{eq:Grn3}) that \emph{the leading terms are independent of $u$}. This implies that the dimension of the quotient ring $\mathbb{C}[s_0,s_1,s_2]/\langle g_1,g_2,g_3\rangle$, or equivalently the number of solutions, is independent of the value of $u$. Of course, the explicit solutions of the BAE will depend on the value of $u$, but there will always be 3 solutions to the ZRC for $N=3$ at any value of $u$.\par

Similarly, we can write down a slightly more non-trivial example for $N=4$. The ideal is given by $\langle g_1,\cdots,g_6\rangle$ where the Gr\"obner basis elements $g_i$ are given by
\small{
\begin{align}
g_1=&\,\mathbf{11520 s_{0}^3}+(-864 - 3840 u^2 - 13824 u^4)s_{0}^2+(3 + 48 u^2 + 288 u^4 + 768 u^6 + 768 u^8)s_{2}^2\\\nonumber
&\,+(-1296 + 1728 u^2 + 2304 u^4 - 3072 u^6) s_{0}s_{2}+(-4 - 64 u^2 - 384 u^4 - 1024 u^6 - 1024 u^8)s_{2}\\\nonumber
&\,+(-251 - 3632 u^2 - 1056 u^4 + 3328 u^6 - 6912 u^8)s_{0}+(-2 - 32 u^2 - 192 u^4 - 512 u^6 - 512 u^8),\\\nonumber
g_2=&\,\mathbf{2304 s_{0}^2s_{2}}-1536s_{0}s_{2}+(768 + 3072 u^2)s_{0}^2+(-1 - 16 u^2 - 96 u^4 - 256 u^6 - 256 u^8)s_{2}\\\nonumber
&\,+(-80 - 2624 u^2 - 768 u^4 + 1024 u^6) s_{0}+(-2 - 32 u^2 - 192 u^4 - 512 u^6 - 512 u^8),\\\nonumber
g_3=&\,\mathbf{768 s_{0}s_{2}^2}+(-2304 + 1024 u^2)s_{0}s_{2}+1280 s_{0}^2+(-96 - 3840 u^2 - 1536 u^4)s_{0}\\\nonumber
&\,+(-3 - 48 u^2 - 288 u^4 - 768 u^6 - 768 u^8),\\\nonumber
g_4=&\,\mathbf{48 s_{2}^3}+(-48 + 64 u^2)s_{2}^2-352 s_{0}s_{2}+(-6 - 112 u^2 - 96 u^4)s_{2}\\\nonumber
&\,+(16 - 576 u^2)s_{0}+(3 + 12 u^2 - 48 u^4 - 192 u^6)\\\nonumber
g_5=&\,\mathbf{s_{1}}\\\nonumber
g_6=&\,\mathbf{s_3}.
\end{align}
}\normalsize
The leading terms are in boldface letters.  We see again these terms
are independent of $u$.  For all the values of $N$ which we compute,
this is true. It would be nice to prove this for general $N$.\par

Therefore from the algebro-geometric computation, we conclude that for
\emph{any value} of $u$, there exist solutions with definite parity (parity even/odd for even/odd $N$).  For fixed $N$, the number of
solutions is the same for any value of $u$, which has been given in
(\ref{closedconj}).

We end this section by the following comment. The conclusion that there always exist solutions with definite parity for any $u$ is far from obvious from the ZRC or original BAE. It is also not easy to see this from numerical computations. On the contrary, it is a straightforward observation from the algebro-geometric computation. This shows again that algebro-geometric approach is a powerful tool to analyze the solution space of BAE.

%%%%%%%%%%%%%%%%%%%%%%%%%%%%%%%%%%%%%%%%%%%%%%
\section{Analytical results in closed form}
\label{sec:analyticalResult}
%%%%%%%%%%%%%%%%%%%%%%%%%%%%%%%%%%%%%%%%%%%%%%
In this section, we discuss the analytical results which can be
written in closed forms for arbitrary $N$ and $M$ in the open and
closed channel respectively.

\subsection{Open channel} We first discuss the open channel. The partition function takes the same form as the torus case, which is written as the trace of the $N$-th power of the transfer matrix. If the eigenvalues of the transfer matrix can be found analytically, we can write down the partition function for any $N$. Here by analytical we mean more precisely expressible in terms of \emph{radicals}. In the algebro-geometric approach, we first compute the companion matrix of the eigenvalue of the transfer matrix. The dimension of the companion matrix equals the number of physical solutions of the open channel BAE/Q-system. The eigenvalues of the companion matrix give the eigenvalues of the transfer matrices evaluated at each solution. From Galois theory, if the dimension of the companion matrix is less than 5, the eigenvalues can be expressed in terms of radicals. Therefore for the values of $M$ where all the companion matrices have dimension less than 5, we can obtain the analytical expression for any $N$. This requirement is only met by $M=1$. Already for $M=2$, we need to consider the sectors $K=0,1,2$, and for $K=1$ and $2$ the dimensions of the companion matrices are $4$ and $6$ respectively. For larger $M$, the dimensions of the companion matrices are even larger. We give the closed form expression for $M=1$ and any $N$ in what follows.
\paragraph{The $M=1$ case.} We need to consider $K=0,1$. For $K=0$, the eigenvalue of the transfer matrix is given by
\begin{align}
\Lambda_{\rD,0}(u)=(u+i)^2.
\end{align}
For $K=1$, the solution of BAE takes the form $\{u_1,-u_1\}$. The companion matrix is 2-dimensional. The two eigenvalues of the transfer matrix in this sector are given by
\begin{align}
\lambda_1(u)=&\,-\frac{u^2}{2}-1-\frac{iu}{2}\sqrt{3u^2+4} \,,\\\nonumber
\lambda_2(u)=&\,-\frac{u^2}{2}-1+\frac{iu}{2}\sqrt{3u^2+4} \,.
\end{align}
The closed-form expression of the partition function, taking into account the $su(2)$ multiplicities \eqref{su2degeneracy}, is then
\begin{align}
\label{eq:closedformM1}
Z(u,1,N)=4(u+i)^{2N}+2\left(\lambda_1(u)^N+\lambda_2(u)^N\right) \,.
\end{align}

Let us make one comment on the comparison with the torus case.  The
closed-form results have been found up to\footnote{Note that in the
torus case, $M$ denoted the length of the spin chain \cite{Jacobsen:2018pjt}, while here, in the cylinder
case, the length of the spin chain is given by $2M+1$.} $M=6$ in the
torus case \cite{Jacobsen:2018pjt}.  There we also used the fact that certain companion
matrices can be further decomposed into smaller blocks, which implies
the existence of non-trivial primary decompositions over the field
$\mathbb{Q}$.  Physically, this primary decomposition is related to
decomposing the solutions of BAE/Q-system according to the total
momentum.  In the cylinder case, however, the total momentum is
automatically zero for all allowed solutions, due to the presence of the
boundary.  Therefore, further decomposition according to the total
momentum is not possible in the current case.

\subsection{Closed channel}
The situation is more interesting in the closed channel. The expression for the partition function is qualitatively different from the torus case, since we have a new ingredient: the non-trivial overlap between Bethe states and the boundary state. To find the analytical expressions, we first compute the companion matrices, both for the transfer matrix and the overlaps. For the values of $N$ where the dimensions of the companion matrices are less than 5, we can express the final result in terms of radicals for any $M$. This is satisfied by $N=1,2,3$. The dimensions of the companion matrices are $1,2,3$ respectively. We present the analytical results for these cases in what follows.
\paragraph{The $N=1$ case.} This case is somewhat trivial, but we give it here for completeness. There is only one allowed solution to the BAE, which is $\{0\}$. The eigenvalue of the transfer matrix is given by
\begin{align}
\Lambda_c=-i(u+2i).
\end{align}
The contribution from the overlaps only comes from $\det H_{jk}$ which is given by
\begin{align}
\det H_{jk}=\frac{8}{u(u+2i)}.
\end{align}
The partition function is thus given by \eqref{eq:oddNZ}.
\begin{align}
\label{eq:cfN1}
Z(u,M,1)=\frac{i(-1)^M}{4}u\,\left(-i(u+2i)\right)^{2M+1}\,\frac{8}{u(u+2i)}=2(u+2i)^{2M}.
\end{align}

\paragraph{The $N=2$ case.} This is the simplest non-trivial case where $N$ is even. The Bethe roots take the form $\{u_1,-u_1\}$. There are two such solutions, which can be found straightforwardly by directly solving Bethe equations. Let us denote the companion matrices of $\Lambda_c\big(\tfrac{\tilde{u}}{2},\{\tfrac{\tilde{u}}{2}\}\big)$ by $\mathbb{T}_2(u)$ and companion matrix of the following factor
\begin{align}
\frac{1}{u_1(u_1^2+\frac{1}{4})}\frac{\det G^+_{jk}}{\det G^-_{jk}}
\end{align}
by $\mathbb{F}_2(u)$. The partition function is given by \eqref{eq:evenNZ}
\begin{align}
\label{eq:cfN2}
Z(u,M,2)=-\frac{1}{16}u^2\,\tr\left[\mathbb{T}^{2M+1}_2(u) \cdot \mathbb{F}_2(u)\right] \,.
\end{align}
Let us denote the eigenvalues of $\mathbb{T}_2(u)$ and $\mathbb{F}_2(u)$ by $\lambda_{\mathrm{T},i}(u)$ and $\lambda_{\mathrm{F},i}$, $i=1,2$ respectively. They are given by
\begin{align}
\lambda_{\mathrm{T},1}(u)=&\,1-iu-\sqrt{1+(u+2i)(u+i)^2u} \,,\\\nonumber
\lambda_{\mathrm{T},2}(u)=&\,1-iu+\sqrt{1+(u+2i)(u+i)^2u}
\end{align}
and
\begin{align}
\lambda_{\mathrm{F},1}(u)=&\,\frac{32}{u^2(u+2i)^2}+\frac{16(u^2+2i u-2)\sqrt{1+(u+2i)(u+i)^2u}}{u^2(u+2i)^2(u^4+4iu^3-5u^2-2iu+1)} \,,\\\nonumber
\lambda_{\mathrm{F},2}(u)=&\,\frac{32}{u^2(u+2i)^2}-\frac{16(u^2+2i u-2)\sqrt{1+(u+2i)(u+i)^2u}}{u^2(u+2i)^2(u^4+4iu^3-5u^2-2iu+1)} \,.
\end{align}
Collecting all the results, the explicit closed-form expression for $N=2$ and any $M$ is given by
\begin{align}
Z(u,M,2)=-\frac{1}{16}u^2\left([\lambda_{\mathrm{T},1}(u)]^{2M+1}\lambda_{\mathrm{F},1}(u)+ [\lambda_{\mathrm{T},2}(u)]^{2M+1}\lambda_{\mathrm{F},2}(u)\right) \,.
\end{align}
One may check that this agrees in particular with \eqref{Zexamples} for $M=2$.
We see here that the eigenvalues take rather complicated forms in terms of radicals whose arguments are polynomials of $u$. Nevertheless, the final result is a polynomial, as it should be.

\paragraph{The ${N=3}$ case.} This is the simplest non-trivial case where $N$ is odd. The results are bulky, therefore it is more convenient to write them in terms of smaller building blocks. To this end, we recall the solution for cubic polynomial equations. Let us consider the following generic cubic equation
\begin{align}
\label{eq:cubicE}
a x^3+b x^2+c x+d=0 \,,
\end{align}
where $a\ne 0$. We define
\begin{align}
\Delta_0=b^2-3ac,\qquad \Delta_1=2b^3-9abc+27a^2 d
\end{align}
and
\begin{align}
\label{eq:CT}
C=\left(\frac{\Delta_1+\sqrt{\Delta_1^2-4\Delta_0^3}}{2}\right)^{1/3} \,.
\end{align}
Then the three solutions of the cubic equation (\ref{eq:cubicE}) are given by
\begin{align}
\label{eq:cubicSol}
x_k=-\frac{1}{3a}\left(b+\xi^k\,C+\frac{\Delta_0}{\xi^k C}\right),\qquad k=1,2,3
\end{align}
where $\xi=(-1+i\sqrt{3})/2$.

For $N=3$, the solutions of the Bethe equations take the form $\{u_1,-u_1,0\}$. There are three physical solutions. Let us denote the companion matrix of $\Lambda_c\big(\tfrac{\tilde{u}}{2},\{\tfrac{\tilde{u}}{2}\}\big)$ by $\mathbb{T}_3(u)$ and the companion matrix of the following factor
\begin{align}
\frac{1}{u_1(u_1^2+\frac{1}{4})}\frac{\det H_{jk}}{\det G^-_{jk}}
\end{align}
by $\mathbb{F}_3(u)$. The partition function is given by
\begin{align}
Z(u,M,3)=-\frac{i(-1)^{3M}}{64}u^3\,\tr\left[\mathbb{T}^{2M+1}_3(u) \cdot \mathbb{F}_3(u)\right] \,,
\end{align}
where the trace is over the 3-dimensional quotient ring. One can check explicitly that $\mathbb{T}_3(u)$ and $\mathbb{F}_3(u)$ commute with each other and can thus be diagonalized simultaneously. Let us denote their eigenvalues by $\lambda_{\mathrm{T},i}(u)$ and $\lambda_{\rF,i}(u)$, with $i=1,2,3$. The characteristic equations of $\mathbb{T}_3(u)$ and $\mathbb{F}_3(u)$ take cubic forms
\begin{align}
x^3+b_{\mathrm{T}}x^2+c_{\mathrm{T}}x+d_{\mathrm{T}}=0 \,, \qquad x^3+b_{\rF}x^2+c_{\rF}x+d_{\rF}=0 \,,
\end{align}
where the coefficients are rational functions of $u$. The characteristic equations can be solved by radicals using (\ref{eq:cubicSol}). The relevant quantities are given as follows. For the eigenvalues of $\mathbb{T}_3(u)$, we have
\begin{align}
a_{\mathrm{T}}=1,\qquad b_{\mathrm{T}}=2u^2+3i u-2
\end{align}
and
\begin{align}
\Delta_0^{\mathrm{T}}=&\,-6 i u^5+31 u^4+54 i u^3-41 u^2-12 i u+4 \,, \\\nonumber
\Delta_1^{\mathrm{T}}=&\,27 i u^9-216 u^8-684 i u^7+1096 u^6+1035 i u^5-840 u^4-666 i u^3+300 u^2+72 i u-16 \,.
\end{align}
The eigenvalues $\{\lambda_{\mathrm{T},1},\lambda_{\mathrm{T},2},\lambda_{\mathrm{T},3}\}$ are given by
\begin{align}
\label{eq:lambdaT}
\lambda_{\mathrm{T},i}(u)=-\frac{1}{3}\left(b_{\mathrm{T}}+\xi^i\,C_{\mathrm{T}}+\frac{\Delta_0^{\mathrm{T}}}{\xi^i C_{\mathrm{T}}}\right) \,, \qquad i=1,2,3
\end{align}
where $C_{\mathrm{T}}$ is defined as in (\ref{eq:CT}). For the eigenvalues of $\mathbb{F}_3$, we have
\begin{align}
a_{\rF}=1 \,, \qquad b_{\rF}=-\frac{512}{u^3(2i+u)^3}
\end{align}
and
\begin{align}
\Delta_0^{\rF}=\frac{16384}{u^6(2i+u)^6}\frac{P_0(u)}{P(u)}\,, \qquad \Delta_1^{\rF}=-\frac{4194304}{u^9(2i+u)^9}\frac{P_1(u)}{P(u)} \,,
\end{align}
where
\begin{align}
P(u)=&\,27 u^{10}+270 i u^9-1125 u^8-2520 i u^7+3345 u^6+2934 i u^5 \nonumber \\\nonumber
&\,-1875 u^4-420 i u^3-468 u^2-96 i u+64 \,,\\
P_0(u)=&\,27 u^{10}+270 i u^9-1395 u^8-4680 i u^7+10680 u^6+16704 i u^5\\\nonumber
&\,-16608 u^4-7872 i u^3-576 u^2-1536 i u+1024 \,,\\\nonumber
P_1(u)=&\,27 u^{10}+270 i u^9-1530 u^8-5760 i u^7+15360 u^6+29664 i u^5\\\nonumber
&\,-39648 u^4-33792 i u^3+11520 u^2-6144 i u+4096 \,.
\end{align}
The eigenvalues are given by\footnote{Notice that the powers of $\xi$
in (\ref{eq:lambdaF}) are slightly different from (\ref{eq:lambdaT}).
The reason for this convention is to make sure that
$\lambda_i^{\mathrm{T}}$ and $\lambda_i^{\mathrm{F}}$ correspond to
the same eigenvector.  Working directly with characteristic equations,
it is not immediately clear which eigenvalues correspond to the same
eigenvector.  We establish the correspondence by making numerical checks.  We choose $u$
to be some purely imaginary numbers such that the arguments in the
radicals are real and positive.}
\begin{align}
\label{eq:lambdaF}
\lambda_{\mathrm{F},i}(u)=-\frac{1}{3}\left(b_{\mathrm{F}}+\xi^{i-1}\,C_{\mathrm{F}}+\frac{\Delta_0^{\mathrm{F}}}{\xi^{i-1} C_{\mathrm{F}}}\right),\qquad i=1,2,3
\end{align}
Finally, combining all the results, the closed form expression is given by
\begin{align}
\label{eq:cfN3}
Z(u,M,3)=-\frac{i(-1)^{3M}}{64}u^3\sum_{i=1}^3(\lambda_{\mathrm{T},i})^{2M+1}\lambda_{\mathrm{F},i}(u) \,.
\end{align}

%%%%%%%%%%%%%%%%%%%%%%%%%%%%%%%%%%%%%%%%%%%%%%
\section{Zeros of partition functions}
\label{sec:zero}
%%%%%%%%%%%%%%%%%%%%%%%%%%%%%%%%%%%%%%%%%%%%%%
The study of partition function zeros is a well-known tool to access the phase diagram of
models in statistical physics. The seminal works by Lee and Yang \cite{Lee-Yang}
and by Fisher \cite{Fisher:zeros} studied the zeros of the Ising model partition function,
respectively with a complex magnetic field (at the critical temperature) and at a
complex temperature (in zero magnetic field). But more generally, any statistical model depending
on one (or more) parameters can be studied in the complex plane of the corresponding variable(s).
In particular, the chromatic polynomial with $Q \in \mathbb{C}$ colors has been used as a test bed
to develop a range of numerical, analytical and algebraic tools for computing partition function zeros
and analyzing their behavior as the (partial) thermodynamic limit is approached
\cite{Pottszeros1,Pottszeros2,Pottszeros3,Pottszeros4,Pottszerostorus,Berahanonplanar,Trilattpotts}. Further information about the physical relevance of studying partition function zeros can be found in \cite{Itzykson1991} and the extensive list of references in \cite{Pottszeros1}.

In the case at hand, we are interested in zeros of the partition function $Z(u,M,N)$ of the six-vertex model,
in the complex plane of the spectral parameter, $u \in \mathbb{C}$. As explained in section~\ref{sec:setup},
the algebro-geometric approach permits us to efficiently compute $Z(u,M,N)$ close to the partial thermodynamic limits
$N \gg M$ (open channel) or $M \gg N$ (closed channel), and more precisely for aspect ratios $\rho := N/M$
of the order $\sim 10^{3}$ and $\sim 10^{-3}$, respectively.

\subsection{Condensation curves}

An important result for analyzing these cases is the Beraha-Kahane-Weiss (BKW) theorem \cite{BKW}. When applied to
partition functions of the form \eqref{eq:explicitZ1} for the open channel, respectively \eqref{eq:evenNZ} or \eqref{eq:oddNZ}
for the closed channel, it states that the partition function zeros in the partial thermodynamic limits ($\rho \to \infty$ or
$\rho \to 0$, respectively) will condense on a set of curves in the complex $u$-plane that we shall refer to as {\em condensation curves}.
In particular, the condensation set cannot comprise isolated points, or areas.
By standard theorems of complex analysis, each closed region delimited by these curves constitutes a thermodynamic
phase (in the partial thermodynamic limit).

To be more precise, let $\Lambda_i(u)$ denote the eigenvalues of the relevant transfer matrix (for the open or
closed channel, respectively) that effectively contributes to $Z(u,M,N)$. For a given $u$, we order these eigenvalues by norm,
so that $|\Lambda_1(u)| \ge |\Lambda_2(u)| \ge \cdots$, and we say that an eigenvalue $\Lambda_i(u)$ is dominant
at $u$ if there does not exist any other eigenvalue having a strictly greater norm. Under a mild non-degeneracy
assumption (which is satisfied for the expressions of interest here), the BKW theorem \cite{BKW} states that the condensation curves
are given by the loci where there are (at least) two dominant eigenvalues, $|\Lambda_1(u)| = |\Lambda_2(u)|$. It is intuitively
clear that this defines curves, since the relative phase $\phi(u) \in \mathbb{R}$
defined by $\Lambda_2(u) = {\rm e}^{i \phi(u)} \Lambda_1(u)$ is allowed to vary along the curve. Moreover, a closer analysis \cite{Pottszeros1}
shows that the condensation curves may have bifurcation points (usually called T-points) or higher-order crossings when more
than two eigenvalues are dominant. They may also have end-points under certain conditions; see \cite{Pottszeros1} for more details.

A numerical technique for tracing out the condensation curves has been outlined in our previous paper on the toroidal
geometry \cite{Jacobsen:2018pjt}. It builds on an efficient method for the numerically exact diagonalization of the relevant transfer matrix,
and on a direct-search method that allows us to trace out the condensation curves. We refer the reader to \cite{Jacobsen:2018pjt}
for more details, and focus instead on a technical point that is important (especially in the closed channel) for correctly computing the condensation curves for
the cylindrical boundary conditions studied in this paper.

One might of course choose to obtain the eigenvalues by solving the BAE, either analytically or numerically.
However, the Bethe ansatz does not provide a general principle to order the eigenvalues by norm.
It is of course well known that in many, if not most, Bethe-ansatz solvable models, for ``physical'' values of
the parameters the dominant eigenvalue and its low-lying excitations are characterized by particularly
nice and symmetric arrangements of the Bethe roots, and hence one can easily single out those eigenvalues. However, we here wish to examine our model for {\em all}
complex values of the parameter $u$, and it is quite possible---and in fact true, as we shall see---that there will be a
complicated pattern of crossings (in norm) of eigenvalues throughout the complex $u$-plain.
To apply the BAE one would therefore have to make sure to obtain all the physical eigenvalues and compare their norm for each value of $u$.
By contrast, the numerical scheme (Arnoldi's method) that we use for the direct numerical diagonalization of the transfer matrix
is particularly well suited for computing only the first few eigenvalues (in norm), so we shall rely on it here. We shall later
compare the computed condensation curves with the zeros of partition functions obtained using Bethe ansatz and algebraic geometry.

The reader will have noticed that above we have twice referred to the diagonalization of a ``relevant'' transfer matrix. By this we mean a
transfer matrix whose spectrum contains only the eigenvalues that provide non-zero contributions to $Z(u,M,N)$, after taking
account of the boundary conditions via the trace \eqref{eq:openZ} in the open channel, or the sandwich between boundary states \eqref{eq:closedZ}
in the closed channel. These contributing eigenvalues correspond to the physical solutions in \eqref{eq:explicitZ1} for the open channel,
or in \eqref{eq:evenNZ} and \eqref{eq:oddNZ} for the closed channel. A ``relevant'' transfer matrix is thus not only a linear
operator that can build up the partition function $Z(u,M,N)$, but it must also have the correct dimension, namely
$\sum_{K} {\cal N}(M,K)$ given by \eqref{dimsol-open} in the open channel, or ${\cal N}(N)$ given by \eqref{closedconj}
in the closed channel. Ensuring this is an issue of representation theory. We begin by discussing it in the open channel, which is easier.

\subsection{Open channel}

The defining ingredient of the transfer matrix is the
$\check{R}$-matrix. Using \eqref{defcheckR}--\eqref{Rmat}, it reads
\begin{align}
\check{R}(u)=\left(
            \begin{array}{cccc}
              a(u) & 0 & 0 & 0 \\
              0 & c(u) & b(u) & 0 \\
              0 & b(u) & c(u) & 0 \\
              0 & 0 & 0 & a(u) \\
            \end{array}
          \right) \,,
\label{Rcheckmat}
\end{align}
with $a(u) = u+i$, $b(u)=u$ and $c(u)=i$. The most immediate transfer matrix approach is to let the diagonal-to-diagonal transfer
matrix $t_D(u)$ given by \eqref{eq:openT} act in the 6-vertex model representation, that is, on the space
$\{ |\uparrow\rangle, |\downarrow\rangle\}^{\otimes 2M+1}$ of dimension $2^{2M+1}$.

If we constrain to a fixed magnon number $K$, the dimension reduces to ${2M+1 \choose K}$. This is larger than ${\cal N}(M,K)$
given by \eqref{dimsol-open}, because we have not restricted to $su(2)$ highest-weight states. Therefore, each eigenvalue
would appear with a multiplicity given by \eqref{su2degeneracy}. Since each eigenvalue actually does contribute to $Z(u,M,N)$,
dealing with this naive representation provides a feasible route to computing the condensation curves (and this was actually the approach
used in \cite{Jacobsen:2018pjt}). However, the appearance of multiplicities is cumbersome and impedes the efficiency of the computations.

\subsubsection{Temperley-Lieb algebra}
\label{sec:TL-open}

To overcome this problem, notice that in the more general XXZ model with quantum-group
deformation parameter $q$, the integrable $\check{R}$-matrix may be taken as
\begin{equation}
\label{Rcheckalphabeta}
 \check{R}_{i,i+1}(u) = \alpha I + \beta E_i \,,
\end{equation}
for certain coefficients $\alpha$, $\beta$ depending on $u$ and $q$.
Here $I$ denotes the identity operator and $E_i$ is a generator of the Temperley-Lieb (TL) algebra.
The defining relations of this algebra, acting on $L=2M+1$ sites, are
\begin{eqnarray}
 E_i E_i &=& \delta E_i \,, \nonumber \\
 E_i E_{i \pm 1} E_i &=& E_i \,, \label{TLrels} \\
 E_i E_j &=& E_j E_i \mbox{ for } |i-j|>1 \,, \nonumber
\end{eqnarray}
where $i,j = 1,2,\ldots,L-1$ and the parameter $\delta := q + q^{-1}$. A representation of $E_i$, written in the same
6-vertex model representation as \eqref{Rcheckmat}, reads
\begin{align}
 E_i = \left(
            \begin{array}{cccc}
              0 & 0 & 0 & 0 \\
              0 & q^{-1} & 1 & 0 \\
              0 & 1 & q & 0 \\
              0 & 0 & 0 & 0 \\
            \end{array}
          \right) \,.
		  \label{TLrep}
\end{align}
By taking tensor products, one may check that this satisfies the relations \eqref{TLrels}.
We can match with \eqref{Rcheckalphabeta} by taking
\begin{equation}
 \alpha = u+i \,, \qquad \beta = u \,, \qquad q = -1 \,, \qquad \delta = -2 \,.
\end{equation}

The trick is now that there exists another representation of the TL algebra having exactly the required dimension
${\cal N}(M,K)$. The basis states of this representation are link patterns on $L$ sites with $d := L-2K$ defects. A link
pattern consists of a pairwise matching of $L-d = 2K$ points (usually depicted as $K$ arcs) and $d$ defect points,
subject to the constraint of planarity: two arcs cannot cross, and an arc is not allowed to straddle a defect point.
We show here two possible link patterns for $L=5$ and $d=1$ (hence $M=2$ and $K=2$):
\begin{equation}
\label{sample-link-patterns}
\begin{tikzpicture}[scale=0.6]
 \draw[fill] (0,0) circle (2 pt);
 \draw[thick] (1,0) arc(180:360:1.5 cm);
 \draw[thick] (2,0) arc(180:360:0.5 cm);
 \draw (5.5,0) node{and};
 \draw[fill] (9,0) circle (2 pt);
 \draw[thick] (7,0) arc(180:360:0.5 cm);
 \draw[thick] (10,0) arc(180:360:0.5 cm);
\end{tikzpicture}
\end{equation}

The TL generator $E_i$
acts on sites $i$ and $i+1$ by first contracting them, then adding a new arc between $i$ and $i+1$.
This can be visualized by placing the graphical representation
$E_i = \raisebox{-0.1 em}{\begin{tikzpicture}[scale=0.3]
 \draw[thick] (0,0) arc(180:360:0.5 cm and 0.4 cm);
 \draw[thick] (1,-1) arc(0:180:0.5 cm and 0.4 cm);
\end{tikzpicture}}$
on top of the link pattern.
If a loop is formed in
the contraction, it is removed and replaced by the weight $\delta$. If a contraction involves an arc and a defect point,
the defect point moves to the other extremity of the arc. If a contraction involves two distinct arcs, the opposite ends of
those arcs become paired by an arc.
For instance, the action of $E_1$ on the two link patterns in \eqref{sample-link-patterns} produces
\begin{equation}
\begin{tikzpicture}[scale=0.6]
 \draw[fill] (4,0) circle (2 pt);
 \draw[thick] (0,0) arc(180:360:0.5 cm);
 \draw[thick] (2,0) arc(180:360:0.5 cm);
 \draw (5.5,0) node{and};
 \draw (7.3,0) node{$\delta \times$};
 \draw[fill] (10,0) circle (2 pt);
 \draw[thick] (8,0) arc(180:360:0.5 cm);
 \draw[thick] (11,0) arc(180:360:0.5 cm);
\end{tikzpicture}
\end{equation}

Recall from (\ref{eq:szminusK}) that the spin $s$ associated with the $K$-magnon sector in the chain of $L = 2M+1$ sites reads $s = \frac{L}{2} - K$.
The generators $E_i$ can decrease $s$ by contracting a pair of defects and replacing them by an arc.
It is however possible to define a representation of the TL algebra in which $s$ is fixed, by defining the
action of $E_i$ to be zero whenever there is a pair of defects at sites $i$ and $i+1$. In the literature on
the TL algebra, these representations in terms of link patterns with a conserved number of defects
are known as {\em standard modules} and denoted ${\mathcal W}_s$.
%In the literature on the TL algebra, the representations in terms of link patterns as just described are known as {\em standard modules}
%and denoted ${\mathcal W}_s$.
Meanwhile, in TL representation theory, the partition function in the open channel is no longer written
in terms of a trace as in (\ref{eq:openZ}). Instead it is written as a so-called Markov trace
\begin{equation}
 Z(u,M,N) = {\rm Mtr} \, \left[ t_D(u)^N \right] \,,
\end{equation}
which can be interpreted diagramatically as the stacking of $N$ rows of diagrams, followed by a gluing operation in which the top and
the bottom of the system are identified and each resulting loop replaced by the corresponding weight $\delta$. It is a remarkable fact
that this Markov trace can be computed as a linear combination of ordinary matrix traces over the standard modules, as follows:
\begin{equation}
 Z(u,M,N) = \sum_{s=1/2,3/2,\ldots}^{M+1/2} (1+2s)_q {\rm tr}_{{\mathcal W}_s} \left[ t_D(u)^N \right] \,.
\label{Markov_trace}
\end{equation}
We have here defined the $q$-deformed numbers
\begin{equation}
 (n)_q = \frac{q^n - q^{-n}}{q - q^{-1}} = U_{n-1} \left( \frac{\delta}{2} \right) \,,
\end{equation}
where $U_p(x)$ denotes the $p$-th order Chebyshev polynomial of the second kind.
This result \eqref{Markov_trace} can be proved by using the quantum group symmetry $U_q(su(2))$ enjoyed by the spin chain in the
open channel \cite{Saleur:1988zx}, or alternatively by purely combinatorial means \cite{Richard:2006fjr}.

The factors $(1+2s)_q$ appearing in \eqref{Markov_trace} account for the multiplicities in the problem. In the limit $q \to -1$ corresponding
to the XXX case of interest, the $q$-deformed numbers become $(n)_q = n$ for $n$ odd, and $(n)_q = -n$ for $n$ even. The latter minus sign can be
eliminated at the price of an overall sign change of the partition function \eqref{Markov_trace}, since only even $n=1+2s$ occur in the
problem. This corresponds to $q \mapsto -q$, so that the quantum group symmetry $U_q(su(2))$ becomes just ordinary $su(2)$ in the limit.
The multiplicities then become $(1+2s)_q = 1+2s = 2M+2-2K$, in agreement with (\ref{eq:szminusK}).

In conclusion, we see that not only do the link pattern representations of the TL algebra lead to the correct dimensions ${\mathcal N}(M,K)$,
but they also account for the correct $su(2)$ multiplicities $2M+2-2K$ of eigenvalues in the XXX spin chain.

\subsubsection{Results}

We have computed the condensation curves by applying the numerical methods of \cite{Jacobsen:2018pjt} to the
transfer matrix $t_D(u)$ given by \eqref{eq:openT}. The latter is taken to act on the representation given by the
union of link patterns on $L=2M+1$ sites with $K \in \{0,1,\ldots,M\}$ arcs and $d=L-2K$ defects.

\begin{figure}[h!]
\begin{center}
\includegraphics[scale=0.28]{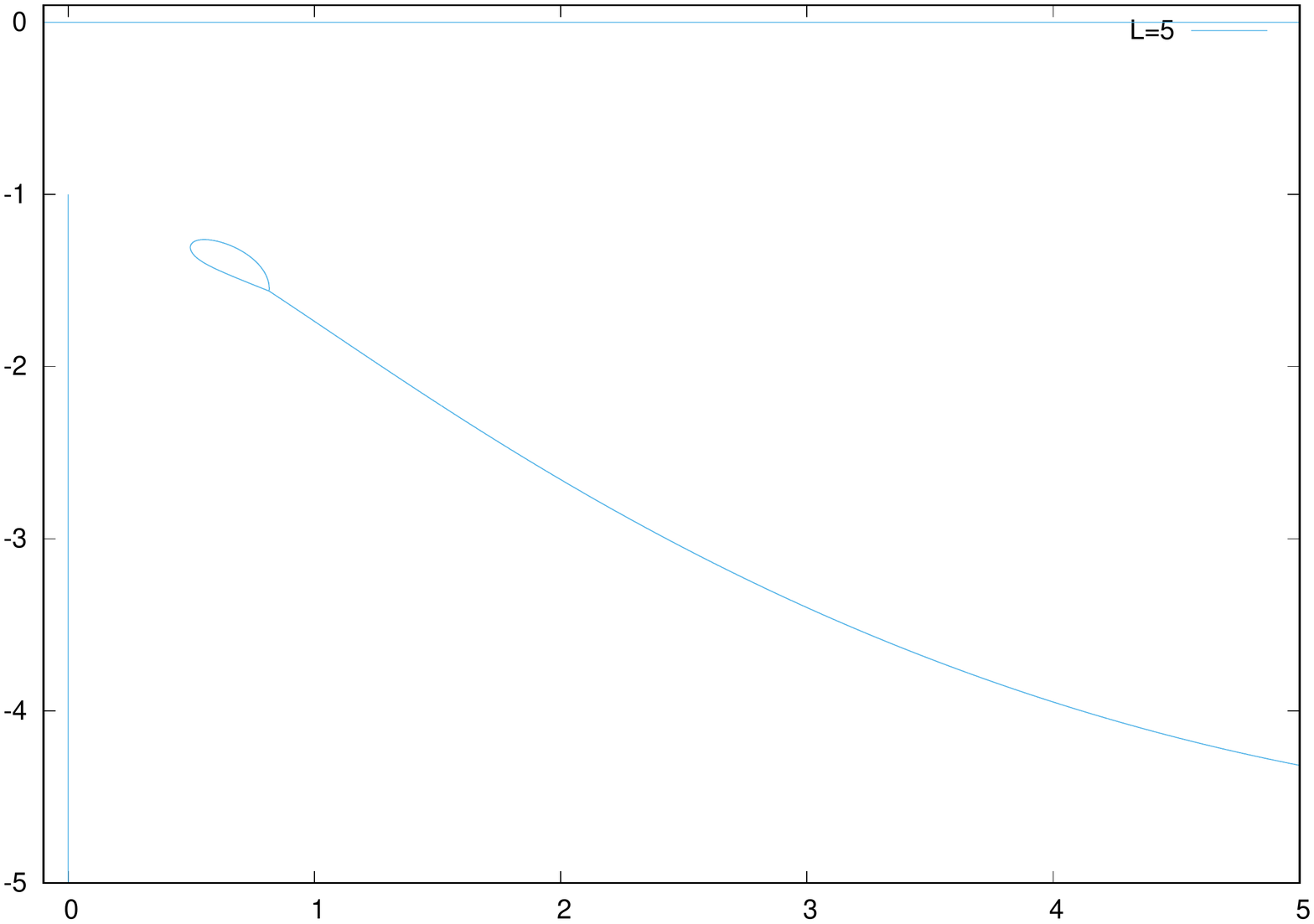}\includegraphics[scale=0.28]{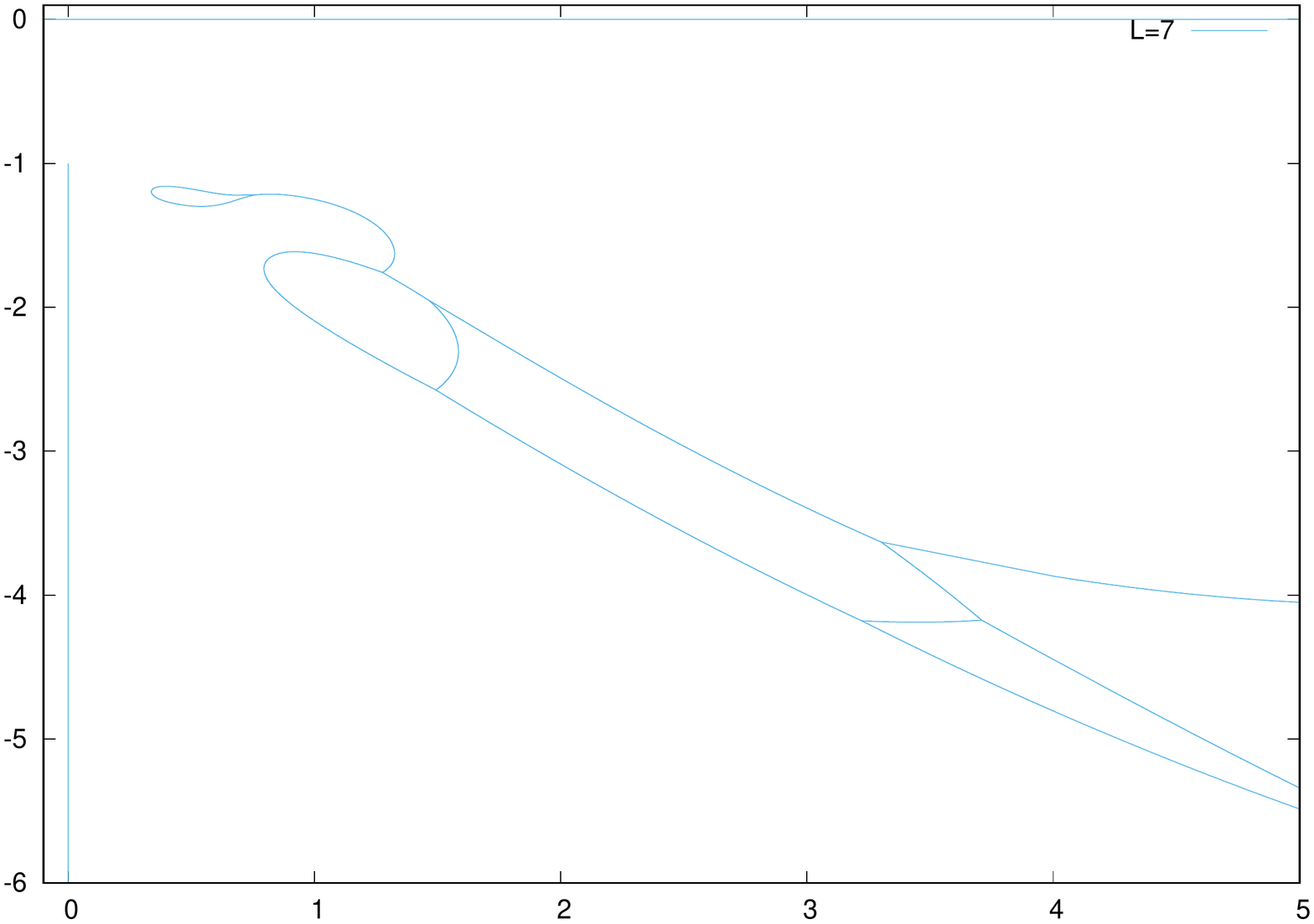} \\
\includegraphics[scale=0.28]{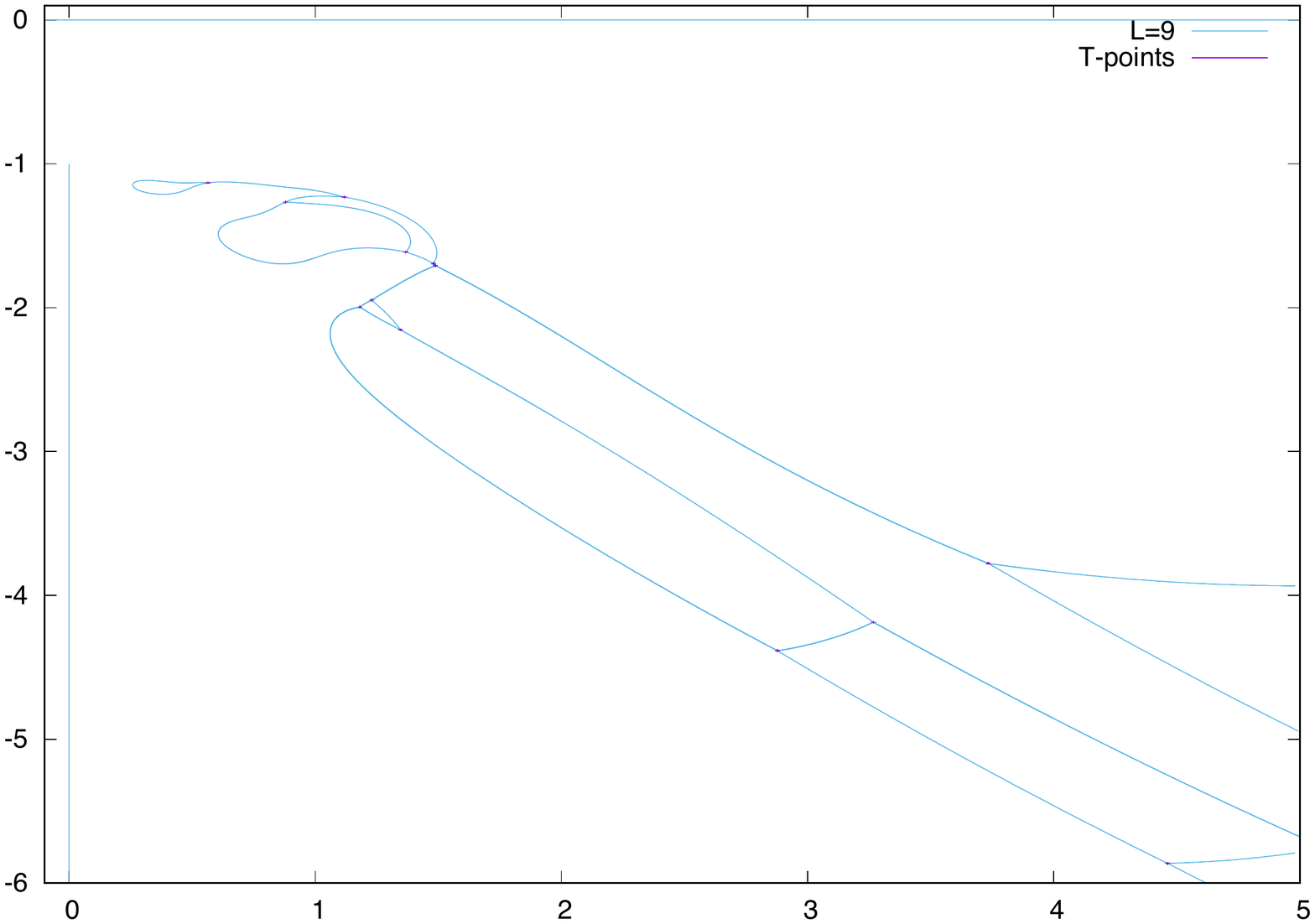}\includegraphics[scale=0.28]{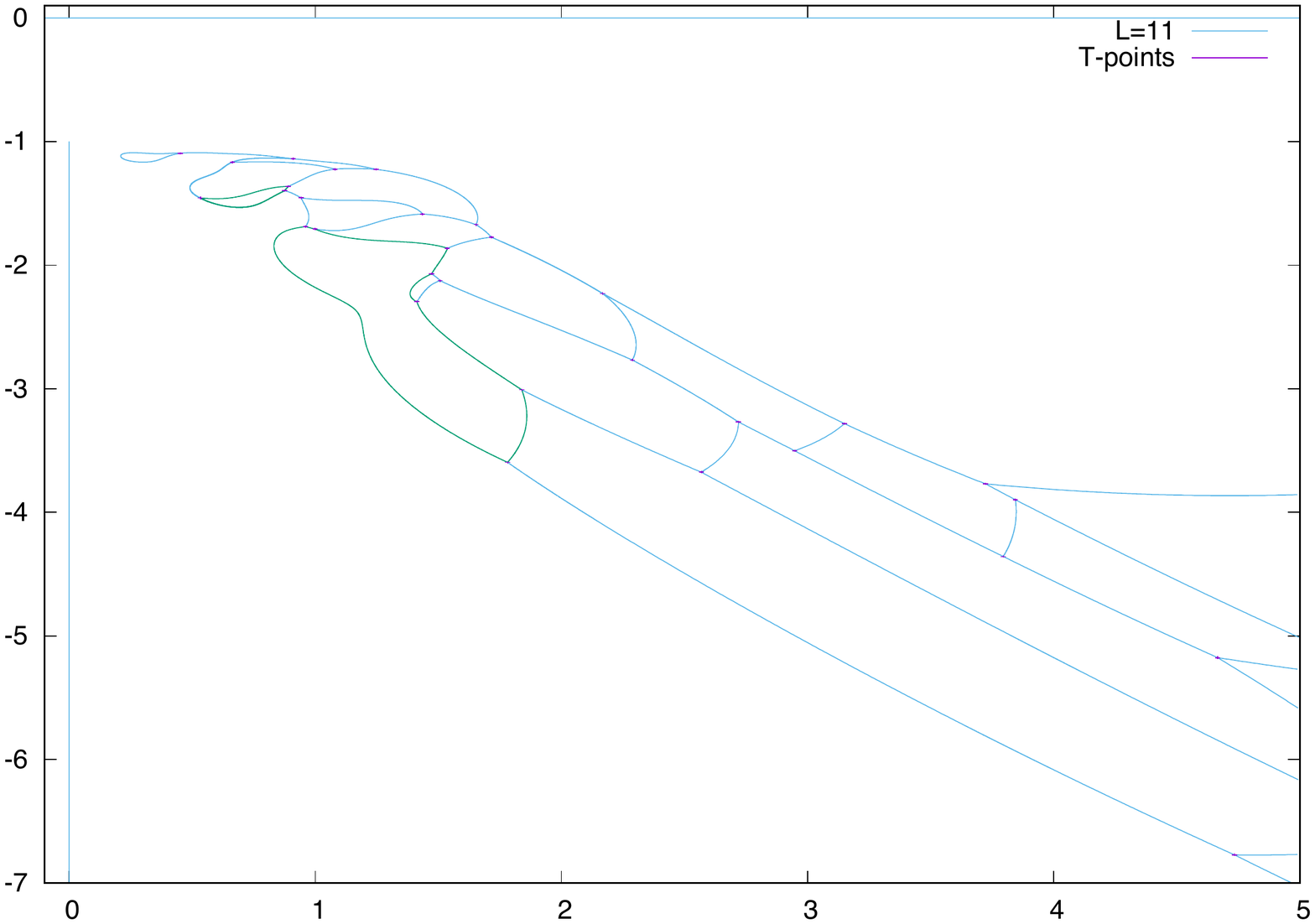}
\caption{Condensation curves for partition function zeros on a $(2M+1)\times 2N$ cylinder, in the limit
$N \to \infty$ (aspect ratio $\rho \to \infty$, open channel). The panels show, in reading direction, the
cases $M=2,3,4,5$.}
\label{fig:cond-open}
\end{center}
\end{figure}

The results for the condensation curves with $M=2,3,4,5$ are shown in Figure~\ref{fig:cond-open}.
The curves are confined to the half-space $\Im\, u \le 0$, and they are invariant under changing the
sign of $\Re\, u$. Therefore it is enough to consider them in the fourth quadrant: $\Re \,u \ge 0$, $\Im \,u \le 0$.
The condensation curves display several noteworthy features:
\begin{enumerate}
 \item Outside the curves and in the enclosed regions delimited by blue curves, the dominant eigenvalue belongs to the $K=M$ magnon sector
(i.e., $d=1$ defect in the TL representation). For the largest size $M=5$ there are also enclosed regions delimited by green curves: in this
case the dominant eigenvalue belongs to the $K=M-1$ sector ($d=3$).
 \item The whole real axis forms part of the curve. In fact, when $u \in \mathbb{R}$, {\em all} the eigenvalues are equimodular and have
norm $(u^2+1)^M$. Above the real axis ($\Im\, u > 0$) the dominant eigenvalue is the unique eigenvalue in the $K=0$ sector.
 \item There is a segment of the imaginary axis, $\Re\, u = 0$ and $\Im\, u \le u_{\rm c}(M)$ which also belongs to the condensation curve.
 Along this segment, the two dominant eigenvalues come from the $K=M$ sector. For the end-point $u_{\rm c}(M)$
 we find the following results: \\[2mm]
 \begin{tabular}{l|rrrrrr}
 $M$ & 2 & 3 & 4 & 5 & 6 & 7 \\ \hline
 $u_{\rm c}(M)$ & -1.091487 & -1.065097 & -1.050552 & -1.041328 & -1.034954 & -1.030285 \\
 \end{tabular} \\[2mm]

 \item It seems compelling from these data that
 \begin{equation}
  u_{\rm c}(M) \to -1 \mbox{ as } M \to \infty \,,
 \end{equation}
 with a finite-size correction proportional to $1/M$.
 We also note that at this asymptotic end-point, $u = -i$, for all finite $M$ there is a
 unique dominant eigenvalue which belongs to the $K=M$ sector and has norm 1, while all other eigenvalues have norm 0.
 \item The remainder of the condensation curve forms a single connected component with no end-points. It however has
 a number of T-points that grows fast with $M$. Notice that we have taken great care to determine all of these T-points, some
 of which are very close and thus hard to distinguish in the figures. To help the reader identifying them, they have been marked
 by small crosses.
 \item For the leftmost point $u_\star$ of this connected component (i.e., the point with the smallest imaginary part)
 we find the following results: \\[2mm]
 \begin{tabular}{l|rrrrrr}
 $M$ & 2 & 3 & 4 & 5 & 6 & 7 \\ \hline
 $\Re\, u_\star(M)$ &  0.496489 &  0.338134 &  0.258384 &  0.209593 &  0.176490 &  0.152498 \\
 $\Im\, u_\star(M)$ & -1.307913 & -1.196739 & -1.146652 & -1.117510 & -1.098264 & -1.084539 \\
 \end{tabular} \\[2mm]

 \item It seems compelling from these data that
 \begin{eqnarray}
  \Re\, u_\star(M) \to 0 \mbox{ as } M \to \infty \,, \nonumber \\
  \Im\, u_\star(M) \to -1 \mbox{ as } M \to \infty \,,
 \end{eqnarray}
 both with finite-size corrections proportional to $1/M$. We conclude that the leftmost point of the connected component converges
 to the same value as the end-point, namely $u=-i$. This kind of ``pinching'' is characteristic of a phase transition
 \cite{Lee-Yang,Fisher:zeros}; note however that the limit $u \to -i$ of the XXX model is singular and does not present a critical
 point in the usual sense.

\end{enumerate}

We now compare the condensation curves with the partition function zeros.
The partition functions $Z(u,M,N)$ were first computed from the algebro-geometric approach,
for $M=2,3,4,5$ and $N=1024$, which corresponds to a very large aspect ratio $\rho \sim 10^3$. The
zeros of $Z(u,M,N)$ were then computed by the program {\sc MPSolve} \cite{MR1772050}\cite{MPSolve}, which is a multiprecision
implementation of the Ehrlich-Aberth method \cite{Ehrlich,Aberth}, an iterative approach to finding all zeros
of a polynomial simultaneously.

\begin{figure}[h!]
\begin{center}
\includegraphics[scale=0.28]{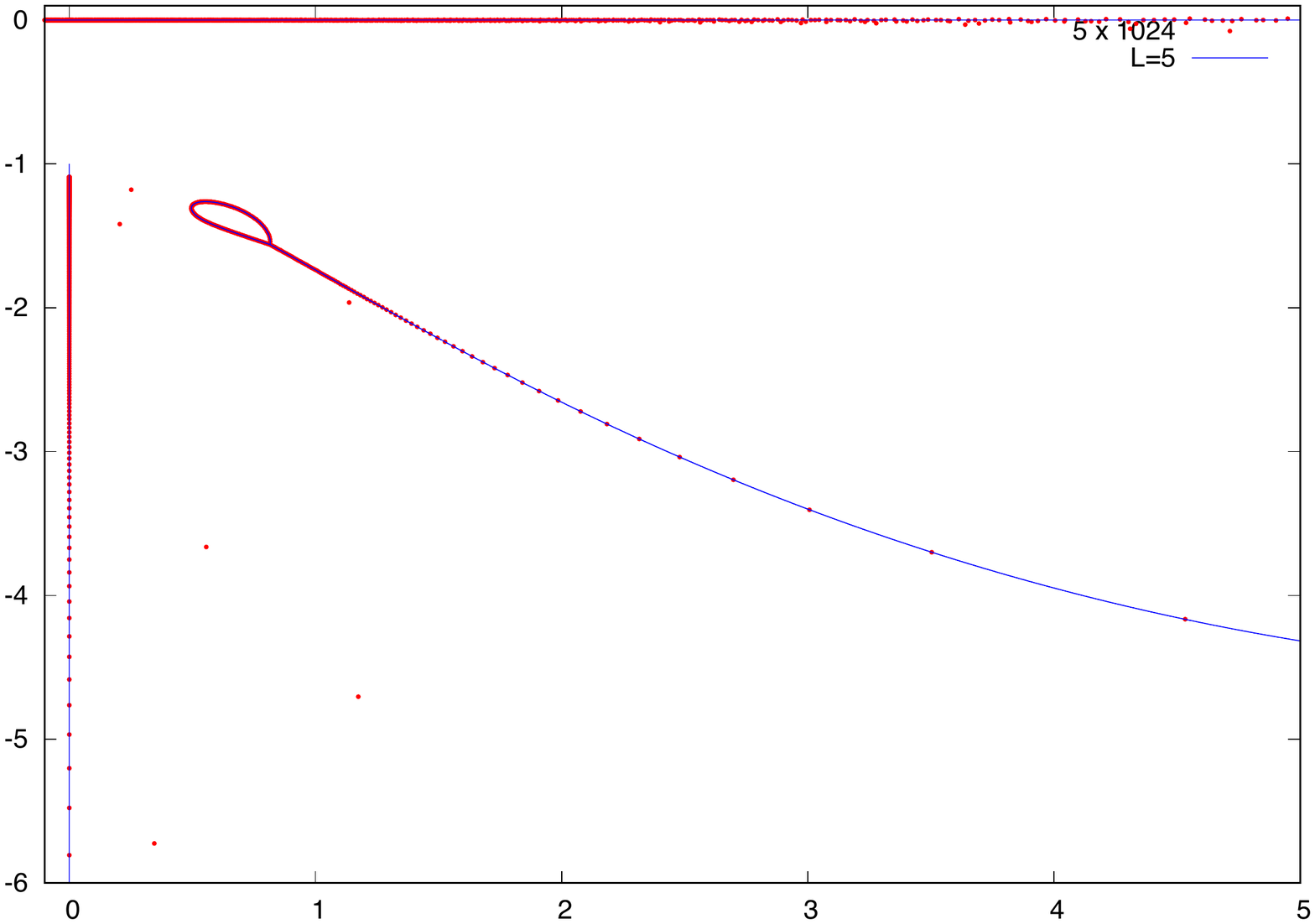}\includegraphics[scale=0.28]{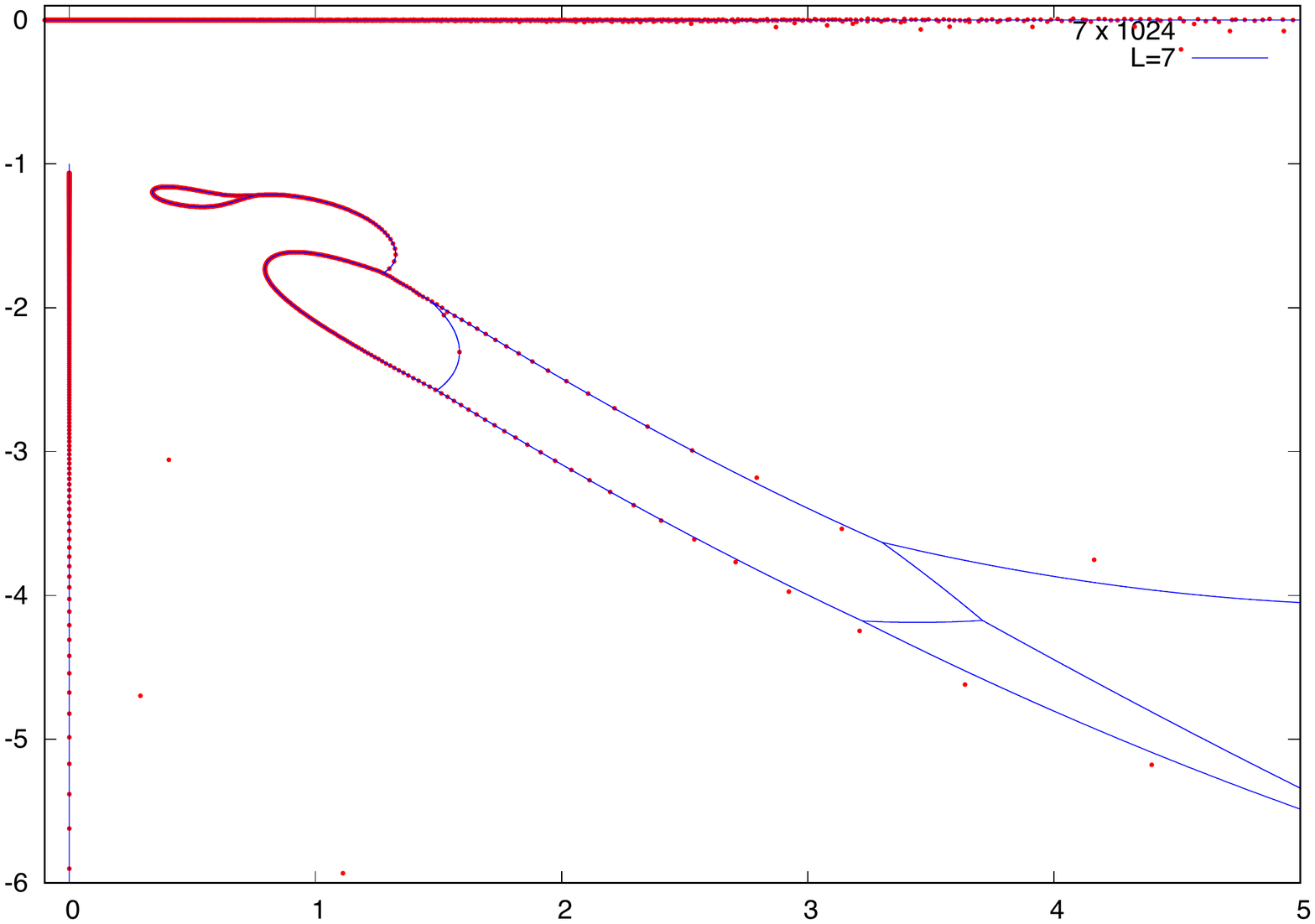} \\
\includegraphics[scale=0.28]{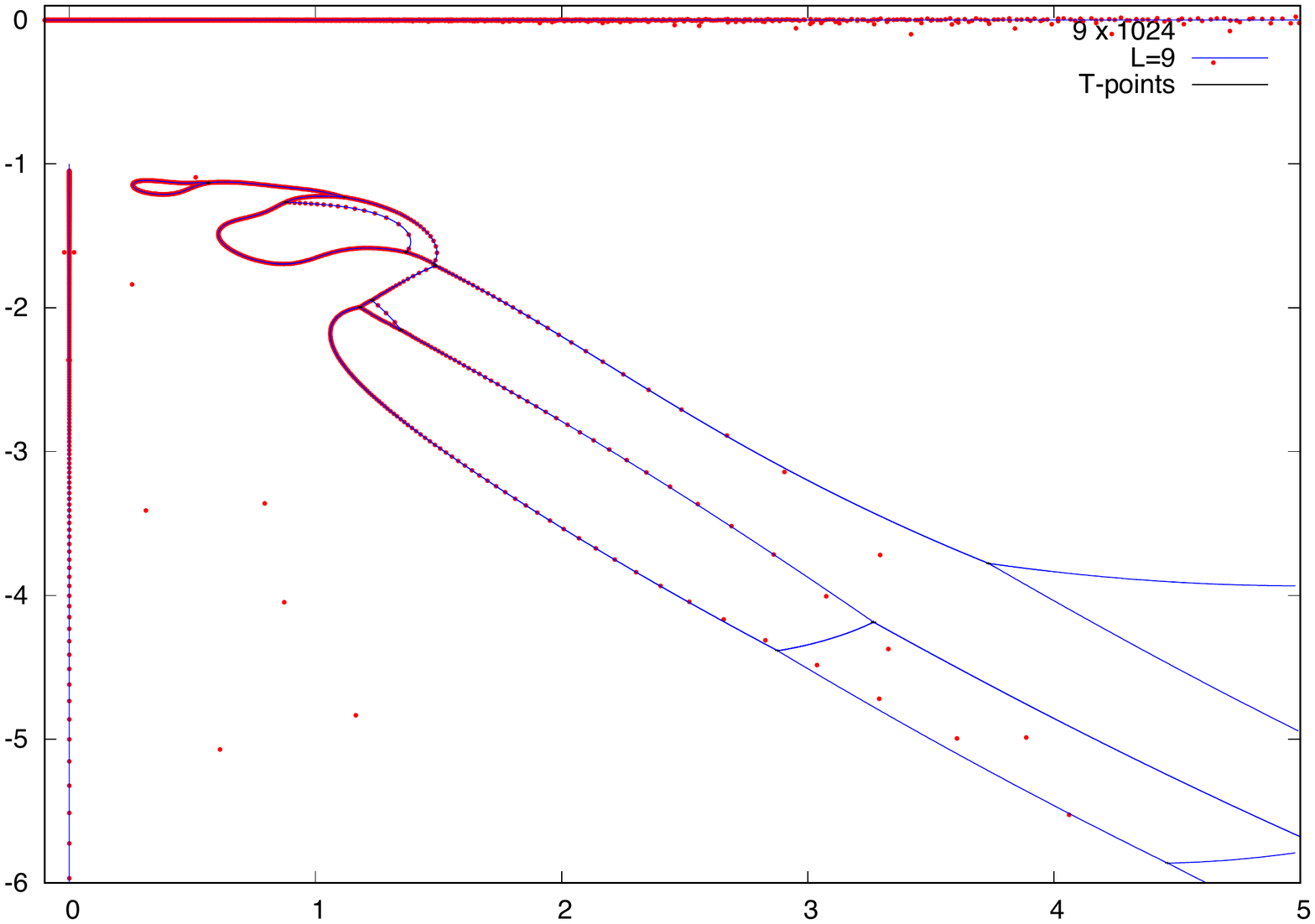}\includegraphics[scale=0.28]{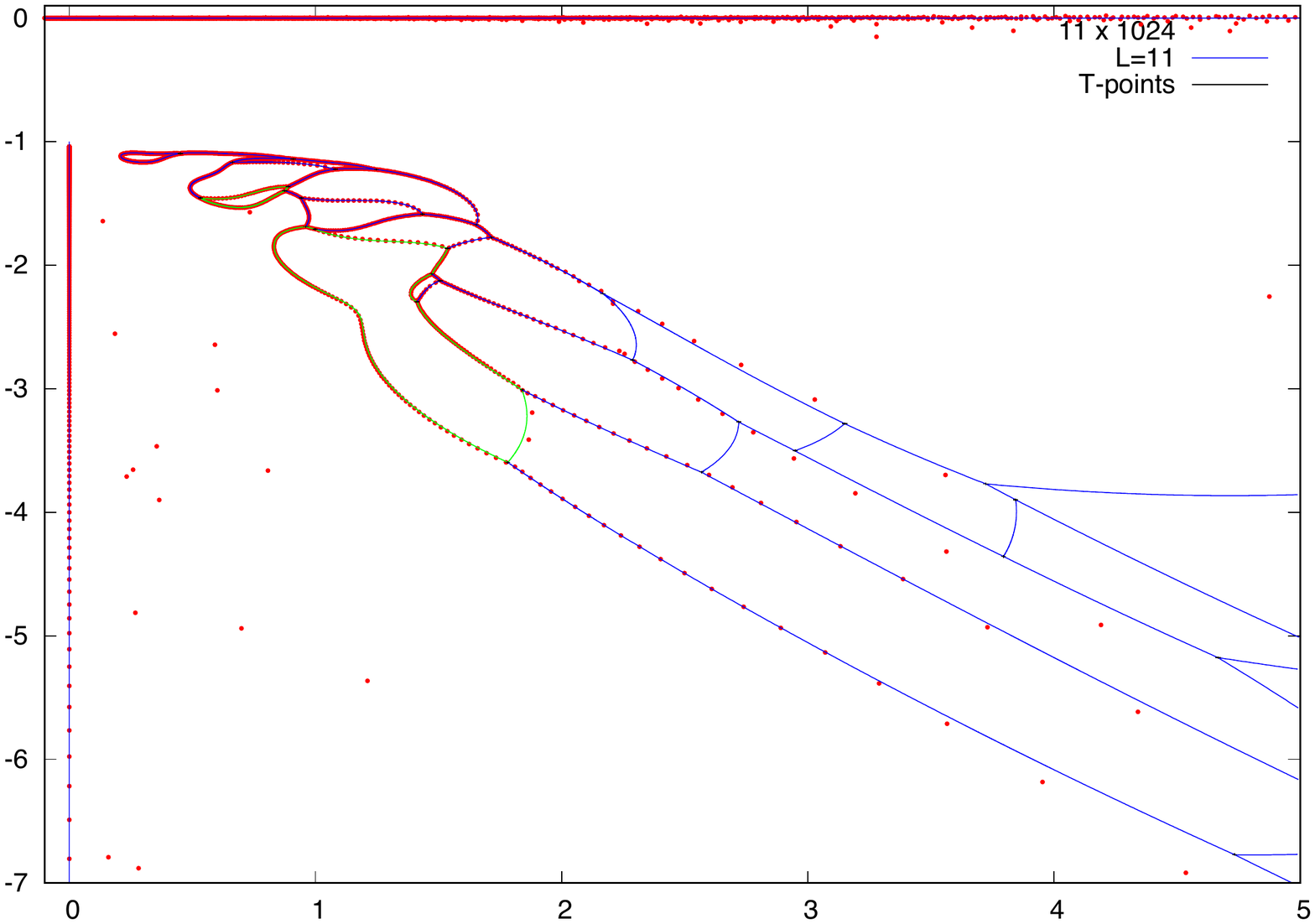}
\caption{Comparison between the partition function zeros on a $(2M+1)\times 2N$ cylinder, with $N=1024$,
and the corresponding condensation curves in the $N \to \infty$ limit (open channel).
The panels show, in reading direction, the cases $M=2,3,4,5$.}
\label{fig:cond-open-comp}
\end{center}
\end{figure}

The resulting zeros are shown in Figure~\ref{fig:cond-open-comp}, as red points superposed on the condensation curves
of Figure~\ref{fig:cond-open}. The agreement is in general very good, although some portions of the condensation curves
are very sparsely populated with zeros; in those cases the zeros are still at a discernible distance from the curves, in spite
of the large aspect ratio. We have verified that the agreement improves upon increasing $\rho$.%
\footnote{One may notice that some of the roots (in particular for $M=5$, in the region $0 < \Re\, u < 1$) stray off the
curves in a seemingly erratic fashion. We believe that this is an artifact of {\sc MPSolve} when applied to polynomials
of very high degree.}

\subsection{Closed channel}

In the closed channel the $\check{R}$-matrix can be inferred from \eqref{Rtilde} and \eqref{defcheckR}. It reads
\begin{align}
\check{R}^{\rm c}(u)=\left(
            \begin{array}{cccc}
              b(u) & 0 & 0 & 0 \\
              0 & c(u) & a(u) & 0 \\
              0 & a(u) & c(u) & 0 \\
              0 & 0 & 0 & b(u) \\
            \end{array}
          \right) \,,
\label{Rtildecheckmat}
\end{align}
still with $a(u) = u+i$, $b(u)=u$ and $c(u)=i$. However, as we shall
soon see, it is convenient to apply a diagonal gauge transformation
$D={\rm diag}(1,-1)$ in the left in-space and the right out-space of
$\check{R}^{\rm c}$; that is, $\check{R}_{12}^{\rm c} \mapsto D_{1}\,
\check{R}_{12}^{\rm c}\, D_{1}$.
This has the effect of changing
the sign of $c(u)$ while leaving the partition function unchanged: the gauge matrices square to the identity
at the intersections between $\check{R}$-matrices when taking powers of the transfer matrix $\tilde{\rT}_D(u)$
given by \eqref{eq:closedT}. To complete the transformation, the first and last row of gauge transformations have
to be absorbed into a redefinition of the boundary states $\langle \Psi_0 |$ and $| \Psi_0 \rangle$ appearing in
\eqref{eq:closedZ}.

\subsubsection{Temperley-Lieb algebra}

As in the open channel, we can rewrite the $\check{R}$-matrix in
terms of TL generators \eqref{TLrep}:
\begin{equation}
\label{Rtildecheckalphabeta}
 \check{R}^{\rm c}_{i,i+1}(u) = \alpha I + \beta E_i \,.
\end{equation}
To match \eqref{Rtildecheckmat}, with $c(u) = -i$ after the gauge transformation, we must now set
\begin{equation}
 \alpha = u \,, \qquad \beta = u+i \,, \qquad q = -1 \,, \qquad \delta = -2 \,.
\end{equation}

In the closed channel, the TL algebra is defined on $L = 2N$ sites. The goal is now to find a representation
having the same dimension ${\cal N}(N)$, see \eqref{closedconj}, as the number of physical solutions
appearing in the closed-channel expressions of the partition
function, \eqref{eq:evenNZ} and \eqref{eq:oddNZ}.
This issue is more complicated than in the open channel.

As a first step, we let the TL generators act on the basis of link patterns, as before. Since the boundary states
restrict to zero total spin, the only allowed number of Bethe roots is $K=N$ (see section~\ref{sec:close}).
This implies that the link patterns are free of defects ($d=0$). The transfer matrix $\tilde{\rT}_D(u)$ is then
given by \eqref{eq:closedT} with \eqref{Rtildecheckalphabeta}, where the TL generators $E_i$ act on the
link patterns as described in section~\ref{sec:TL-open}. To reproduce the partition function \eqref{eq:closedZ}
we also need to interpret the boundary state \eqref{bdrystate} within the TL representation. The natural
object is the quantum-group singlet of two neighboring sites
\begin{equation}
 | \psi_0 \rangle_{\rm TL} = q^{1/2} | \uparrow \rangle \otimes | \downarrow \rangle + q^{-1/2} | \downarrow \rangle \otimes | \uparrow \rangle \,,
\end{equation}
which is represented in terms of link patterns as a short arc joining the neighboring sites. We should however
remember at this stage the gauge transformation that allowed us to switch the sign of $c(u)$. To compensate
this, we need to insert a minus sign for a down-spin in the second tensorand, to obtain
\begin{equation}
 | \tilde{\psi}_0 \rangle_{\rm TL} = -q^{1/2} | \uparrow \rangle \otimes | \downarrow \rangle + q^{-1/2} | \downarrow \rangle \otimes | \uparrow \rangle \,,
\end{equation}
With $q=-1$, this is proportional to $| \psi_0 \rangle$ of \eqref{bdrystate}.

On the other hand it is easy to check from \eqref{TLrep}
that the TL generator $E_i$ is nothing but the (unnormalized) projector onto the quantum-group singlet.
Therefore, just as
$E_i = \raisebox{-0.1 em}{\begin{tikzpicture}[scale=0.3]
 \draw[thick] (0,0) arc(180:360:0.5 cm and 0.4 cm);
 \draw[thick] (1,-1) arc(0:180:0.5 cm and 0.4 cm);
\end{tikzpicture}}$,
the initial boundary state $| \Psi_0 \rangle$
can be represented graphically by the defect-free link pattern in which sites $2j-1$ and $2j$ are connected by an arc, for each
$j=1,2,\ldots,N$. Similarly, the final boundary state $\langle \Psi_0 |$ is interpreted as the TL contraction of the corresponding
pairs of sites. With these identifications, we have explicitly verified for small $N$ and $M$ that the TL formalism
produces the correct partition functions, such as \eqref{Zexamples}.

With the spin-zero constraint imposed, the TL dimension is thus equal to the number of defect-free link patterns on
$L=2N$ sites. This is easily shown to be given by the Catalan numbers
\begin{equation}
 {\rm Cat}(N) = \frac{1}{N+1} {2N \choose N} \,,
\end{equation}
for which the first 10 values are given by
\begin{equation}
 \{1, 2, 5, 14, 42, 132, 429, 1430, 4862, 16796 \} \,.
\end{equation}
Although this is smaller than the dimension of the 6-vertex-model representation constrained to the $S^z = 0$
sector, viz.\ ${2N \choose N}$, it is not as small as \eqref{closedconj}, so further work is needed.

The transfer matrix and the boundary states are also symmetric under cyclic shifts (in units of two lattice spacings) of the
$L=2N$ sites. This symmetry can be used to further reduce the dimension of the transfer matrix. Indeed, after acting
with $\tilde{\rT}_D(u)$ we project each link pattern obtained onto a suitably chosen image under the cyclic group $\mathbb{Z}_N$.
In this way each orbit under $\mathbb{Z}_N$ is mapped onto a unique representative link pattern. The dimension of the
corresponding rotation invariant transfer matrix then reduces to \cite{Chang04}
\begin{equation}
 {\rm dim}_{\mathbb{Z}_N}(N) =
 \frac{1}{N} \sum_{m | N} \varphi(N/m) {2 m \choose m} - {\rm Cat}(N) \,,
\end{equation}
where the sum is over the divisors of $N$, and $\varphi(x)$ denotes the Euler totient function.
The first 10 values are given by
\begin{equation}
 \{1, 2, 3, 6, 10, 28, 63, 190, 546, 1708 \} \,.
\end{equation}

But one can go a bit further, since the transfer matrix and boundary states are also invariant under reflections.
This gives rise to a symmetry under the dihedral group $\mathbb{D}_N$. The dimension of the rotation-and-reflection
invariant transfer matrix then becomes \cite{Chang04}
\begin{equation}
 {\rm dim}_{\mathbb{D}_N}(N) =
 \frac12 \left( \frac{1}{N} \sum_{m | N} \varphi(N/m) {2 m \choose m} - {\rm Cat}(N) + { N \choose \lfloor N/2 \rfloor} \right) \,,
\end{equation}
of which the first 10 values are
\begin{equation}
 \{1, 2, 3, 6, 10, 24, 49, 130, 336, 980 \} \,.
\end{equation}

The process of imposing more and more symmetries and reducing the dimension of the relevant transfer matrices might be realized at the ZRC level by imposing more and more constraints on the $Q$-functions. Consider the ZRC of a closed spin chain with length $L=2N$ and $N$ magnons. The corresponding Bethe states are in the $S^z=0$ sector. In order to restrict to the parity symmetric solutions, we need to impose the condition $Q(u)=Q(-u)$ for any $u$. This leads to further constraints to the ZRC and reduces the number of allowed solutions down to $\mathcal{N}(N)$. Since $Q(u)$ is a polynomial of order $N$, the constraint $Q(u)=Q(-u)$ can also be imposed by $Q(x_k)=Q(-x_k)$ at $N$ different values. Now the main observation is that at certain values of $x_k$, the constraints have a clear physical meaning. For example, taking $x_1=i/2$, the constraint $Q(i/2)=Q(-i/2)$ is equivalent to
\begin{align}
\frac{Q(-i/2)}{Q(+i/2)}=\prod_{k=1}^N\frac{u_k+\tfrac{i}{2}}{u_k-\tfrac{i}{2}}=1,
\end{align}
which restricts to the solutions with zero total momentum. It is therefore an interesting question to see whether the dihedral symmetry can be realized in this way. If so, at which further value(s) of $x_k$ would we need to impose $Q(x_k) = Q(-x_k)$ ?
%JJ: I think Yunfeng had a nice remark on the dihedral symmetry corresponding to the opposite signs of just one
%pair of Bethe roots. This would be a nice place to explain and comment on this.

We do not presently know if and how one can identify a TL representation whose dimension equals
${\cal N}(N) = {N \choose \lfloor N/2 \rfloor}$ given by \eqref{closedconj}. It certainly appears remarkable at this stage
that
\begin{equation}
 2 \, {\rm dim}_{\mathbb{D}_N}(N) - {\rm dim}_{\mathbb{Z}_N}(N) = {\cal N}(N) \,,
\end{equation}
as already noticed in \cite{Chang04}. It is also worth pointing out that ${\cal N}(N)$ can be interpreted as the number
of defect-free link patterns on $2N$ sites which are symmetric around the mid-point. We leave the further
investigation of this question for future work.

\subsubsection{Results}

We have computed the condensation curves in the closed channel, using the TL link-pattern representations identified above,
namely using: 1) spin-zero (i.e., defect free) link patterns, 2) spin-zero link patterns with cyclic symmetry $\mathbb{Z}_N$, and 3) spin-zero
link patterns with dihedral symmetry.

\begin{figure}[h!]
\begin{center}
\includegraphics[scale=0.28]{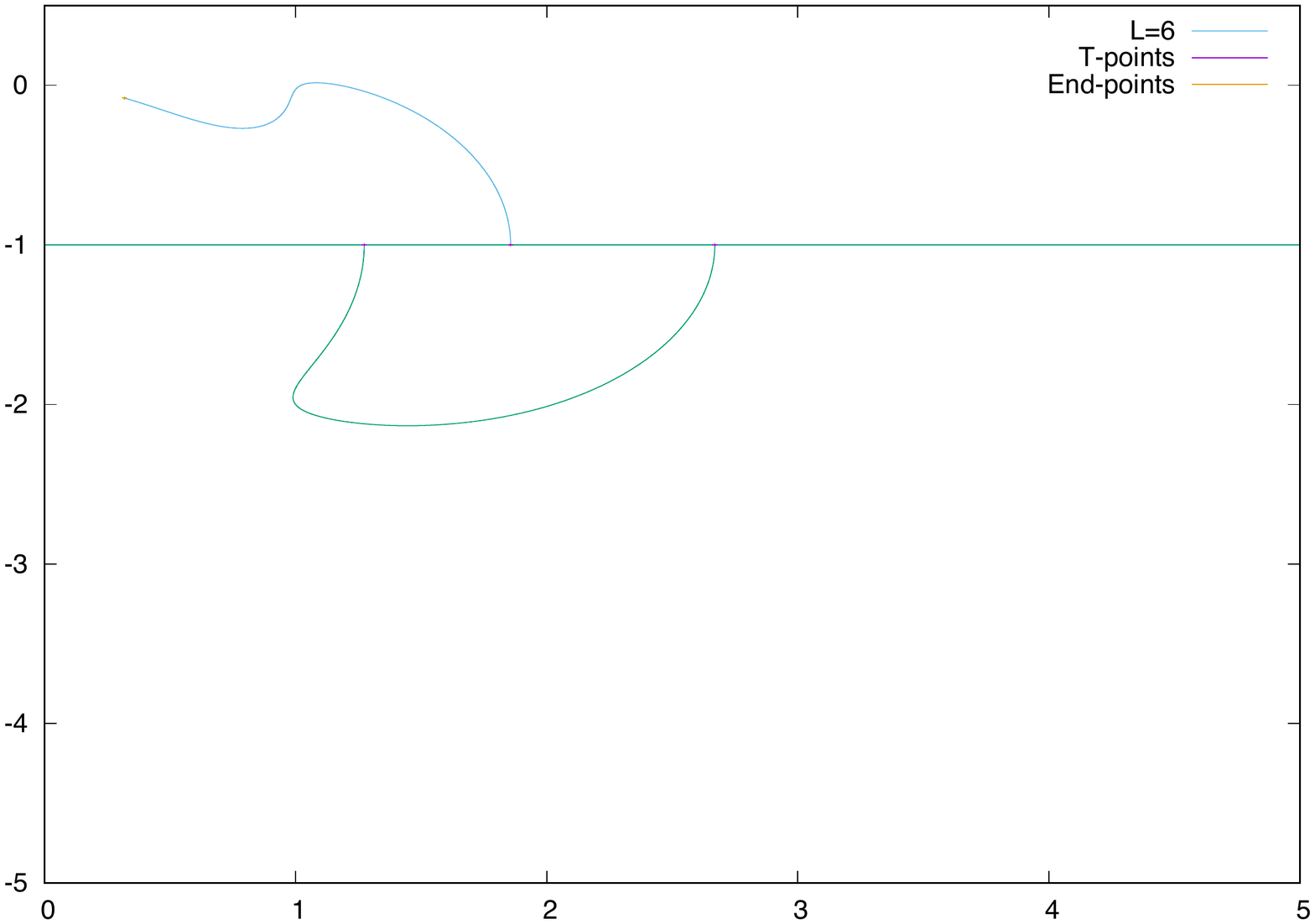}\includegraphics[scale=0.28]{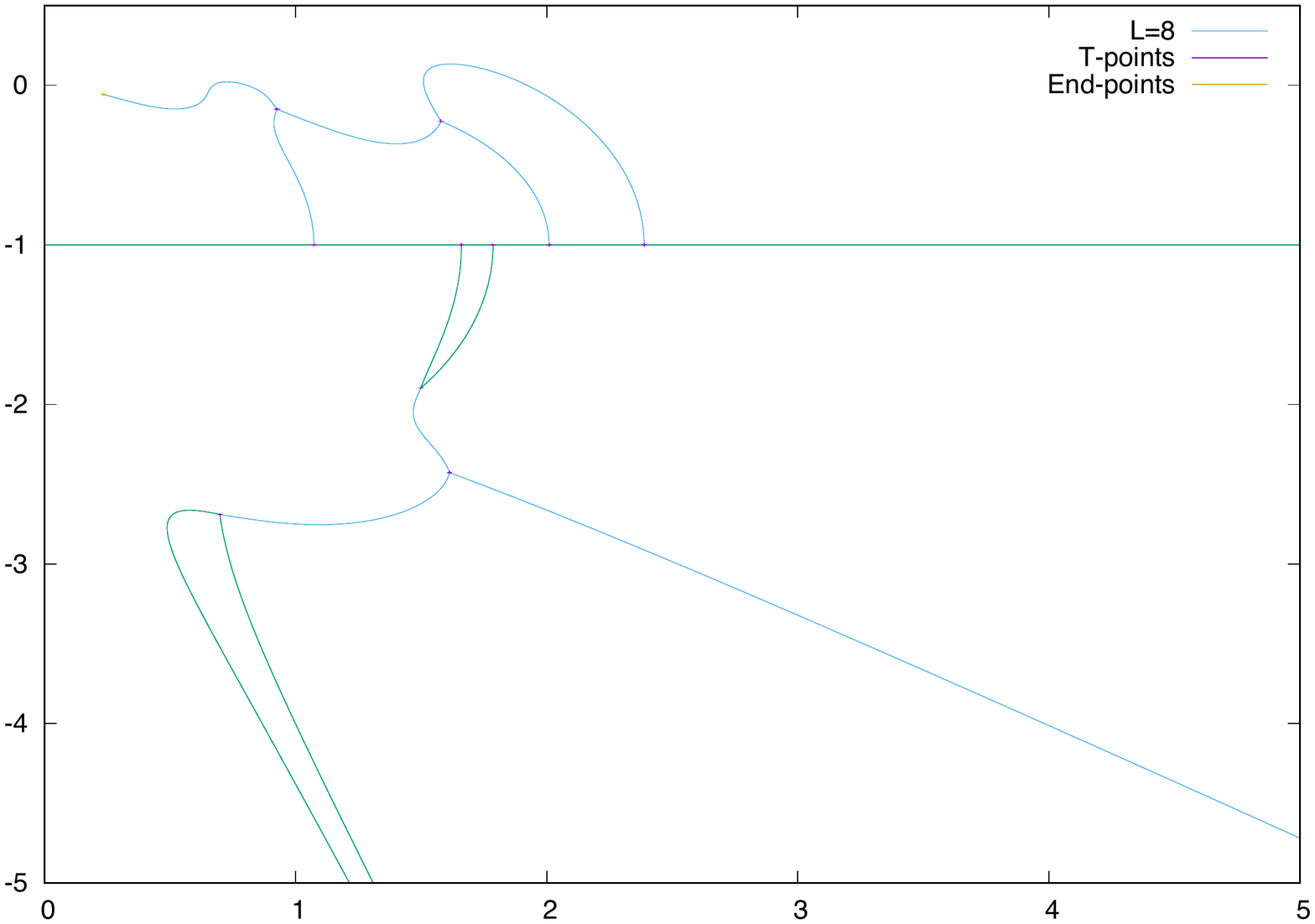} \\
\includegraphics[scale=0.28]{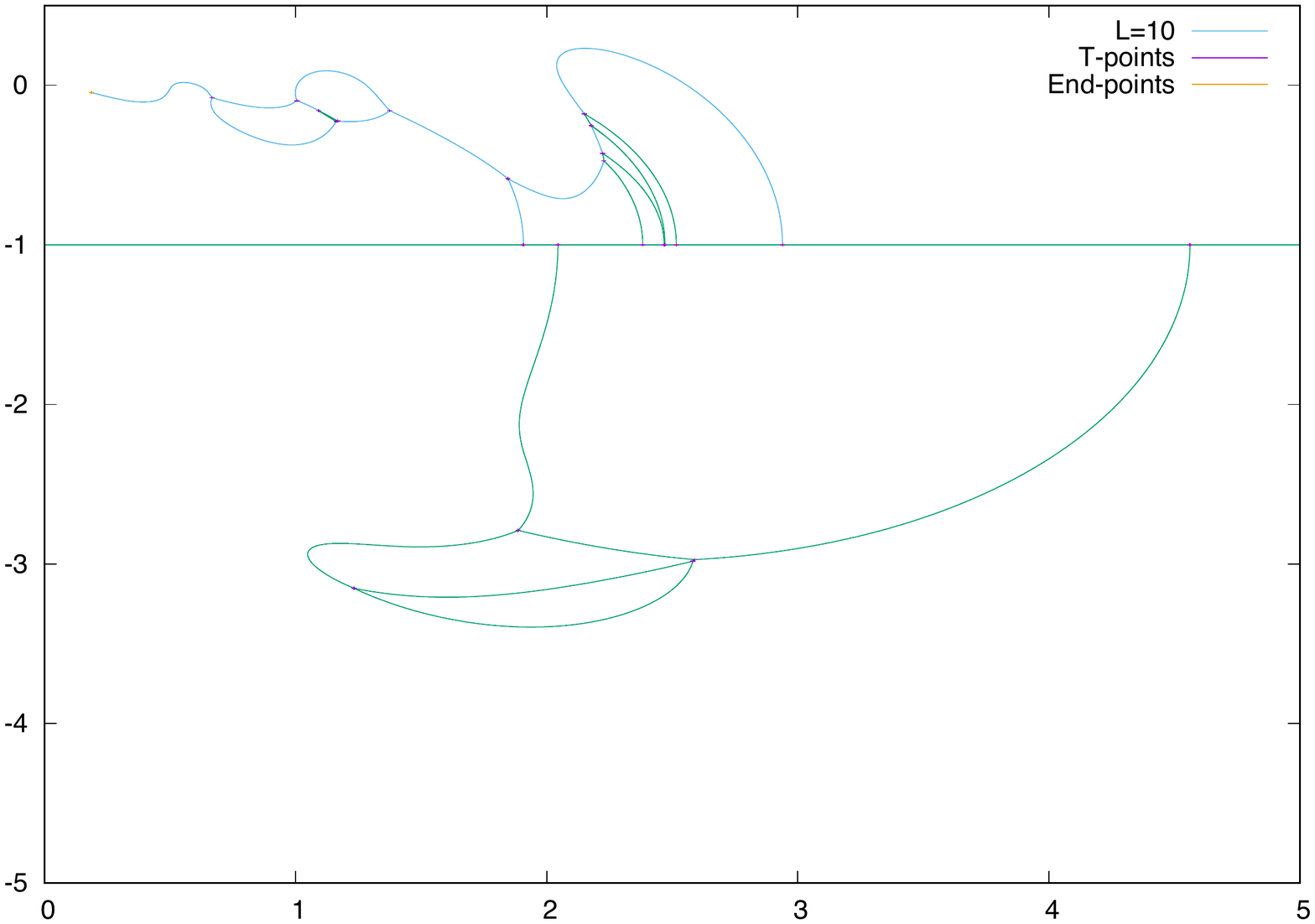}\includegraphics[scale=0.28]{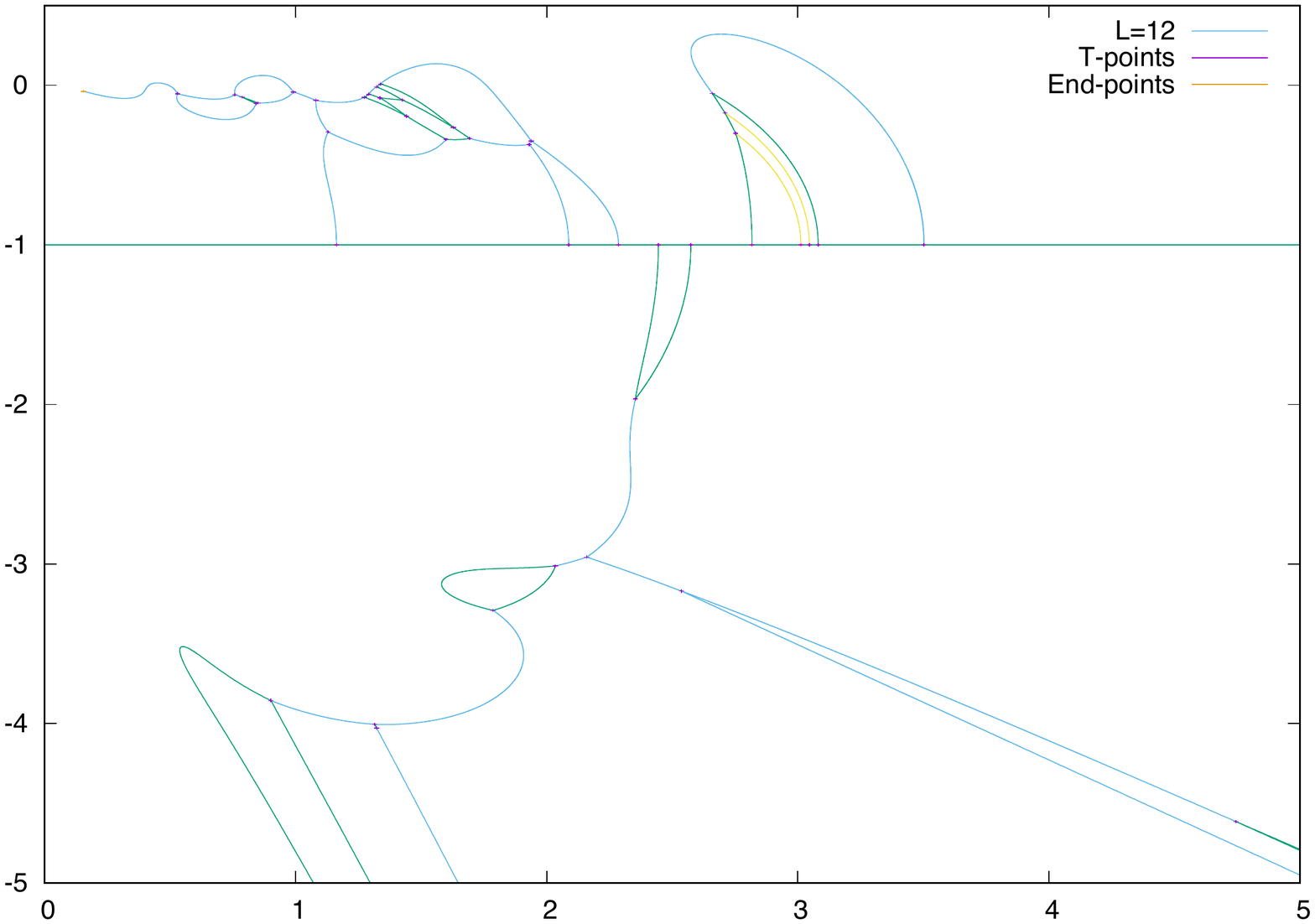}
\caption{Condensation curves for partition function zeros on a $(2M+1)\times 2N$ cylinder, in the limit
$M \to \infty$ (closed channel), for a system exhibiting only spin-zero symmetry. The panels show, in reading direction, the
cases $N=3,4,5,6$.}
\label{fig:cond-spin-zero}
\end{center}
\end{figure}

Figure~\ref{fig:cond-spin-zero} shows the results using only the spin-zero constraint. In the regions enclosed by curves of
blue color there is a unique dominant eigenvalue, whereas in the regions enclosed by green (resp.\ yellow) color the
dominant eigenvalue has multiplicity two (resp.\ three). Obviously the sought-after representation of dimension ${\cal N}(N)$
is expected to be multiplicity-free, so the corresponding condensation curve should be free of green and yellow branches.
Nevertheless, the curves in Figure~\ref{fig:cond-spin-zero} are expected to correctly produce the condensation curves of
partition function zeros for any system described by the transfer matrix $\tilde{\rT}_D(u)$ and with boundary states that
impose only the spin-zero symmetry, while breaking any other symmetry (e.g., by imposing spatially inhomogeneous weights).

\begin{figure}[h!]
\begin{center}
\includegraphics[scale=0.28]{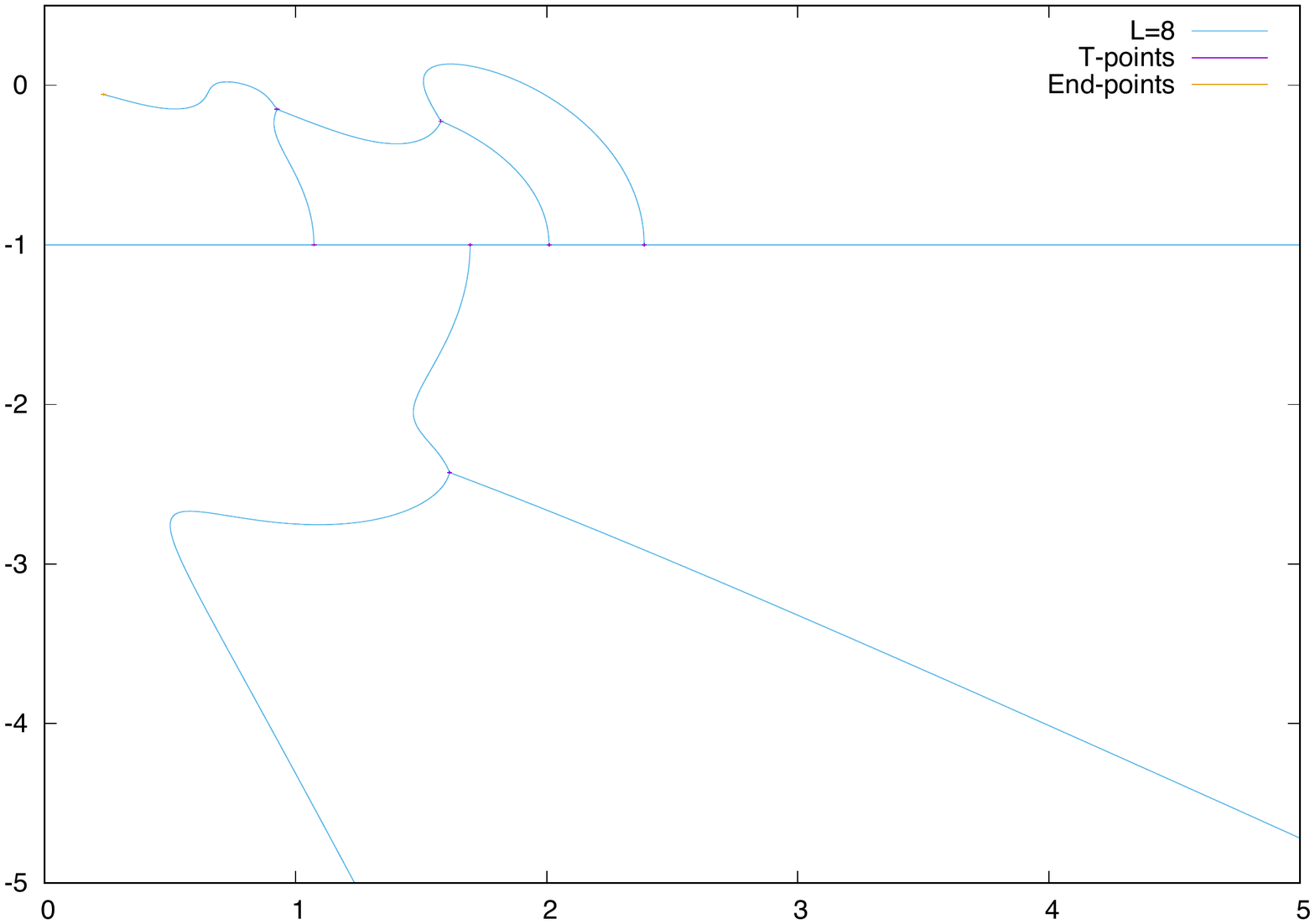}\includegraphics[scale=0.28]{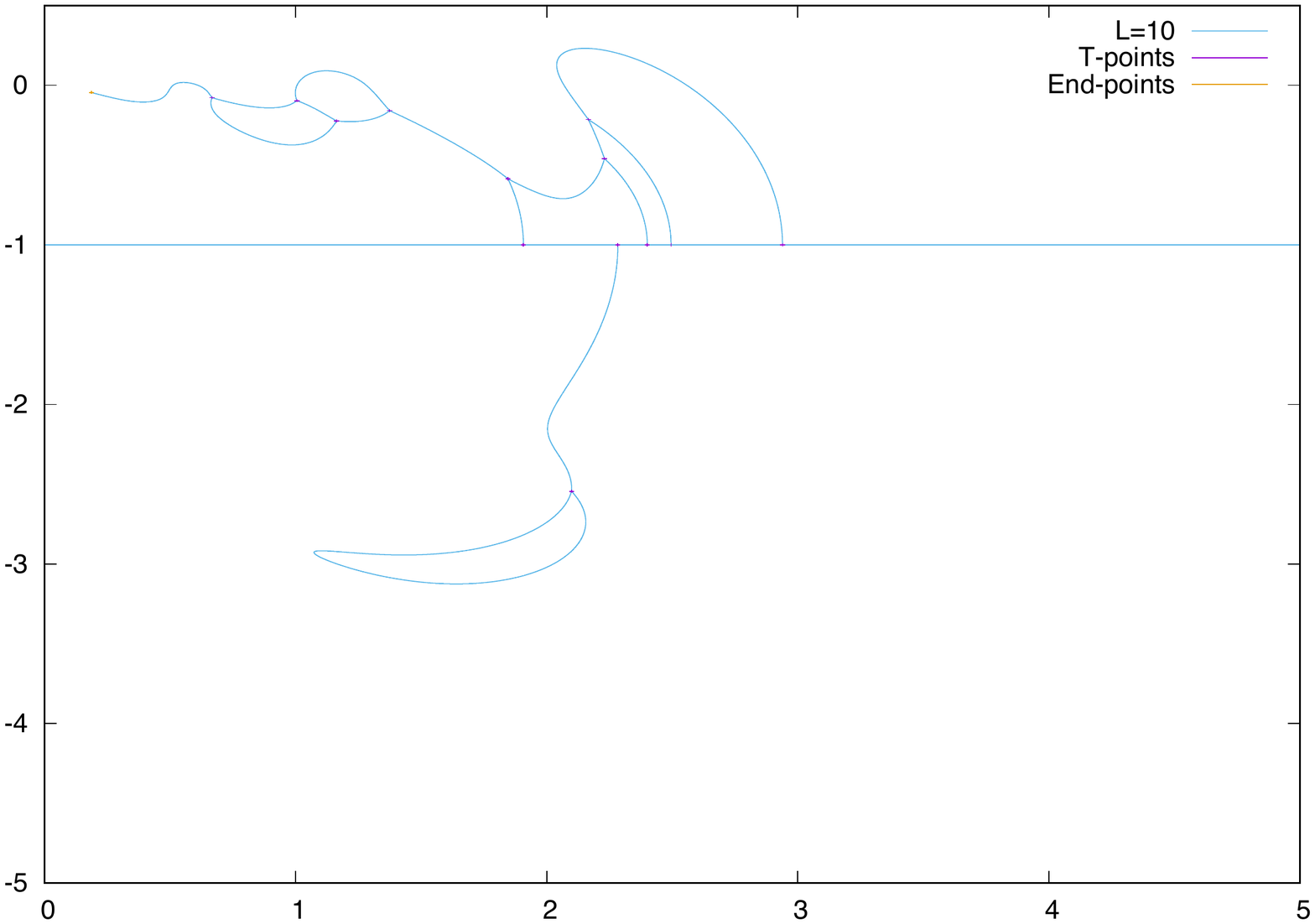} \\
\includegraphics[scale=0.28]{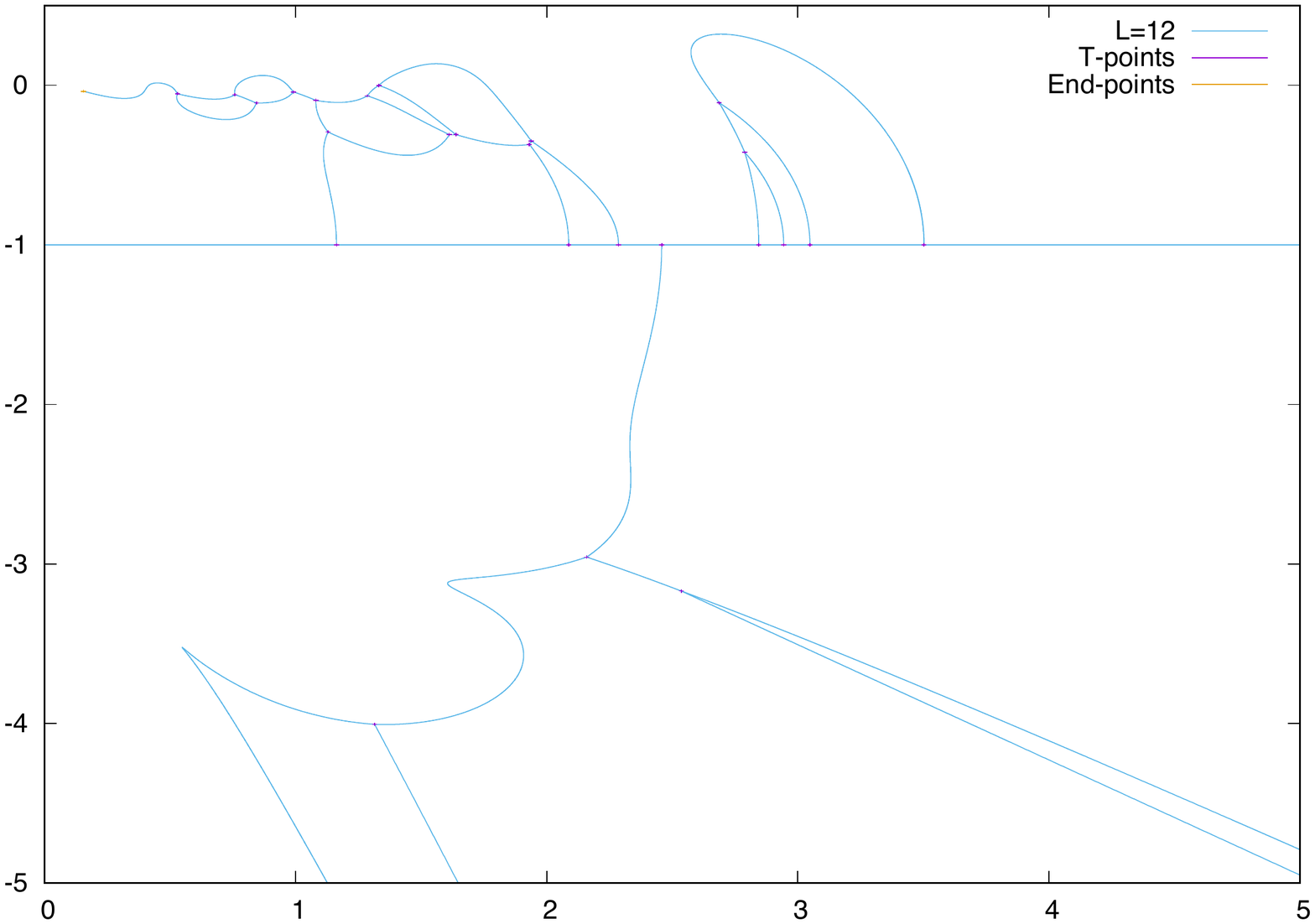}\includegraphics[scale=0.28]{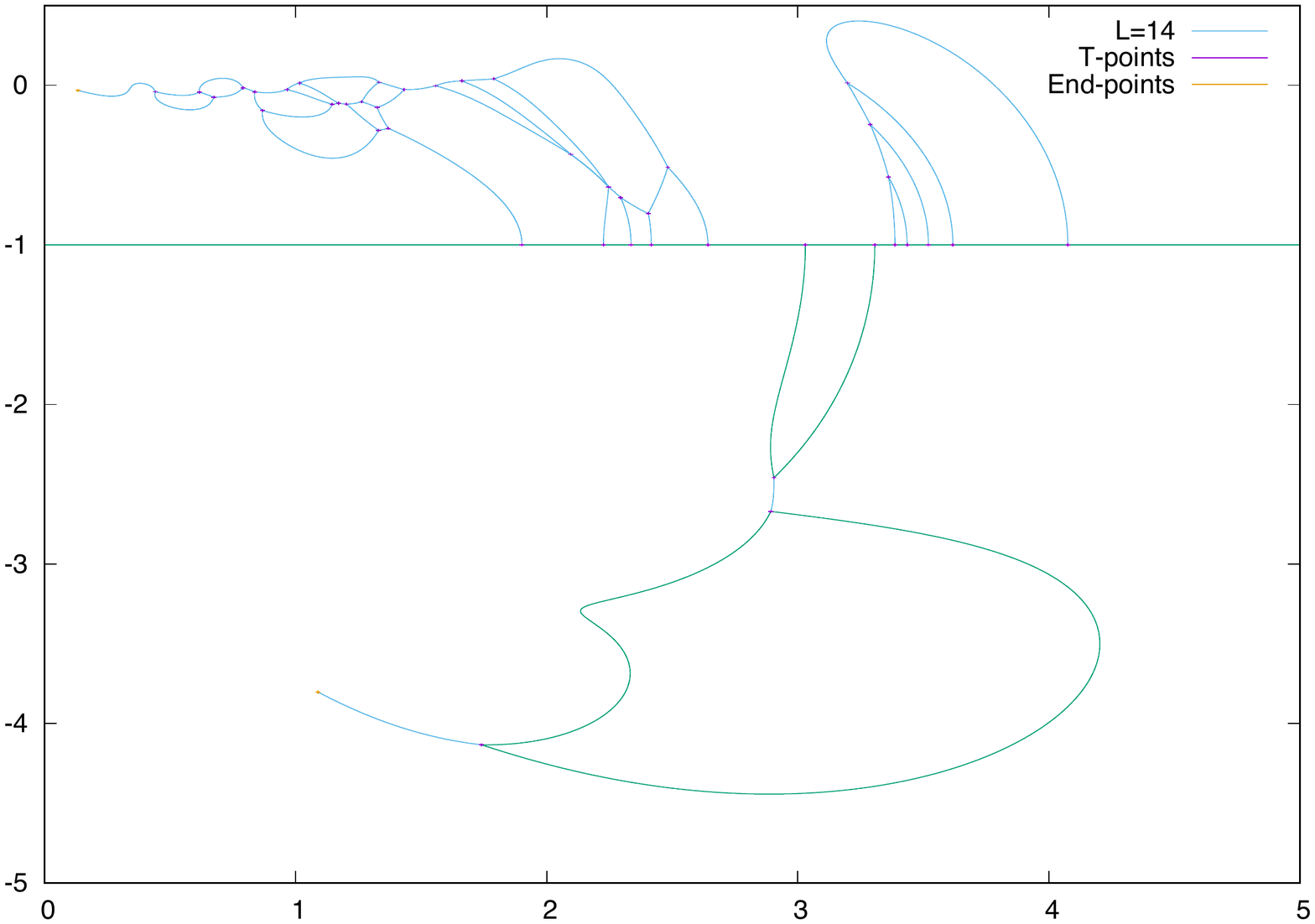}
\caption{Condensation curves for partition function zeros on a $(2M+1)\times 2N$ cylinder, in the limit
$M \to \infty$ (closed channel), for a system exhibiting spin-zero and cyclic symmetry $\mathbb{Z}_N$. The panels show, in reading direction, the
cases $N=4,5,6,7$.}
\label{fig:cond-cyclic}
\end{center}
\end{figure}

Next we show in Figure~\ref{fig:cond-cyclic} the results using both the spin-zero and the cyclic symmetry $\mathbb{Z}_N$. When compared
to Figure~\ref{fig:cond-spin-zero} it can be seen that many branches of the curves are unchanged. However all of the yellow and
some of the green curves have now disappeared, reflecting the fact that the eigenvalues which were formerly dominant inside the
regions enclosed by green and yellow colors have now been eliminated from the spectrum, since they do not correspond to
$\mathbb{Z}_N$ symmetric eigenstates. The curves in Figure~\ref{fig:cond-cyclic} should give the correct condensation curves
for systems having the spin-zero and cyclic symmetries. But those with $N=4,5$ should even provide the correct results for the
full ``Cooper-pair'' symmetry \eqref{paired}, since the TL dimension is then equal to the number of physical solutions.

\begin{figure}[h!]
\begin{center}
\includegraphics[scale=0.28]{12-cyclic.pdf}\includegraphics[scale=0.28]{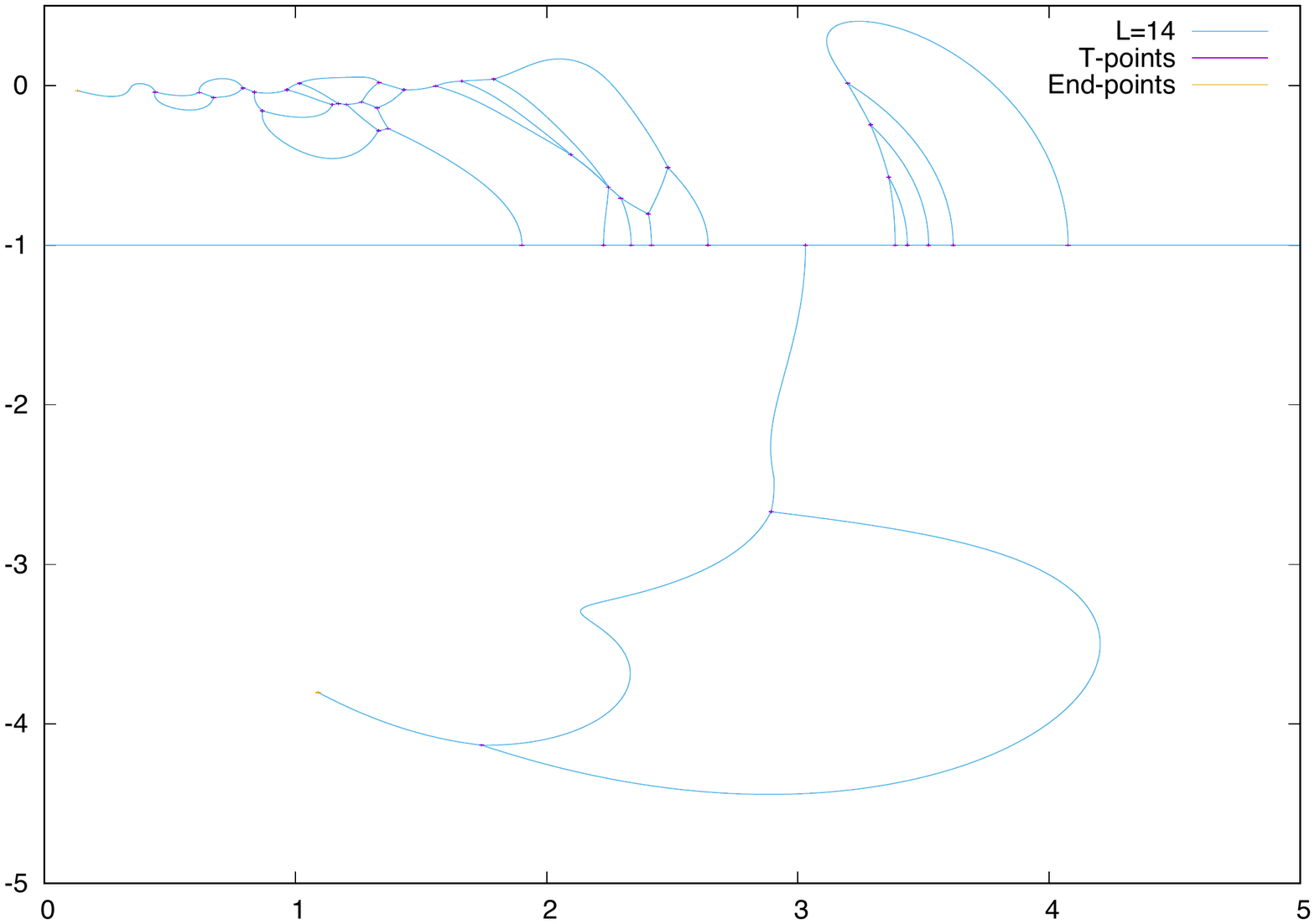}
\caption{Condensation curves for partition function zeros on a $(2M+1)\times 2N$ cylinder, in the limit
$M \to \infty$ (closed channel), for a system exhibiting spin-zero and dihedral symmetry $\mathbb{D}_N$. The panels show the
cases $N=6,7$.}
\label{fig:cond-dihedral}
\end{center}
\end{figure}

Finally we depict in Figure~\ref{fig:cond-dihedral} results using the spin-zero and dihedral symmetry $\mathbb{D}_N$. For $N=6$ the
condensation curve is identical to the one found with cyclic symmetry, meaning that none of the four eliminated eigenvalues
(when going from ${\rm dim}_{\mathbb{Z}_N}(6) = 28$ to ${\rm dim}_{\mathbb{D}_N}(6) = 24$) was dominant anywhere in the complex $u$-plane.
It should provide the correct result for the full ``Cooper-pair'' symmetry \eqref{paired} if the elimination of four more eigenvalues
(going from ${\rm dim}_{\mathbb{D}_N}(6) = 24$ to ${\cal N}(6) = 20$) turned out to be equally innocuous.
The $N=7$ curve with dihedral symmetry has only branches corresponding to multiplicity-free eigenvalues, so it may also apply
to the full paired symmetry, although a greater amount of eigenvalues are redundant in this case.

\begin{figure}[h!]
\begin{center}
\includegraphics[scale=0.28]{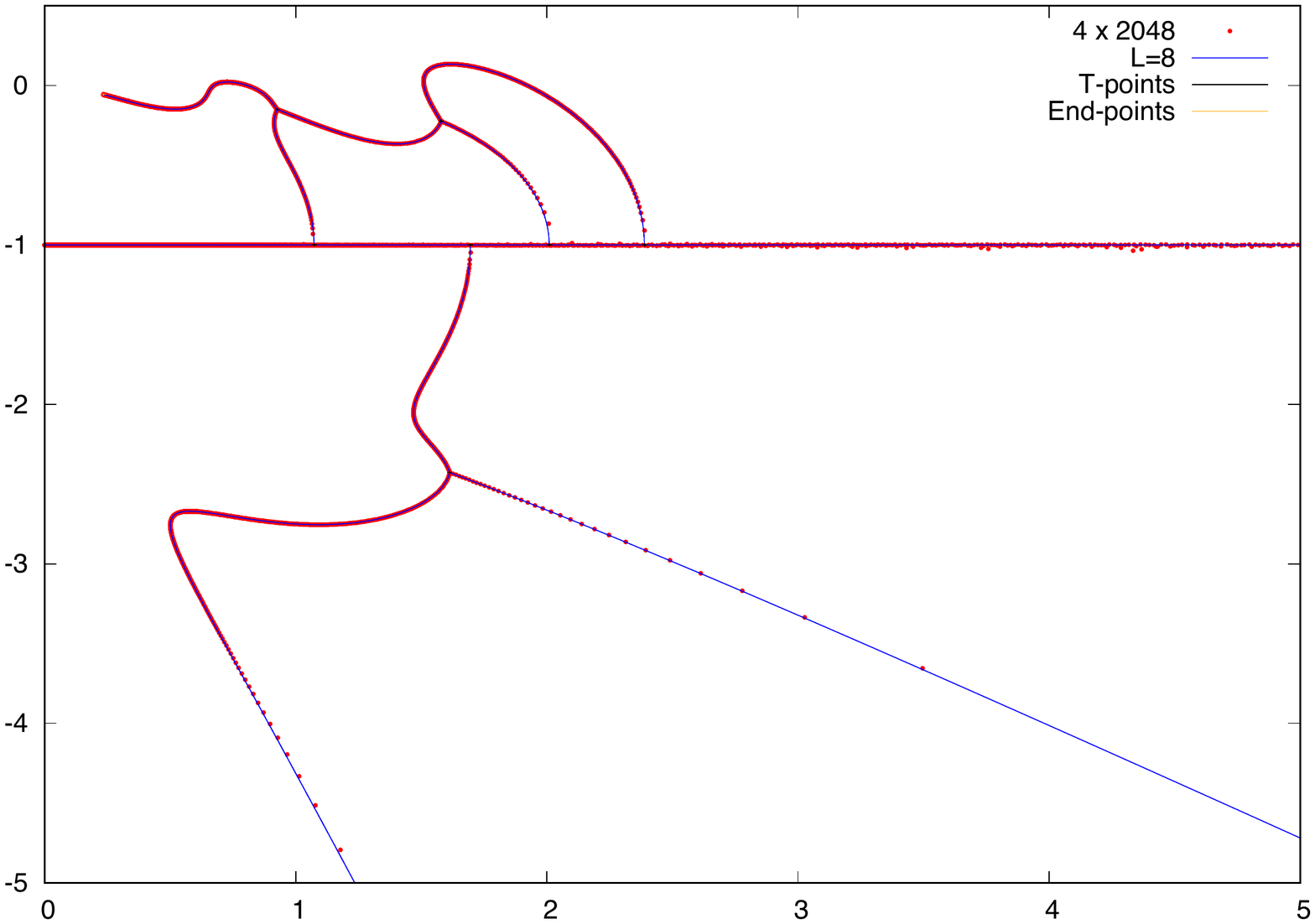}\includegraphics[scale=0.28]{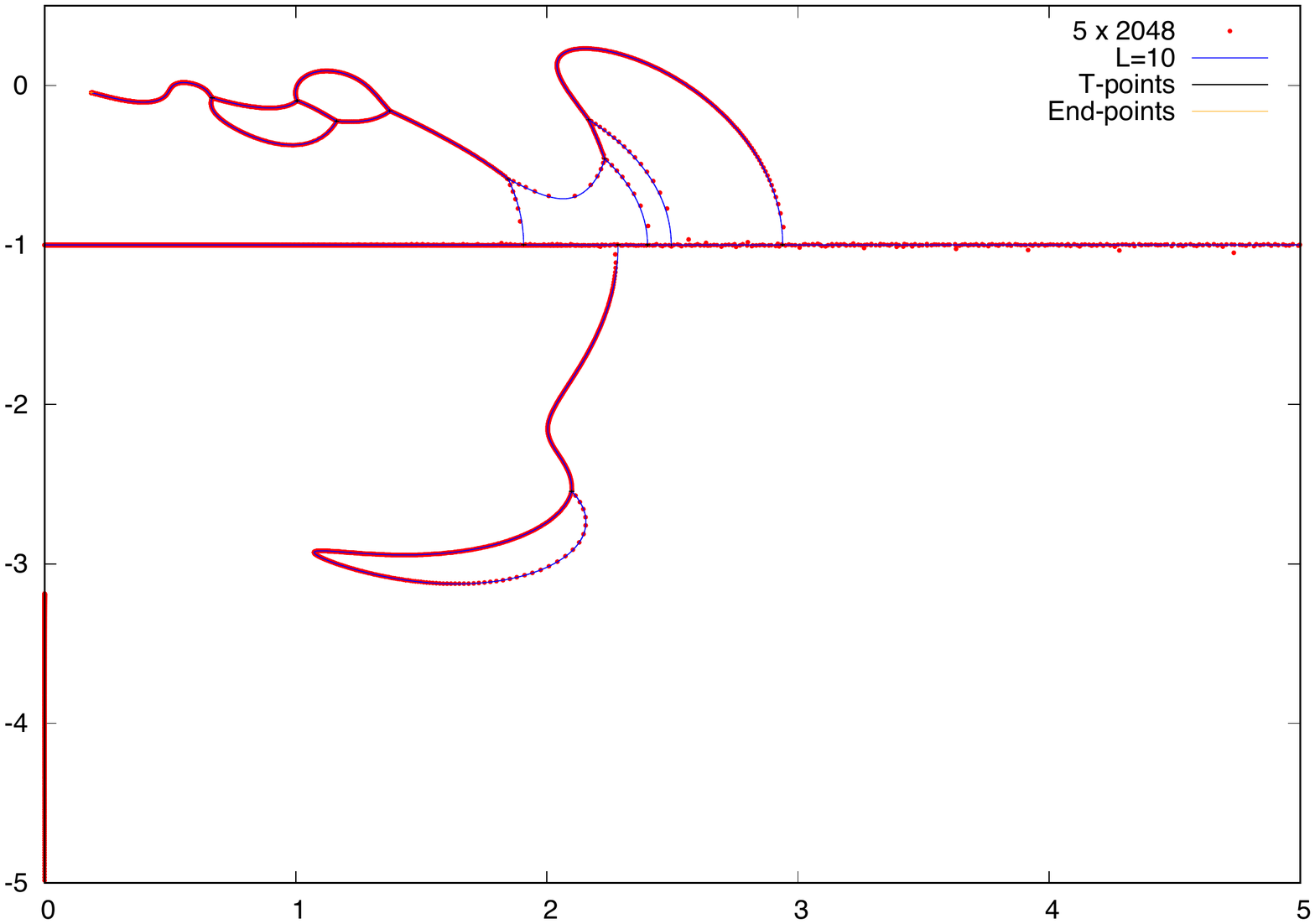} \\
\includegraphics[scale=0.28]{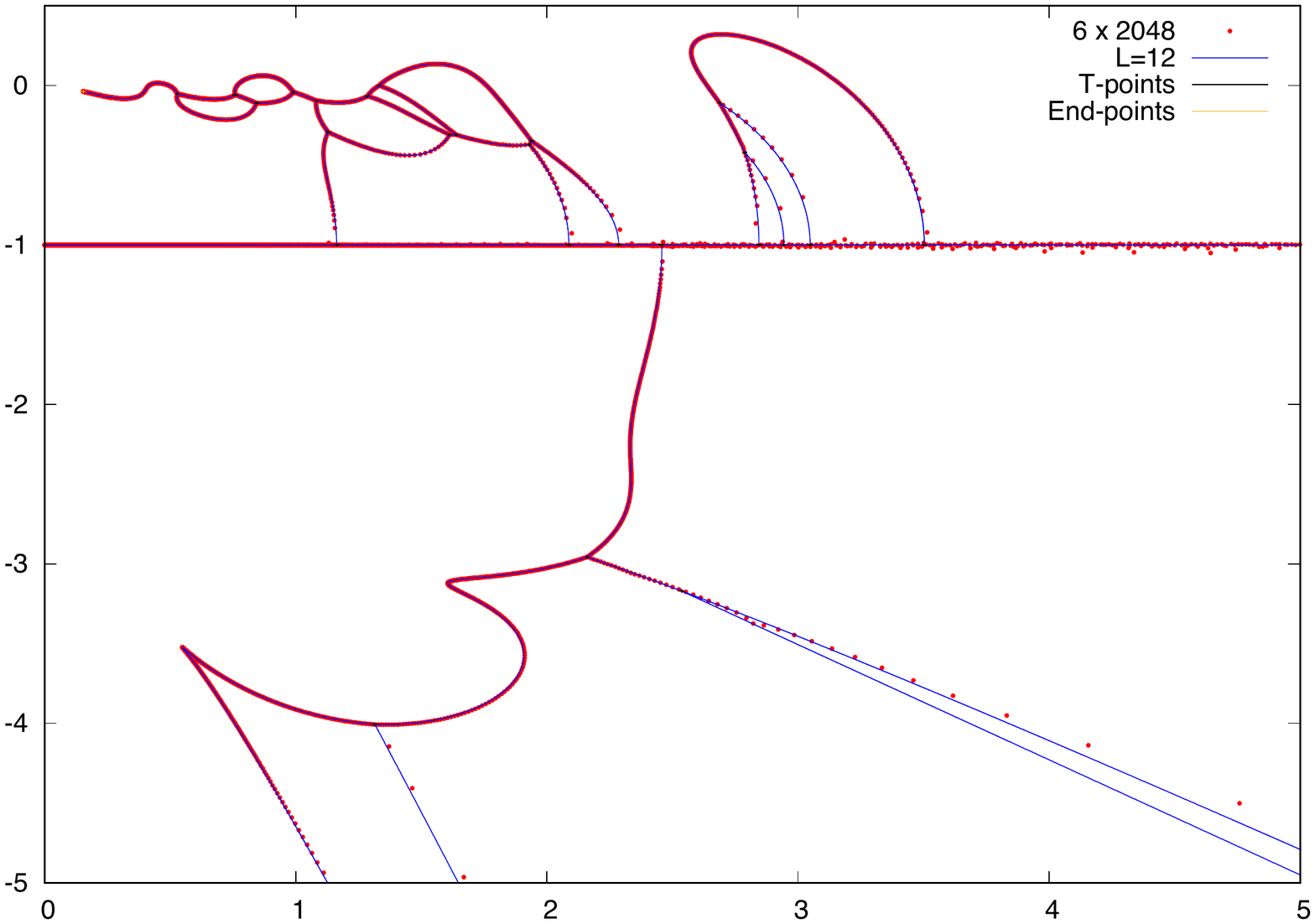}\includegraphics[scale=0.28]{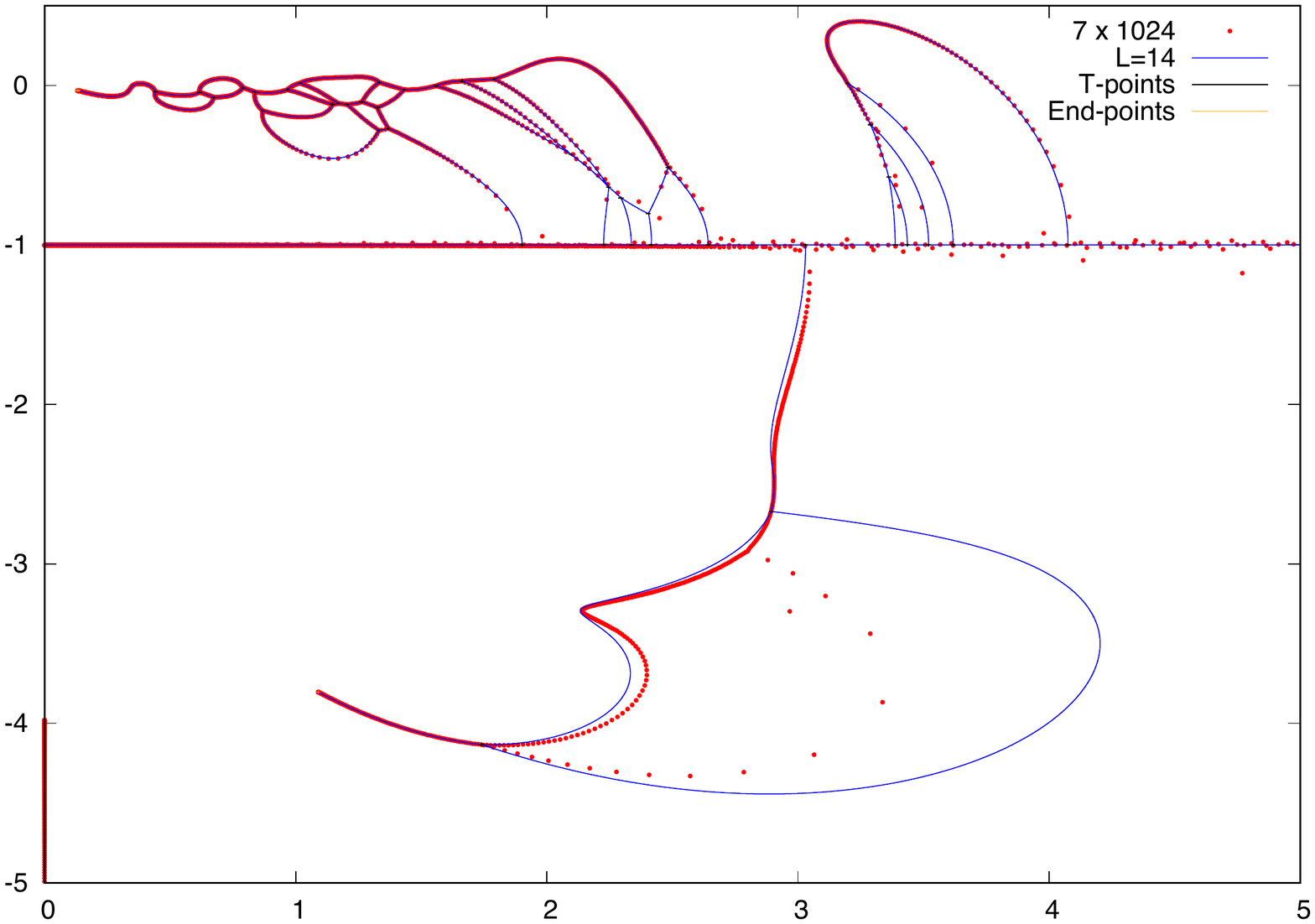}
\caption{Comparison between the partition function zeros on a $(2M+1)\times 2N$ cylinder, with $M=2048$,
and the corresponding condensation curves in the $M \to \infty$ limit (closed channel).
The panels show, in reading direction, the cases $N=4,5,6,7$.}
\label{fig:cond-closed-comp}
\end{center}
\end{figure}

All the curves contain an end-point $u_{\rm e}$ close to the origin for which we have found
the following results: \\[2mm]
 \begin{tabular}{l|rrrrrr}
 $N$ & 4 & 5 & 6 & 7 & 8 & 9 \\ \hline
 $\Re\, u_{\rm e}(N)$ &  0.234690 &  0.186435 &  0.154775 &  0.132364 &  0.115649 &  0.102696 \\
 $\Im\, u_{\rm e}(N)$ & -0.057271 & -0.045012 & -0.037154 & -0.031665 & -0.027604 & -0.024475 \\
 \end{tabular} \\[2mm]
It seems compelling from these data that
 \begin{equation}
  u_{\rm e}(N) \to 0 \mbox{ as } N \to \infty \,,
 \end{equation}
 with finite-size correction in both the real and imaginary parts proportional to $1/N$.

\medskip

To finish this section, we now compare the condensation curves with
the actual partition function zeros.  This is done for $N=4,5,6,7$ in
Figure~\ref{fig:cond-closed-comp}.  For the partition function zeros,
we have $M=2048$, except for $N=7$ where we have only $M=1024$; this
ensures an aspect ratio $\rho < 10^{-2}$ in all cases.  The agreement
with the condensation curves appears excellent, with the possible
exception of the bubble-shaped region with $-5 < \Im\, u < -2$ in the
$N=7$ case.

\section{Conclusions and discussions}
\label{sec:conlusion}
We have computed the \emph{exact} partition functions for the 6-vertex
model on \emph{intermediate} size lattices with periodic boundary
condition in one direction and free open boundary conditions in the other.
This work is a natural continuation of a previous work
\cite{Jacobsen:2018pjt} by three of the authors on the partition
function of the 6-vertex model where periodic boundary conditions were
imposed in both directions.  The presence of free open boundary
conditions brings new features and challenges.  To overcome these
challenges, we have further developed the application of algebro-geometric methods to the Bethe
ansatz equations in various directions.  We have incorporated recent developments in
integrability such as rational $Q$-systems for open spin chains
\cite{Bajnok:2019zub} and the exact formulae for overlaps between
integrable boundary states and Bethe states \cite{Pozsgay:2013,
Brockmann:2014a, Brockmann:2014b, Piroli:2017sei, Pozsgay:2018ybn}.
We have also developed powerful algorithms to perform the algebraic
geometry computations, such as the construction of Gr\"obner bases and companion matrices,
in the presence of a \emph{free parameter}.

Equipped with these new developments, we obtained the following exact results for the cylinder partition function $Z(u,M,N)$.
\begin{itemize}
\item \textbf{Open channel.} In the open channel, for $M=1$, we obtained a closed-form expression \eqref{eq:closedformM1} valid for any $N$. For $M=2,3,4,5,6$, we have computed the partition function for fixed $N$, both for small values and large values. For small values, we computed $N=2,3,4,5,6$. These results are given in appendix~\ref{eq:exactsmallMN}. For large values, we computed $N=128, 256, 512, 1024,2048$. These results were used to generate the zeros of partition functions in the partial thermodynamic limit. The exact results for large $N$ are not suitable to be put in the paper, so we have uploaded them as ancillary data files.
\item \textbf{Closed channel.} In the closed channel, for $N=1,2,3$, we obtained closed-form expressions valid for any $M$. The results are given in (\ref{eq:cfN1}), (\ref{eq:cfN2}) and (\ref{eq:cfN3}), respectively. For $N=4,5,6,7$ we have computed partition functions for fixed $M$, both for small and large values. For small values of $M$, we computed $M=2,3,4,5,6$ for $N=4,5,6$ because for these values we can compare the results in both channels and make non-trivial consistency checks. For large values of $M$, we computed $M=128,256,512,1024,2048$ for $M=4,5,6$ and $M=128,256,512,1024$ for $M=7$. The results for large values of $M$ have been used to obtain the zeros of the partition function in the closed channel and are uploaded as ancillary files.
\end{itemize}
We studied the partial thermodynamic limit of the partition function in both channels using the exact results. In particular, we computed the zeros of the partition functions in these limits and found that they condense on certain curves. The condensation curves in the partial thermodynamic limit can be found by a numerical approach based on the BKW theorem. This numerical approach has been applied in the torus case and was further developed in the current context by taking into account the new features, especially in the closed channel. Comparing the distribution of the zeros obtained from the exact partition function and the condensation curve obtained from the numerical approach, we found nice agreement and were able to shed light on several interesting features. The condensation curves in both the open and closed channels were found to involve very intricate features with multiple bifurcation points and enclosed regions. We believe that the further study of these curves might be of independent interest.\par

There are many other questions which deserve further investigation.

One of the most interesting directions is to compute the partition
function for the $q$-deformed case.  For generic values of $q$
(namely, when $q$ is not a root of unity), $Q$-systems for both closed
and open chains have been formulated in a recent work
\cite{Bajnok:2019zub}.  This should provide a good starting point for
developing an algebro-geometric approach, since $QQ$-relations are
more efficient than Bethe equations and give only physical solutions.
It would presumably be easier to first study the torus case where the relevant
Bethe equations are those of the periodic XXZ spin chain.  After that,
one could move to the more complicated cylinder case.

We have focused here on the cylinder geometry with free boundary
conditions. The case of fixed boundary conditions (with two arbitrary
boundary parameters) may now also be in
reach, using the new $Q$-system \cite{Nepomechie:2019gqt}.

From the perspective of algebro-geometric computations, it will be desirable to sharpen the computational power of
our method. For instance, we will try to apply the modern implements of Faug\`ere's F4 algorithm \cite{FAUGERE199961}, which is in general more efficient than Buchberger's algorithm.

\section*{Acknowledgements}
ZB and RN are grateful for the hospitality extended to them at the
University of Miami and the Wigner Research Center, respectively.  ZB
was supported in part by the NKFIH grant K116505. RN was supported
in part by a Cooper fellowship.
YZ thanks Janko Boehm for help on applied algebraic geometry.  YJ and
YZ acknowledge support from the NSF of China through Grant No.~11947301.
JLJ acknowledges support from the European Research Council through
the advanced grant NuQFT.

\appendix
%%%%%%%%%%%%%%%%%%%%%%%%%%%%%%%%%%%%%%%%%%%%%%%%%%%%%%%%%%%%%%
\section{Basic notions of computational algebraic geometry}
\label{app:AGbasic}
%%%%%%%%%%%%%%%%%%%%%%%%%%%%%%%%%%%%%%%%%%%%%%%%%%%%%%%%%%%%%%
In this appendix, we give a brief introduction to some basic notions of computational algebraic geometry which are used in the main text.

\subsection{Polynomial ring and ideal}
\label{sec:Ideal}
\paragraph{Polynomial ring} Let us start with the notion of \emph{polynomial ring} which is denoted by $A_K[z_1,\ldots,z_n]$ or $A_K$ for short. It is the set of all polynomials in $n$ variables $z_1,z_2,\ldots,z_n$ whose coefficients are in the field $K$. In our case, the field is often taken to be the set of complex numbers $\mathbb{C}$ or rational numbers $\mathbb{Q}$.\par

\paragraph{Ideal} An ideal $I$ of $A_K$ is a subset of $A_K$ such that
\begin{enumerate}
\item $f_1+f_2\in I$, if $f_1\in I$ and $f_2\in I$,
\item $g f\in I$, for $f\in I$ and $g\in A_K$.
\end{enumerate}
Importantly, any ideal $I$ of the polynomial ring $A_K$ is \emph{finitely generated}. This means, for any ideal $I$, there exists a finite number of polynomials $f_i\in I$ such that any polynomial $F\in I$ can be written as
\begin{align}
F=\sum_{i=1}^k f_i g_i,\qquad g_i\in A_K.
\end{align}
We can write $I=\langle f_1,f_2,\ldots,f_k\rangle$. Here the polynomials $\{f_k\}$ are called a \emph{basis} of the ideal.

\subsection{Gr\"obner basis}
As mentioned before, an ideal is generated by a set of basis $\{f_1,\ldots,f_k\}$. The choice of the basis is not unique. Namely, the same ideal can be generated by several different choices of bases
\begin{align}
I=\langle f_1,\ldots,f_k\rangle=\ldots=\langle g_1,\ldots,g_s\rangle.
\end{align}
Notice that in general $k$ does not have to be the same as $s$. For many cases, a convenient basis is needed. For solving polynomial equations and the polynomial reduction problem, it is most convenient to work with a \emph{Gr\"obner basis}.\par

We can introduce the Gr\"obner basis by considering \emph{polynomial reduction}. A polynomial reduction of a given polynomial $F$ over a set of polynomials $\{f_1,\ldots,f_k\}$ is given by
\begin{align}
F=\sum g_i f_i+r,\qquad g_i\in A_K
\end{align}
where $r$ is a polynomial that cannot be reduced further by any of the $f_i$. The polynomial $r$ is called the \emph{remainder} of the polynomial reduction. One important fact is that the polynomial reduction is \emph{not unique} for a generic basis $\{f_1,\ldots,f_k\}$. As a simple example, let us take $f_1=y^2-1$ and $f_2=xy-1$, and consider $F(x,y)=x^2 y+xy^2+y^2$. The polynomial reduction of $F(x,y)$ over $f_1$ and $f_2$ can be performed in two different ways
\begin{align}
F(x,y)=&\,(x+1)\,f_1+x\,f_2+(2x+1),\\\nonumber
F(x,y)=&\,f_1+(x+y)f_2+(x+y+1).
\end{align}
As we can see, the remainders are $r_1=2x+1$ and $r_2=x+y+1$ respectively. Therefore for a generic basis, the remainder of the polynomial is not well-defined. A basis of an ideal $\{ g_1,\ldots,g_s \}$ is said to be a Gr\"obner basis if the polynomial reduction is well-defined in the sense that the remainder is unique. Now we move to the more formal definition of the Gr\"obner basis.

\paragraph{Monomial ordering} To define a Gr\"obner basis, we first need to define monomial orders in the polynomial ring. A monomial order $\prec$ is specified by the following two rules
\begin{itemize}
\item If $u\prec v$ then for any monomial $w$, we have $u w\prec v w$.
\item If $u$ is non-constant monomial, then $1\prec u$.
\end{itemize}
Some commonly used orders are \texttt{lex} (Lexicographic), \texttt{deglex} (DegreeLexicographic) and \texttt{degrevlex} (DegreeReversedLexicographic).

\paragraph{Leading term} Once a monomial order $\prec$ is specified, for any polynomial $f$, we can define the \emph{leading term} uniquely. The leading term, which is denoted by LT($f$), is defined as the highest monomial of $f$ with respect to the monomial order $\prec$.

\paragraph{Gr\"obner basis} A Gr\"obner basis $G(I)$ of an ideal $I$ with respect to the monomial order $\prec$ is a basis of the ideal $\{g_1,\ldots,g_s\}$ such that for any $f\in I$, there exists a $g_i\in G(I)$ such that LT($f$) is divisible by LT($g_i$). A Gr\"obner basis for a given ideal with a monomial order can by computed by standard algorithms such as Buchberger algorithm \cite{Buchberger:1976:TBR:1088216.1088219} or the F4/F5 algorithms \cite{FAUGERE199961}. For an ideal $I$, given a monomial order $\prec$, the so-called minimal reduced Gr\"obner basis is unique.

\subsection{Quotient ring and companion matrix}
\paragraph{Quotient ring} Given an ideal $I$, we can define the \emph{quotient ring} $A_K/I$ by the equivalence relation: $f\sim g$ if and only if $f-g\in I$.\par

Polynomial reduction and Gr\"obner basis provide a canonical representation of the elements in the quotient ring $A_K/I$. Two polynomials $F_1$ and $F_2$ belong to the same element in the quotient ring $A_K/I$ if and only if their remainders of the polynomial reduction are the same. In particular, $f\in I$ if and only if its remainder of the polynomial reduction is zero. This gives a very efficient method to determine whether a polynomial $f$ is in the ideal $I$ or not.

\paragraph{Dimension of quotient ring} To our purpose, the dimension of the quotient ring is important. Consider a system of polynomial equations of $n$ variables
\begin{align}
\label{eq:systemff}
f_1(z_1,\ldots,z_n)=\cdots=f_k(z_1,\ldots,z_n)=0,
\end{align}
we can define the ideal and the quotient ring as
\begin{align}
I=\langle f_1,\ldots, f_k\rangle,\qquad Q_I=\mathbb{C}[z_1,\ldots,z_n]/\langle f_1,\ldots,f_k\rangle.
\end{align}
One crucial result is that the linear dimension of the quotient ring $\text{dim}_K Q_I$ equals the number of the solutions of the system (\ref{eq:systemff}). Therefore, if the number of solutions of the polynomial equations (\ref{eq:systemff}) is finite, $Q_I$ is a finite dimensional linear space. Let $G(I)$ be the Gr\"obner basis of the ideal $I$. The linear space is spanned by monomials which are not divisible by any elements in LT[$G(I)$].

\paragraph{Companion matrix} Another important notion for our applications is the \emph{companion matrix}. The main idea is that we can represent any polynomial $f\in A_K$ as a matrix in the quotient ring which is a finite dimensional linear space. More precisely, let $(m_1,\ldots, m_N)$ be the monomial basis of $A_K/I$, which can be constructed by the Gr\"obner basis $G(I)$. Given any polynomial, we can define an $N\times N$ matrix as follows
\begin{enumerate}
\item Multiply $f$ with one of the basis monomials $m_i$, perform the polynomial reduction with respect to the Gr\"obner basis $G(I)$ and find the remainder $r_i$. It is clear that $r_i$ sits in the quotient ring $A_K/I$.
\item Since $r_i$ is in the quotient ring, we can expand it in terms of the basis $(m_1,\ldots,m_N)$, namely $r_i=c_{ij}m_j$. In terms of formulas, we can write
    \begin{align}
    [f\times m_i]_{G(I)}=\sum_j c_{ij}\,m_j
    \end{align}
    where $[F]_{G(I)}$ means the remainder of the polynomial reduction of $F$ with respect to the Gr\"obner basis $G(I)$.
\item The companion matrix of $f$ is defined by
\begin{align}
(M_f)_{ij}=c_{ij}.
\end{align}
\end{enumerate}

\paragraph{Properties of companion matrix} Let us denote the companion matrix of the polynomials $f$ and $g$ by $M_f$ and $M_g$. It is clear that $M_f=M_g$ if and only if $[f]=[g]$ in $A_K/I$. Furthermore, we have the following properties
\begin{align}
M_{f+g}=M_f+M_g,\qquad M_{fg}=M_f M_g=M_g M_f.
\end{align}
If $M_g$ is an invertible matrix, we can actually define the companion matrix of the \emph{rational function} $f/g$ by
\begin{align}
M_{f/g}=M_f M_g^{-1}.
\end{align}
The companion matrix is a powerful tool for computing the sum over solutions of the polynomial system (\ref{eq:systemff}). As we mentioned before, the dimension of $Q_I$ equals the number of solutions of (\ref{eq:systemff}). Let us denote the $N$ solutions to be $(\vec{\xi}_1,\ldots,\vec{\xi}_N)$. Then we have the following important result
\begin{align}
\sum_{i=1}^N f(\vec{\xi}_i)=\text{Tr}\,M_f.
\end{align}

%%%%%%%%%%%%%%%%%%%%%%%%%%%%%%%%%%%%%%%%%%%%%%%%%%
\section{More details on AG computation}
\label{app:AGdetail}
%%%%%%%%%%%%%%%%%%%%%%%%%%%%%%%%%%%%%%%%%%%%%%%%%%

In this section, we summarize the algorithm of our algebra-geometry
based partition function computation for this paper.

\begin{itemize}
\item We first compute the Gr\"obner basis and quotient ring
linear basis of the TQ
relation equations \eqref{eq:openTQ} and QQ relation equations
\eqref{Q00sclosed}. Note that the TQ and QQ relations contain the free
parameter
$u$. It is possible to compute the corresponding Gr\"obner basis
analytically in $u$ via sophisticated computational algebraic-geometry
algorithms, like ``slimgb'' in the software {\sc Singular} \cite{DGPS}.

However, we
find that it is more efficient to set $u$ to some integral value,
and compute the Groebner basis. The computation is done with the standard
Gr\"obner basis command ``std'' in {\sc Singular}. In this approach, we maximize the power of
parallelization since the Groebner basis running time for different values
of $u$ is quite uniform.
\item Then we compute the power of companion matrices $(\mathbb
  T_{M,K})^N$. Although from the Gr\"obner basis it is
  straightforward to evaluate $(\mathbb
  T_{M,K})^N$, $\mathbb
  T_{M,K}$ is usually a dense matrix and the matrix product is a heavy
  computation. Instead, we postpone the matrix computations to the
  end, and evaluate the polynomial power $F^N$ first. Here $F$ is
  the corresponding polynomial of $\mathbb
  T_{M,K}$. After each polynomial multiplication step, we divide the
  polynomial by the ideal's Gr\"obner basis
  to save RAM usage by trimming high-degree
  terms. To speed up the
  computation, we apply the binary strategy, i.e., $F^N=F^{N/2}
  F^{N/2}$. After $F^N$ is calculated, a standard polynomial division
  computation provides the companion matrix power $(\mathbb
  T_{M,K})^N$. So the partition function for a particular $u$ value is
  obtained.
\item In previous steps, $u$ is set as an integral number. To get
  the analytic partition function in $u$, we have to repeat the
  computation, and then interpolate in $u$. We know that for
  both the closed and open channel partition functions, the maximum
  degree in $u$ is $2 M
  N$. Hence, we compute the partition function with $2MN+1$
  integer values. These results are then interpolated to the analytic partition
  function in $u$. The interpolation is carried out with the Newton
  polynomial method.
\end{itemize}

The whole computation is powered by our codes in {\sc Singular}. The
parallelization is implemented in the Gr\"obner basis and companion matrix
power steps, for different integer values of $u$'s. The interpolation step is
not parallelized, although it is also straightforward to do so in the
future.

We remark that through the computations, the coefficient field is
chosen to be
the rational number field $\mathbb Q$. We observe that the resulting
analytic partition function contains large-integer coefficients, so
finite-field techniques may not speed up the computation.

\section{The overlap (\ref{overlap})}\label{app:overlap}

The overlap (\ref{overlap}) can be deduced from results in
the paper by Pozsgay and R\'akos \cite{Pozsgay:2018ybn} (based on \cite{Pozsgay:2013,
Brockmann:2014a}), to which we refer here by PR. Their $R$-matrix is
given by PR (2.6), which has the same form as ours (\ref{Rmat}),
except with
\begin{equation}
	a(u)=\sinh(u+\eta)\,, \qquad b(u)=\sinh(u)\,, \qquad
	c(u)=\sinh(\eta)\,.
\end{equation}
Moreover, they work with the ``quantum monodromy matrix'' given by PR (2.33)
\begin{align}
T_{QTM}(u) &= R_{2N,0}(u-\eta+\omega)\, R_{2N-1,0}(u-\omega)\,
	\ldots R_{2,0}(u-\eta+\omega)\, R_{1,0}(u-\omega) \nonumber \\
&= \left( \begin{array}{cc}
	A(u) & B(u) \\
    C(u) & D(u) \\
 \end{array} \right) \,.
\end{align}	
By choosing
\begin{equation}
	\omega=\frac{\eta}{2} - u \,,
	\label{set1}
\end{equation}
scaling the variables as
\begin{equation}
	u \mapsto \epsilon\, u \,, \qquad  \eta \mapsto i \epsilon \,,
	\label{set2}
\end{equation}
and keeping the leading order in $\epsilon$, our shifted monodromy matrix
$\widehat{T}_{a}^{(2N)}(u-\frac{i}{2}; \{ \theta_{j}(u)\})$
(\ref{monodromy}) with alternating inhomogeneities (\ref{alt2})
can be obtained.
% Moreover, the transfer matrix eigenvalue PR (2.49) and the norm of the Bethe states PR (2.52)
% give (after suitable shifts) our
% (\ref{Lambdagenclosed}) and (\ref{Gaudin}), respectively.

In order to relate the generic boundary states PR (2.36)--(2.38)
to the dimer state, the boundary parameters in PR (2.15) can be chosen as follows
\begin{equation}
\alpha \rightarrow \infty \,, \qquad \beta=0 \,, \qquad \theta = 0 \,,
\end{equation}
so that the K-matrices are proportional to the identity matrix.
In this limit, the overlap PR (3.4) together with PR (3.3) gives our
overlap (\ref{overlap}).

\section{The relation (\ref{claim})}\label{app:proof}

We show here that the relation (\ref{claim}) follows from two simpler lemmata.

\paragraph{Lemma 1:}
\begin{align}
\langle \Phi_{0} | U^{\dagger} = \left(\frac{i}{2v-i}\right)^{N}
\langle \Phi_{0} | \tau(v; \{\theta_{j}(v)\}) \,.
\label{lemma1}
\end{align}
This lemma follows from the observation
\begin{align}
\tau(v; \{\theta_{j}(v)\}) = i^{N} R_{12}(2v)\, R_{34}(2v)\ldots
R_{2N-1, 2N}(2v)\, P_{2N-3, 2N-1}\, \ldots P_{3,5}\, P_{1,3} \,,
\end{align}
together with
\begin{align}
\langle \Phi_{0} | R_{12}(2v)\, R_{34}(2v)\ldots R_{2N-1, 2N}(2v) =
(2v-i)^{N} \langle \Phi_{0} | \,,
\end{align}
and
\begin{align}
\langle \Phi_{0} | P_{2N-3, 2N-1}\, \ldots P_{3,5}\, P_{1,3} =
(-1)^{N} \langle \Phi_{0} | U^{\dagger} \,.
\end{align}

Taking the scalar product of (\ref{lemma1}) with
transfer-matrix eigenvectors
$| \mathbf{u}  \rangle$ (which are constructed
using B-operators with alternating inhomogeneities
$\{\theta_{j}(v)\}$) and setting $v=\frac{\tilde{u}}{2}$,
we obtain
\begin{align}
\langle \Phi_{0} | U^{\dagger} | \mathbf{u}   \rangle
& = \left(\frac{i}{\tilde{u}-i}\right)^{N}
\langle \Phi_{0} |
\tau(\tfrac{\tilde{u}}{2}; \{\theta_{j}(\tfrac{\tilde{u}}{2})\}) |
\mathbf{u}  \rangle \nonumber \\
& = \left(\frac{i}{\tilde{u}-i}\right)^{N}
\Lambda_{\rm c}(\tfrac{\tilde{u}}{2};\{\theta_{j}(\tfrac{\tilde{u}}{2})\})
\langle \Phi_{0} |  \mathbf{u}  \rangle \,.
\label{result}
\end{align}

\paragraph{Lemma 2:} The following relation is valid off shell
\begin{align}
\langle \Phi_{0} | \mathbf{u}  \rangle =
\left(\frac{2v-i}{2v+i}\right)^{N} \langle \mathbf{u}
| \Phi_{0} \rangle \,.
\end{align}
See e.g. (3.3) in \cite{Pozsgay:2018ybn}, with (\ref{set1}) and
(\ref{set2}).

For our case, with $v=\frac{\tilde{u}}{2}$, we have
\begin{align}
\langle \Phi_{0} | \mathbf{u}  \rangle =
\left(\frac{\tilde{u}-i}{\tilde{u}+i}\right)^{N} \langle \mathbf{u}  | \Phi_{0} \rangle \,.
\label{lemma2}
\end{align}
Inserting (\ref{lemma2}) in the RHS of (\ref{result}), we obtain
\begin{align}
\langle \Phi_{0} | U^{\dagger} | \mathbf{u}  \rangle
= \left(\frac{i}{\tilde{u}+i}\right)^{N}
\Lambda_{\rm c}(\tfrac{\tilde{u}}{2};\{\theta_{j}(\tfrac{\tilde{u}}{2})\})
\langle \mathbf{u}  | \Phi_{0} \rangle \,,
\end{align}
which coincides with (\ref{claim}).

\section{Parity of states with paired Bethe roots}\label{app:parity}

Following \cite{Doikou:1998cz, Doikou:1998jh},
the parity operator $\Pi$ in the closed channel (length $2N$) is defined by
\begin{align}
	\Pi\, X_{n}\, \Pi^{-1} = X_{2N+1-n} \,,
\end{align}
where $X_{n}$ is any operator at site $n \in \{1, 2, \ldots, 2N\}$,
and is given by
\begin{align}
	\Pi = P_{1, 2N}\, P_{2, 2N-1} \ldots P_{N, N+1} \,,
\end{align}	
hence $\Pi = \Pi^{-1} = \Pi^{\dag}$. The parity operator has a simple
and beautiful action on the $B$-operator, namely
\begin{align}
\Pi\, B(u)\, \Pi = -B(-u) \,,
\label{nice}
\end{align}
while the reference state (\ref{refstate}) remains invariant under parity
\begin{align}
\Pi\, |0\rangle = |0\rangle  \,.
\label{parityref}
\end{align}
It follows from (\ref{nice}) and (\ref{parityref}) that Bethe states
(\ref{Bethestates}) corresponding to
the paired Bethe roots (\ref{paired}) (even $N$) are eigenstates of parity
with eigenvalue $+1$
\begin{align}
\Pi | u_{1} \,, -u_{1} \,, \ldots \,, u_{\frac{N}{2}} \,,
-u_{\frac{N}{2}} \rangle = | u_{1} \,, -u_{1} \,, \ldots \,, u_{\frac{N}{2}} \,,
-u_{\frac{N}{2}} \rangle \,.
\label{paritypairedevenN}
\end{align}
Similarly, Bethe states corresponding to
the paired Bethe roots (\ref{pairedodd}) (odd $N$) are eigenstates of parity
with eigenvalue $-1$
\begin{align}
\Pi | u_{1} \,, -u_{1} \,, \ldots \,, u_{\frac{N-1}{2}} \,,
-u_{\frac{N-1}{2}}\,,  0 \rangle = -| u_{1} \,, -u_{1} \,, \ldots \,,
u_{\frac{N-1}{2}} \,,
-u_{\frac{N-1}{2}}\,,  0\rangle \,.
\label{paritypairedoddN}
\end{align}

The dimer state $|\Phi_{0} \rangle$ (\ref{dimer}) is an eigenstate of
parity with eigenvalue $(-1)^{N}$
\begin{align}
	\Pi |\Phi_{0} \rangle = (-1)^{N} |\Phi_{0} \rangle \,,
\end{align}
which is consistent with the fact that the overlaps
$\langle \Phi_{0} | \mathbf{u} \rangle$
are nonzero only for Bethe states with paired
Bethe roots \eqref{paired}, \eqref{pairedodd}.

\section{Exact partition functions}
\label{eq:exactsmallMN}
In this appendix, we list all the exact partition functions for $2\le
M,N\le 6$ apart from the simple ones that have already been given in
the main text (\ref{Zexamples}).  For these values of $M$ and $N$, the partition
function can be computed in both channels.  As a consistency check,
computations in the two channels give the same result, as it should be.
\subsection{$N=6$}
\small{
\begin{align}
Z_{6,6}=&\,2 u^{72}+288 i u^{71}-14328 u^{70}-414000 i u^{69}+8317584 u^{68}+127125504 i u^{67}\\\nonumber
&\,-1559236944 u^{66}-15897457728 i u^{65}+138130609500u^{64}+1041934800608 i u^{63}\\\nonumber
&\,-6921377423424 u^{62}-40952806081344 i u^{61}+217832067312960 u^{60}+1049571874084608 i u^{59}\\\nonumber
&\,-4610142527559840 u^{58}-18559041008388480 i u^{57}+68788394401561470 u^{56}\\\nonumber
&\,+235662122430397008 i u^{55}-748761418763439296 u^{54}-2212828011779983200 i u^{53}\\\nonumber
&\,+6098355792769470156 u^{52}+15707717731534712832 i u^{51}-37888459151633760240 u^{50}\\\nonumber
&\,-85733900199121648320 i u^{49}+182272478003908656432 u^{48}+364591770627856882608 i u^{47}\\\nonumber
&\,-686967220132168504248 u^{46}-1220600244357970313168 i u^{45}+2047050659153089585764 u^{44}\\\nonumber
&\,+3243098453976454684320 i u^{43}-4857133424214480178560 u^{42}-6881082470938635503328 i u^{41}\\\nonumber
&\,+9226096556900912648334 u^{40}+11712585761982875838624 i u^{39}-14083587268444001339280 u^{38}\\\nonumber
&-16044127855808638420656 i u^{37}+17319816398521272862676 u^{36}+17719150885149621701664 i u^{35}\\\nonumber
&\,-17180252444447853950520 u^{34}-15786443250650483906112 i u^{33}+13745313212235461309790 u^{32}\\\nonumber
&\,+11338458117413798884752 i u^{31}-8858516711168371303152 u^{30}-6552673143433555176576 i u^{29}\\\nonumber
&\,+4587036228059022522828 u^{28}+3037167694581592157600 i u^{27}-1900876723794459666576 u^{26}\\\nonumber
&\,-1123749410112064351008 i u^{25}+626974446577452175698 u^{24}+329820882193355585184 i u^{23}\\\nonumber
&\,-163410941489958322920 u^{22}-76159342969170112944 i u^{21}+33342588484553082888 u^{20}\\\nonumber
&\,+13690734769961746560 i u^{19}-5263022902673709824 u^{18}-1890408224237932800 i u^{17}\\\nonumber
&\,+633003570392541120 u^{16}+197095658448168960 i u^{15}-56899838812276224 u^{14}\\\nonumber
&\,-15180285240013824 i u^{13}+3728675863226880 u^{12}+839585772859392 i u^{11}\\\nonumber
&\,-172442609104896 u^{10}-32118023696384 i u^9+5386833271296 u^8\\\nonumber
&\,+806617128960 i u^7-106666801152 u^6-12280172544 i u^5+1206835200 u^4\\\nonumber
&\,+98304000 i u^3-6340608 u^2-294912 i u+8192
\end{align}

\begin{align}
Z_{5,6}=&\,2 u^{60}+240 i u^{59}-10200 u^{58}-247760 i u^{57}+4106064 u^{56}+50944128 i u^{55}\\\nonumber
&\,-501290288 u^{54}-4069332864 i u^{53}+28034725284 u^{52}+167372075712 i u^{51}\\\nonumber
&-879568749504 u^{50}-4117046950656 i u^{49}+17320477694784 u^{48}+65956540414464 i u^{47}\\\nonumber
&\,-228627359471520 u^{46}-724727457569280 i u^{45}+2108994608155482 u^{44}+5652786955676832 i u^{43}\\\nonumber
&\,-13995089127459600 u^{42}-32084625043816128 i u^{41}+68261467147861284 u^{40}\\\nonumber
&\,+135034284911783616 i u^{39}-248788374633283104 u^{38}-427529707088463360 i u^{37}\\\nonumber
&\,+686116309138082856 u^{36}+1029410431077285360 i u^{35}-1445199354794151720 u^{34}\\\nonumber
&\,-1899914808354007664 i u^{33}+2340240191331772812 u^{32}+2702070900176216160 i u^{31}\\\nonumber
&\,-2925317019312145664 u^{30}-2970027518108113056 i u^{29}+2827960377154344960 u^{28}\\\nonumber
&\,+2525039773174991216 i u^{27}-2113704343901047104 u^{26}-1658217850739230320 i u^{25}\\\nonumber
&\,+1218539894552498448 u^{24}+838213811707493088 i u^{23}-539306661016030248 u^{22}\\\nonumber
&\,-324234959917029984 i u^{21}+181939121823624930 u^{20}+95158509487089840 i u^{19}\\\nonumber
&\,-46317104213610216 u^{18}-20941877029818960 i u^{17}+8777163928275864 u^{16}\\\nonumber
&\,+3401716413897600 i u^{15}-1215689053318848 u^{14}-399309726257280 i u^{13}+120092906906976 u^{12}\\\nonumber
&\,+32925815343360 i u^{11}-8187331182336 u^{10}-1835366381568 i u^9+368278769664 u^8+65577553920 i u^7\\\nonumber
&\,-10251309056 u^6-1387327488 i u^5+159490560 u^4+15134720 i u^3-1136640 u^2-61440 i u+2048
\end{align}

\begin{align}
Z_{4,6}=&\,8 u^{48}+336 i u^{47}-7728 u^{46}-123856 i u^{45}+1484064 u^{44}+13860000 i u^{43}\\\nonumber
&\,-104403144 u^{42}-652530240 i u^{41}+3461257458 u^{40}+15858754400 i u^{39}-63627652560 u^{38}\\\nonumber
&\,-225943164384 i u^{37}+716128871152 u^{36}+2039662782720 i u^{35}-5249124545904 u^{34}\\\nonumber
&\,-12261248086848 i u^{33}+26092620286092 u^{32}+50742484822368 i u^{31}-90406264918472 u^{30}\\\nonumber
&\,-147877208599872 i u^{29}+222439906309668 u^{28}+308118607243360 i u^{27}-393433500079944 u^{26}\\\nonumber
&\,-463456204858080 i u^{25}+503911508965434 u^{24}+505852778886192 i u^{23}-468842183109432 u^{22}\\\nonumber
&\,-401107676795728 i u^{21}+316601908242108 u^{20}+230384778687264 i u^{19}-154393257821456 u^{18}\\\nonumber
&\,-95159659506624 i u^{17}+53851823641602 u^{16}+27924428347584 i u^{15}-13235329354320 u^{14}\\\nonumber
&\,-5716979185248 i u^{13}+2242521937456 u^{12}+795411629952 i u^{11}-253808103360 u^{10}\\\nonumber
&\,-72409421696 i u^9+18332792928 u^8+4082245632 i u^7-790647936 u^6-131353344 i u^5+18380160 u^4\\\nonumber
&\,+2105344 i u^3-190464 u^2-12288 i u+512
\end{align}

\begin{align}
Z_{3,6}=&\,2 u^{36}+144 i u^{35}-3240 u^{34}-41760 i u^{33}+371556 u^{32}+2500704 i u^{31}\\\nonumber
&\,-13426440 u^{30}-59457600 i u^{29}+222050034 u^{28}+710461408 i u^{27}-1970740872 u^{26}\\\nonumber
&\,-4783539888 i u^{25}+10235477796 u^{24}+19420445664 i u^{23}-32825856240 u^{22}-49607005056 i u^{21}\\\nonumber
&\,+67208955660 u^{20}+81794924496 i u^{19}-89535239664 u^{18}-88204716144 i u^{17}+78198825096 u^{16}\\\nonumber
&\,+62348094624 i u^{15}-44646023160 u^{14}-28655915040 i u^{13}+16442440386 u^{12}+8405591184 i u^{11}\\\nonumber
&\,-3811764096 u^{10}-1524747232 i u^9+534186864 u^8+162405504 i u^7-42326208 u^6-9318528 i u^5\\\nonumber
&\,+1700640 u^4+249600 i u^3-28800 u^2-2304 i u+128
\end{align}

\begin{align}
Z_{2,6}=&\,2 u^{24}+96 i u^{23}-1488 u^{22}-12464 i u^{21}+67908 u^{20}+267360 i u^{19}\\\nonumber
&\,-818024 u^{18}-2043648 i u^{17}+4286358 u^{16}+7635536 i u^{15}-11600448 u^{14}-15070368 i u^{13}\\\nonumber
&\,+16767884 u^{12}+15970080 i u^{11}-12994512 u^{10}-9014240 i u^9+5311314 u^8+2637792 i u^7-1094360 u^6\\\nonumber
&\,-375504 i u^5+103512 u^4+22016 i u^3-3648 u^2-384 i u+32
\end{align}

%\begin{align}
%Z_{1,6}=&\,8 u^{12}+48 i u^{11}-240 u^{10}\\\nonumber
%&\,-880 i u^9+2016 u^8+3168 i u^7-3720 u^6-3168 i u^5+1890 u^4+880 i u^3-312 u^2-48 i u+8
%\end{align}

}

\subsection{$N=5$}
\small{
\begin{align}
Z_{6,5}=&\,2 u^{60}+240 i u^{59}-10140 u^{58}-248000 i u^{57}+4185570 u^{56}+53344848 i u^{55}\\\nonumber
&\,-542357480 u^{54}-4562838000 i u^{53}+32597238180 u^{52}+201557883000 i u^{51}-1094433249708 u^{50}\\\nonumber
&\,-5278042287600 i u^{49}+22813265943960 u^{48}+89030269264680 i u^{47}-315622898284860 u^{46}\\\nonumber
&\,-1021614345486960 i u^{45}+3032155944987750 u^{44}+8282078769490080 i u^{43}-20883263266224960 u^{42}\\\nonumber
&\,-48740210535148320 i u^{41}+105537643218985020 u^{40}+212434648891515000 i u^{39}\\\nonumber
&\,-398192299717983780 u^{38}-696081663569207040 i u^{37}+1136284932112092120 u^{36}\\\nonumber
&\,+1734011186676432768 i u^{35}-2476033587009228660 u^{34}-3310802034120343280 i u^{33}\\\nonumber
&\,+4148085867425363430 u^{32}+4871950812684428040 i u^{31}-5365880829686871812 u^{30}\\\nonumber
&\,-5543013568038622320 i u^{29}+5370865405192114860 u^{28}+4880969524493532440 i u^{27}\\\nonumber
&\,-4159536448109346180 u^{26}-3322896661947378912 i u^{25}+2487244973239722900 u^{24}\\\nonumber
&\,+1743350380048456200 i u^{23}-1143361265067321060 u^{22}-700989597010688640 i u^{21}\\\nonumber
&\,+401316693215927490 u^{20}+214260526736194800 i u^{19}-106515303269030220 u^{18}\\\nonumber
&\,-49218120113100480 i u^{17}+21095469547685670 u^{16}+8367142598539080 i u^{15}-3062728016717700 u^{14}\\\nonumber
&\,-1031401068624240 i u^{13}+318402566218770 u^{12}+89734056144240 i u^{11}-22976204056848 u^{10}\\\nonumber
&\,-5314648800960 i u^9+1103026376160 u^8+203688568320 i u^7-33111380480 u^6-4671793152 i u^5\\\nonumber
&\,+561016320 u^4+55685120 i u^3-4362240 u^2-245760 i u+8192
\end{align}

\begin{align}
Z_{5,5}=&\,2 u^{50}+200 i u^{49}-6800 u^{48}-133600 i u^{47}+1816720 u^{46}+18696280 i u^{45}\\\nonumber
&\,-153516300 u^{44}-1041341760 i u^{43}+5980862700 u^{42}+29622561920 i u^{41}-128320692620 u^{40}\\\nonumber
&\,-491601126800 i u^{39}+1680469080490 u^{38}+5162580577360 i u^{37}-14337030631180 u^{36}\\\nonumber
&\,-36164934863280 i u^{35}+83189622229650 u^{34}+175077243069600 i u^{33}-338033265929600 u^{32}\\\nonumber
&\,-600140879412320 i u^{31}+981612780573350 u^{30}+1481512519939600 i u^{29}-2065877195101300 u^{28}\\\nonumber
&\,-2664254653480000 i u^{27}+3180146358121680 u^{26}+3515139558674328 i u^{25}-3599034368401500 u^{24}\\\nonumber
&\,-3413530724251600 i u^{23}+2998653880187200 u^{22}+2438862766979800 i u^{21}-1835361722793380 u^{20}\\\nonumber
&\,-1276908650973200 i u^{19}+820389106281650 u^{18}+486066456229800 i u^{17}-265119363721020 u^{16}\\\nonumber
&\,-132847449311520 i u^{15}+61002271510150 u^{14}+25593377506000 i u^{13}-9776149068460 u^{12}\\\nonumber
&\,-3385803508640 i u^{11}+1057993826450 u^{10}+296565427400 i u^9-74059916520 u^8-16339593600 i u^7\\\nonumber
&\,+3151663680 u^6+524266240 i u^5-73811200 u^4-8550400 i u^3+780800 u^2+51200 i u-2048
\end{align}

\begin{align}
Z_{4,5}=&\,2 u^{40}+160 i u^{39}-4440 u^{38}-69280 i u^{37}+728000 u^{36}+5674392 i u^{35}\\\nonumber
&\,-34874340 u^{34}-176140320 i u^{33}+752228880 u^{32}+2769883720 i u^{31}-8912697344 u^{30}\\\nonumber
&\,-25297249440 i u^{29}+63775686230 u^{28}+143577146200 i u^{27}-289901804220 u^{26}\\\nonumber
&\,-526881584016 i u^{25}+864503010900 u^{24}+1283718138360 i u^{23}-1728451887060 u^{22}\\\nonumber
&\,-2113229980800 i u^{21}+2348251006860 u^{20}+2372768013000 i u^{19}-2180177662980 u^{18}\\\nonumber
&\,-1820882530080 i u^{17}+1381238400930 u^{16}+950393607264 i u^{15}-592143230220 u^{14}\\\nonumber
&\,-333299096320 i u^{13}+168978930950 u^{12}+76875351000 i u^{11}-31236020084 u^{10}\\\nonumber
&\,-11269764880 i u^9+3584938290 u^8+996761760 i u^7-239659680 u^6-49170048 i u^5+8454080 u^4\\\nonumber
&\,+1185280 i u^3-130560 u^2-10240 i u+512
\end{align}

\begin{align}
Z_{3,5}=&\,2 u^{30}+120 i u^{29}-2340 u^{28}-25120 i u^{27}+182970 u^{26}+1000824 i u^{25}\\\nonumber
&\,-4342200 u^{24}-15435600 i u^{23}+45931380 u^{22}+116188080 i u^{21}-252735000 u^{20}\\\nonumber
&\,-476824800 i u^{19}+785325960 u^{18}+1134545880 i u^{17}-1442491140 u^{16}-1617100080 i u^{15}\\\nonumber
&\,+1599383250 u^{14}+1395092760 i u^{13}-1072037820 u^{12}-724305120 i u^{11}+428924070 u^{10}\\\nonumber
&\,+221637600 i u^9-99319260 u^8-38260800 i u^7+12523890 u^6+3435384 i u^5-774840 u^4-139840 i u^3\\\nonumber
&\,+19680 u^2+1920 i u-128
\end{align}

\begin{align}
Z_{2,5}=&\,32 u^{20}+320 i u^{19}-1760 u^{18}-7680 i u^{17}+28560 u^{16}+87696 i u^{15}\\\nonumber
&\,-218880 u^{14}-446160 i u^{13}+747090 u^{12}+1031120 i u^{11}-1177868 u^{10}-1117760 i u^9+880230 u^8\\\nonumber
&\,+572520 i u^7-306660 u^6-134064 i u^5+46290 u^4+12240 i u^3-2480 u^2-320 i u+32
\end{align}

%\begin{flalign}
%Z_{1,5}=&\,2 u^{10}+40 i u^9-180 u^8-480 i u^7+870 u^6+1008 i u^5-780 u^4-480 i u^3+210 u^2+40 i u-8\\\nonumber
%&\,
%\end{flalign}

}

\subsection{$N=4$}
\begin{align}
Z_{6,4}=&\,2 u^{48}+192 i u^{47}-6432 u^{46}-125344 i u^{45}+1695768 u^{44}+17380704 i u^{43}\\\nonumber
&\,-142071744 u^{42}-957909984 i u^{41}+5456197416 u^{40}+26727665408 i u^{39}-114171998592 u^{38}\\\nonumber
&\,-430010518272 i u^{37}+1440779616064 u^{36}+4325994981888 i u^{35}-11709940237824 u^{34}\\\nonumber
&\,-28718431904256 i u^{33}+64075191370944 u^{32}+130502828494848 i u^{31}-243315793126400 u^{30}\\\nonumber
&\,-416238684174336 i u^{29}+654563752915968 u^{28}+947664700456960 i u^{27}-1264641410790912 u^{26}\\\nonumber
&\,-1556980053712896 i u^{25}+1769626440158208 u^{24}+1857530670661632 i u^{23}-1801001830855680 u^{22}\\\nonumber
&\,-1612788643440640 i u^{21}+1333448855969280 u^{20}+1017287720361984 i u^{19}-715462372579328 u^{18}\\\nonumber
&\,-463315439579136 i u^{17}+275828260637184 u^{16}+150674393284608 i u^{15}-75348310695936 u^{14}\\\nonumber
&\,-34399109013504 i u^{13}+14290811226112 u^{12}+5382179758080 i u^{11}-1829435990016 u^{10}\\\nonumber
&\,-558255153152 i u^9+151954421760 u^8+36599955456 i u^7-7720648704 u^6-1406828544 i u^5\\\nonumber
&\,+217178112 u^4+27590656 i u^3-2752512 u^2-196608 i u+8192
\end{align}

\begin{align}
Z_{5,4}=&\,2 u^{40}+160 i u^{39}-4512 u^{38}-72352 i u^{37}+789416 u^{36}+6439584 i u^{35}\\\nonumber
&\,-41590944 u^{34}-220803168 i u^{33}+988466328 u^{32}+3799026688 i u^{31}-12701096768 u^{30}\\\nonumber
&\,-37309082496 i u^{29}+97056390560 u^{28}+225028607488 i u^{27}-467409758976 u^{26}-873387327744 i u^{25}\\\nonumber
&\,+1472970252384 u^{24}+2247880822272 i u^{23}-3110257952256 u^{22}-3907451351040 i u^{21}\\\nonumber
&\,+4461712923648 u^{20}+4633176305664 i u^{19}-4376312765952 u^{18}-3759173240832 i u^{17}\\\nonumber
&\,+2934531772416 u^{16}+2079427166208 i u^{15}-1335273784320 u^{14}-775224616960 i u^{13}\\\nonumber
&\,+405738771968 u^{12}+190745063424 i u^{11}-80192927744 u^{10}-29991620608 i u^9+9914835456 u^8\\\nonumber
&\,+2874777600 i u^7-723861504 u^6-156229632 i u^5+28375040 u^4+4218880 i u^3\\\nonumber
&\,-491520 u^2-40960 i u+2048
\end{align}

\begin{align}
Z_{4,4}=&\,2 u^{32}+128 i u^{31}-2720 u^{30}-33088 i u^{29}+275920 u^{28}+1721920 i u^{27}\\\nonumber
&\,-8468640 u^{26}-33967296 i u^{25}+113839344 u^{24}+324510720 i u^{23}-797262912 u^{22}\\\nonumber
&\,-1704729984 i u^{21}+3195196320 u^{20}+5276663040 i u^{19}-7705800576 u^{18}-9976315392 i u^{17}\\\nonumber
&\,+11469719520 u^{16}+11721735168 i u^{15}-10651210752 u^{14}-8600710656 i u^{13}+6162514560 u^{12}\\\nonumber
&\,+3908175360 i u^{11}-2185631232 u^{10}-1072522752 i u^9+458877312 u^8+169820160 i u^7-53830656 u^6\\\nonumber
&\,-14436352 i u^5+3218944 u^4+581632 i u^3-81920 u^2-8192 i u+512
\end{align}

\begin{align}
Z_{3,4}=&\,2 u^{24}+96 i u^{23}-1536 u^{22}-13376 i u^{21}+76656 u^{20}+321984 i u^{19}\\\nonumber
&\,-1060064 u^{18}-2847552 i u^{17}+6375696 u^{16}+12024320 i u^{15}-19217088 u^{14}-26146944 i u^{13}\\\nonumber
&\,+30396896 u^{12}+30257664 i u^{11}-25802112 u^{10}-18824192 i u^9+11701920 u^8+6151680 i u^7\\\nonumber
&\,-2706944 u^6-984576 i u^5+290688 u^4+68096 i u^3-12288 u^2-1536 i u+128
\end{align}

\begin{align}
Z_{2,4}=&\,2 u^{16}+64 i u^{15}-576 u^{14}-2912 i u^{13}+9928 u^{12}+24864 i u^{11}\\\nonumber
&\,-47840 u^{10}-71968 i u^9+85176 u^8+80128 i u^7-60608 u^6-36480 i u^5+16864 u^4+5888 i u^3\\\nonumber
&\,-1536 u^2-256 i u+32
\end{align}

\subsection{$N=3$}

\begin{align}
Z_{6,3}=&\,2 u^{36}+144 i u^{35}-3672 u^{34}-54720 i u^{33}+566172 u^{32}+4421808 i u^{31}\\\nonumber
&\,-27359952 u^{30}-138421152 i u^{29}+585564318 u^{28}+2106246952 i u^{27}-6526250244 u^{26}\\\nonumber
&\,-17601308640 i u^{25}+41667849936 u^{24}+87178071096 i u^{23}-162100318500 u^{22}\\\nonumber
&\,-269073694512 i u^{21}+400109679918 u^{20}+534347671248 i u^{19}-642039336300 u^{18}\\\nonumber
&\,-694728826992 i u^{17}+677171433726 u^{16}+594340399176 i u^{15}-469220702148 u^{14}\\\nonumber
&\,-332676139248 i u^{13}+211359220146 u^{12}+119998562832 i u^{11}-60673432464 u^{10}\\\nonumber
&\,-27204293440 i u^9+10758322656 u^8+3726162432 i u^7-1119674880 u^6-288110592 i u^5+62295552 u^4\\\nonumber
&\,+10997760 i u^3-1511424 u^2-147456 i u+8192
\end{align}

\begin{align}
Z_{5,3}=&\,2 u^{30}+120 i u^{29}-2544 u^{28}-30784 i u^{27}+254340 u^{26}+1570464 i u^{25}\\\nonumber
&\,-7633440 u^{24}-30170016 i u^{23}+99104634 u^{22}+274915920 i u^{21}-652019508 u^{20}\\\nonumber
&\,-1335239136 i u^{19}+2379520968 u^{18}+3712304088 i u^{17}-5091867900 u^{16}-6157254672 i u^{15}\\\nonumber
&\,+6573306816 u^{14}+6196463352 i u^{13}-5152840800 u^{12}-3772407312 i u^{11}+2424284226 u^{10}\\\nonumber
&\,+1362333816 i u^9-666211176 u^8-281734272 i u^7+102169152 u^6+31401216 i u^5-8038656 u^4\\\nonumber
&\,-1669120 i u^3+268800 u^2+30720 i u-2048
\end{align}

\begin{align}
Z_{4,3}=&\,8 u^{24}+168 i u^{23}-1920 u^{22}-15344 i u^{21}+91506 u^{20}+418992 i u^{19}\\\nonumber
&\,-1508768 u^{18}-4370976 i u^{17}+10394316 u^{16}+20622512 i u^{15}-34552644 u^{14}\\\nonumber
&\,-49294128 i u^{13}+60175370 u^{12}+62991480 i u^{11}-56528844 u^{10}-43384016 i u^9+28353042 u^8\\\nonumber
&\,+15688800 i u^7-7297568 u^6-2825088 i u^5+896448 u^4+227840 i u^3-44544 u^2-6144 i u+512
\end{align}

\begin{align}
Z_{3,3}=&\,2 u^{18}+72 i u^{17}-828 u^{16}-5328 i u^{15}+22914 u^{14}+71568 i u^{13}-170652 u^{12}\\\nonumber
&\,-321264 i u^{11}+488124 u^{10}+605480 i u^9-614664 u^8-509040 i u^7+341826 u^6+184392 i u^5-78840 u^4\\\nonumber
&\,-26304 i u^3+6624 u^2+1152 i u-128
\end{align}

\begin{align}
Z_{2,3}=&\,2 u^{12}+48 i u^{11}-348 u^{10}\\\nonumber
&\,-1264 i u^9+2934 u^8+4968 i u^7-6420 u^6-6192 i u^5+4338 u^4+2224 i u^3-816 u^2-192 i u+32
\end{align}

\subsection{$N=2$}
\begin{align}
Z_{6,2}=&\,2 u^{24}+96 i u^{23}-1704 u^{22}-17936 i u^{21}+131208 u^{20}+717696 i u^{19}\\\nonumber
&\,-3057152 u^{18}-10408704 i u^{17}+28851648 u^{16}+66013184 i u^{15}-126014976 u^{14}\\\nonumber
&\,-202398720 i u^{13}+275340800 u^{12}+318861312 i u^{11}-315426816 u^{10}-266989568 i u^9+193300992 u^8\\\nonumber
&\,+119365632 i u^7-62486528 u^6-27439104 i u^5+9934848 u^4+2883584 i u^3-638976 u^2-98304 i u+8192
\end{align}

\begin{align}
Z_{5,2}=&\,2 u^{20}+80 i u^{19}-1160 u^{18}-9840 i u^{17}+57240 u^{16}+245376 i u^{15}\\\nonumber
&\,-806208 u^{14}-2081664 i u^{13}+4300128 u^{12}+7203584 i u^{11}-9888512 u^{10}-11208704 i u^9\\\nonumber
&\,+10543104 u^8+8245248 i u^7-5351424 u^6-2863104 i u^5+1245696 u^4+430080 i u^3-112640 u^2\\\nonumber
&\,-20480 i u+2048
\end{align}

\begin{align}
Z_{4,2}=&\,2 u^{16}+64 i u^{15}-720 u^{14}-4640 i u^{13}+20080 u^{12}+62592 i u^{11}\\\nonumber
&\,-145856 u^{10}-260224 i u^9+361824 u^8+397312 i u^7-347264 u^6-241920 i u^5+133504 u^4+57344 i u^3\\\nonumber
&\,-18432 u^2-4096 i u+512
\end{align}

\begin{align}
Z_{3,2}=&\,2 u^{12}+48 i u^{11}-384 u^{10}-1696 i u^9\\\nonumber
&\,+4848 u^8+9600 i u^7-13632 u^6-14208 i u^5+11040 u^4+6400 i u^3-2688 u^2-768 i u+128
\end{align}

%\bibliographystyle{utphys}
%\bibliography{refs}

\providecommand{\href}[2]{#2}\begingroup\raggedright\endgroup

\end{document}